\documentclass[11pt]{elsarticle}

\usepackage{lineno,hyperref}
\modulolinenumbers[5]
\journal{Journal of XXX}
\usepackage{amsmath,accents,amssymb,amsthm,amsfonts,afterpage,float,bm,bbm,stmaryrd,paralist,wrapfig,bbm,soul}

\usepackage{cancel}
\usepackage{color}
\usepackage{fancyhdr}
\usepackage{graphics}
\usepackage{hyperref}
\usepackage{graphicx,epsf}
\usepackage{makeidx}
\usepackage{epsfig}
\usepackage{lscape}
\usepackage{soul,ulem}

\usepackage{empheq,xcolor}

\pssilent

\newtheorem{remark}{Remark}[section]

\numberwithin{equation}{section}
\numberwithin{figure}{section}
\numberwithin{table}{section}

\def\XXint#1#2#3{{\setbox0=\hbox{$#1{#2#3}{\int}$}
\vcenter{\hbox{$#2#3$}}\kern-.51\wd0}}

\newcommand{\reff}[1]{{\rm (\ref{#1})}}

\graphicspath{{Figure/}}

\begin{document}

\setlength{\pdfpageheight}{\paperheight}
\setlength{\pdfpagewidth}{\paperwidth}
\title{Ohta–Kawasaki Model Reveals Patterns on Multicomponent Vesicles}
% \author{Wangbo Luo, Zhonghua Qiao}
% \address{Department of Applied Mathematics, The Hong Kong Polytechnic University, Hung Hom, Kowloon,
% Hong Kong}
% \author{Yanxiang Zhao}
% \address{Department of Mathematics, George Washington University, Washington D.C., 20052}

\author{Wangbo Luo, Zhonghua Qiao\fnref{cor1}}
\address{Department of Applied Mathematics, Hong Kong Polytechnic University, Hung Hom, Kowloon, Hong Kong}

\author{Yanxiang Zhao\fnref{cor2}}
\address{Department of Mathematics, George Washington University, Washington D.C., 20052}

\fntext[cor1]{Corresponding author. Email: zhonghua.qiao@polyu.edu.hk}
\fntext[cor2]{Corresponding author. Email: yxzhao@gwu.edu}

\begin{abstract}

We present a new mechanochemical modeling framework to explore the shape deformation and pattern formation in multicomponent vesicle membranes. In this framework, the shape of the membrane is described by an elastic bending model, while phase separation of membrane-bound activator proteins is determined by an Ohta-Kawasaki (OK) model. The coupled dynamics consist of an overdamped force-balanced equation for the membrane geometry and an OK-type advection-reaction-diffusion equation on the deformable membrane. We implement efficient spectral methods to simulate these dynamics in both two- and three-dimensions. Numerical experiments show that the model successfully reproduces a wide range of experimentally observed membrane morphologies \cite{baumgart2003imaging}. Taken together, the framework unifies curvature mechanics, microphase separation, and active forcing, providing new insight into membrane-bounded multicomponent vesicle dynamics and a practical platform for studying multicomponent biomembrane morphology.

\end{abstract}


\date{\today}
\maketitle

\section{Introduction}\label{Intro}

\subsection{Biological and Modeling Background}

Vesicle membranes are lipid bilayers that self-assemble from amphipathic lipids and incorporate specific membrane proteins, acting as a barrier to separate the vesicle's contents from the surrounding cellular environment. These membranes are critical for transporting cargo within cells, facilitating communication between cells, and even playing roles in disease progression. The membrane can bud off from other cellular membranes \cite{kas1991shape,lipowsky1992budding,lipowsky1995morphology,seifert1993curvature}, merge with them to release contents \cite{dobereiner1993budding}, or be derived from bacteria or viruses. 

Over the past decades, the Helfrich-based elastic bending energy model has become one of the most successful theoretical frameworks to model lipid bilayer membranes, which is in the form of a surface integral of different curvature terms on the membrane \cite{seifert1993curvature,helfrich1973elastic}:
\begin{align}
    E = \int_{\Gamma}\left[a_1+a_2(H-c_0)^2+a_3G\right]\mathrm{d} s,
\end{align}
where $a_1$ denotes the surface tension, $H$ is the mean curvature, $G$ is the Gaussian curvature, $a_2$ is the bending rigidity, and $a_3$ is the stretching rigidity. $c_0$ denotes the spontaneous mean curvature. It reflects up-down bilayer asymmetry arising from leaflet composition or area difference, curvature-inducing proteins, or asymmetric environmental conditions, and thus sets the membrane’s preferred curvature. This model is closely related to the Willmore energy and Willmore flow \cite{muller2014confined} and provides a natural framework for determining the equilibrium shape of membranes by minimizing the elastic bending energy \cite{ciarlet2021mathematical,elliott2017small}. 

Du et al. \cite{du2004phase} proposed a variational phase field formulation of the elastic bending energy to study equilibrium configurations of the vesicle membranes. Later, they studied the effects of the spontaneous curvature $c_0$ on the deformation of the vesicle membrane under this framework \cite{du2005modeling} and extended the approach to three-dimensional simulations \cite{du2006simulating}. Its ability to capture interfacial dynamics makes this formulation suitable to study complex cellular processes. Recently, this formulation has become a critical component for modeling biophysical processes such as cell migration \cite{shao2010computational,shao2012coupling,camley2013periodic} and cell chemotaxis \cite{zhang2025phase}. A detailed description of the phase field formulation is provided in Sections \ref{subsec:phase field} and \ref{subsec:deform_main}.

Meanwhile, experiments on giant unilamellar vesicles (GUV) \cite{dietrich2001lipid,veatch2002organization} show that GUV membranes can demix into coexisting liquid-ordered and liquid-disordered phases that form compositionally distinct domains, thus forming multicomponent membranes. The formation of the multicomponent vesicles is primarily driven by thermodynamic lipid–lipid interactions, line tension at domain boundaries, and curvature-induced lipid sorting, with additional regulation from protein–lipid coupling and cytoskeletal interactions \cite{schmid2017physical,meinhardt2013monolayer,tian2009sorting}. Multicomponent membranes therefore play an important role in processes such as vesicle separation, budding, and fusion. In \cite{benvegnu1992line,julicher1996shape,gozdz1999shapes,baumgart2005membrane}, authors theoretically modeled multicomponent vesicles by minimizing a free energy functional composed of the elastic bending energy and the line tension energy at the interface between different components \cite{julicher1996shape}:
\begin{align}
    E = E_1 + E_2 +E_l, \label{eqn:two_comp_ene}
\end{align}
where $E_1,E_2$ are the bending elastic energies for the individual components and $E_l$ accounts for the interfacial line tension energy between them. Building on this mechanism, Wang and Du \cite{wang2008modelling} proposed a phase field framework for two-component membranes. They introduced two phase field variables: one represents the membrane as a thin interface in three-dimensional space, and the other encodes lipid composition to distinguish coexisting liquid phases. The coupled model links curvature elasticity to composition through phase-dependent bending rigidities and spontaneous curvature, and captures line tension at domain boundaries. Lowengrub et al. \cite{lowengrub2009phase} further extended this work by coupling the framework in \cite{wang2008modelling} with a surface mass conservation equation, enabling detailed simulations of vesicle shape changes and phase separation for multicomponent vesicle membranes. Later studies by Zhao and Du \cite{zhao2010adhesion,zhao2011diffuse} investigated adhesion and fusion effects, demonstrating their influence on lipid phase separation and vesicle morphology across various equilibrium states.

In this work, we propose a new mechanochemical framework for the deformation and pattern formation of multicomponent vesicles. Instead of assigning distinct bending rigidities to liquid-ordered and liquid-disordered phases \cite{julicher1996shape}, our model assumes that the membrane shapes are driven by membrane-associated proteins that laterally segregate on the surface and generate active biochemical force with phase-dependent strength. Recent experiments and modeling support this view. For example, phase separation of proteins such as the low-complexity domain of fused in sarcoma (FUS LC) on membrane surfaces generates compressive stress that induces inward bending and tubule formation \cite{yuan2021membrane}. Curvature-sensing BAR-domain proteins, like Sorting Nexin 33 (Snx33), respond to membrane topology changes and regulate actin polymerization to guide cell polarity during migration \cite{sitarska2023sensing}. Tsai et al. developed a multiscale chemical-mechanical model of yeast budding to show that locally polarized insertion of new surface material, guided by cell division control protein 42 homolog (Cdc42), determines tubular shape and aspect ratio of the budding \cite{tsai2024study}. Johnson et al. reviewed how protein–membrane interactions can both detect and generate curvature, demonstrating their role in processes such as endocytosis, vesicle budding, and organelle shaping \cite{johnson2024protein}. A comprehensive review of the multiscale modeling of biological membranes, including phase separation, diverse lipid and protein composition, and protein-driven shape transition, is provided in \cite{enkavi2019multiscale}. These biological studies suggest a unifying mechanism in which certain membrane-associated proteins dynamically segregate on the membrane, creating localized biochemical protrusion or retraction forces that regulate membrane curvature and shape deformation. Building on this viewpoint, we propose our new mechanochemical model to study the pattern formation and morphology of the multicomponent vesicles. In this framework, an Ohta-Kawasaki energy (reviewed in Section \ref{subsec:OK}), defined on multicomponent vesicle membrane, captures microphase separation of membrane-associated proteins, and the vesicle shape evolves by an overdamped equation derived from force balance between mechanical membrane forces and protein-dependent active biochemical force. In the slow dynamics limit, the model reproduces equilibrium shapes observed in experiments \cite{baumgart2003imaging}. Besides, this novel framework can be coupled to external chemical cues or inhomogeneous focal adhesions to potentially simulate various cell motility such as cell chemotaxis or cell durotaxis, and the concomitant deformations.

\subsection{Microphase Separation: Ohta-Kawasaki Model}\label{subsec:OK}

In biological studies, experimental results have shown the microphase separation of multicomponent vesicle membranes \cite{baumgart2003imaging,dietrich2001lipid,veatch2002organization,schmid2017physical,su2024kinetic}. To model such phenomena, Ohta-Kawasaki (OK) model, first introduced in \cite{ohta1986equilibrium}, is particularly useful because of its ability to predict self-assembled patterns of diblock copolymers \cite{bates1990block,hamley2004developments} and display periodic structures such as lamellar, spherical, and bicontinuous gyroids \cite{barnhart2011adhesion}. Diblock copolymers consist of two distinct polymer segments $A$ and $B$, which are connected by covalent bonds. Due to their chemical incompatibility, these segments tend to segregate into microdomains, leading to microphase separation. This mechanism makes the OK model an ideal tool for simulating the phase separation of membrane-associated proteins. An early work \cite{schmid2017physical} explored the phase separation using the OK-type model on flat membranes.

The free energy functional of the OK model is given by \cite{Xu_Zhao2019,Xu_Zhao2020}
\begin{align}\label{eqn:OK}
        E^{\mathrm{OK}}[u] = \int_{\Omega}\left[\frac{\epsilon_u}{2}|\nabla u|^2+\frac{1}{\epsilon_u}W(u)\right] \mathrm{d}x + \frac{\gamma}{2}\int_{\Omega}|(-\Delta)^{-\frac{1}{2}}(u-\omega)|^2 \ \mathrm{d}x,
\end{align}
with a volume constraint
\begin{align*}
    \int_{\Omega}u\ \text{d}x = \omega|\Omega|,
\end{align*}
where $0<\epsilon_u\ll 1$ is an interface parameter and $\Omega$ is the spatial domain. Here, a phase field labeling function $u = u(x)$ represents the fraction of species $A$, while the species $B$ is given implicitly by $1-u(x)$. The double-well potential $W(u) = 18(u^2-u)^2$ enforces $u$ to be $0$ and $1$ in the bulk, while allowing a smooth but rapid transition between these values across the interfacial region. The first integral in (\ref{eqn:OK}) represents the local surface energy, modeling the short-range interaction between chain molecules and favoring large domains. In contrast, the second integral accounts for the long-range repulsive interaction with $\gamma >0$ being the strength of the repulsive force. This term favors the smaller domain sizes and leads to the microphase separation. 

To enforce the prescribed volume constraint, we introduce a modified energy functional \eqref{eqn:OK} by adding a penalty term. This results in unconstrained formulations, denoted as the penalized OK (pOK) model  \cite{LuoZhao_NumPDE2024}:
\begin{align}\label{eqn:pOK}
    E^{\text{pOK}}[u] & = \int_{\Omega}\left[\frac{\epsilon_u}{2}|\nabla u|^2+\frac{1}{\epsilon_u}W(u)\right] \text{d}x + \frac{\gamma}{2}\int_{\Omega}|(-\Delta)^{-\frac{1}{2}}(u-\omega)|^2 \ \text{d}x \nonumber\\
    &\qquad + \frac{M}{2}\left(\int_{\Omega}u\ \text{d}x - \omega|\Omega|\right)^2, 
    \end{align}
where $M\gg 1$ is the penalty constant. 

In our previous work \cite{luo2025fourier}, we applied the OK model on a spherical domain and obtained numerical results that closely resembled experimental patterns of multicomponent vesicles observed in \cite{baumgart2003imaging}, see Figure 1.1 in \cite{luo2025fourier}. These findings indicate that the OK model can effectively capture the microphase separation on a fixed spherical membrane. By coupling the OK model with a deformable vesicle membrane in our new mechanochemical framework, we aim to systematically explore various patterns on deformable multicomponent vesicle membranes.

% \begin{figure}[htbp]
% \begin{center}
% \includegraphics[width=.25\textwidth]{Figures/Nature_sphere.png} \quad\quad\quad
% \includegraphics[width=.28\textwidth]{Figures/PC_04_FtB.png}
% \end{center}
% \caption{Left: Experimental observation of an multicomponent vesicle membrane in \cite{baumgart2003imaging}; Right: Numerical simulation for the OK model on sphere \cite{luo2025fourier}.}
% \label{fig:real_num}
% \end{figure}

\subsection{Membrane Deformation: Phase Field Model Framework}\label{subsec:phase field}

Biological experiments have shown that cell membrane deformation is often accompanied by complex morphological changes \cite{baumgart2003imaging,mcmahon2005membrane,levental2016continuing}. Over the past decades, phase field modeling has been widely used as a powerful approach for the study of interfacial problems in biology and biophysics \cite{du2011phase,du2020phase}. In this framework, the membrane is represented as a diffuse interface with a small width, creating a smooth transition between the cell interior and exterior, which provides the natural incorporation of biophysical properties such as curvature elasticity, surface tension, and interactions with the surrounding environments. This approach is particularly well-suited for capturing membrane deformations driven by the interplay of chemical signaling, cytoskeletal dynamics, and mechanical surface forces. 

In \cite{du2004phase,du2005modeling,wang2008modelling}, Du, Liu, and Wang introduced a diffuse interface formulation for the elastic bending energy of the cell membranes, providing a mathematically rigorous and computationally efficient framework for the study of the membrane deformation and various equilibrium shapes. Later, Shao et al. \cite{shao2010computational} developed a quantitative phase field model to study cell shape dynamics and simulate the morphology of the motile fish keratocytes. This work extended the original phase field method \cite{langer1986directions,collins1985diffuse} to investigate the dynamics of cell migration. Subsequent studies further enriched this framework by incorporating additional biological mechanisms. For example, several works investigated actin cytoskeleton 'flow' in one- and two-dimensional cells \cite{gracheva2004continuum,larripa2006transport,carlsson2011mechanisms,rubinstein2009actin}. Some models prescribed cell boundary dynamics using phenomenological protrusion rates \cite{barnhart2011adhesion,wolgemuth2011redundant}, while others incorporated mechanical forces along the membrane to capture cell migration velocity without explicitly modeling actin flow or adhesion \cite{shao2010computational,ziebert2012model}. Additional approaches focused on specific aspects of cell motility, such as the influence of the cell–substrate adhesion dynamics \cite{buenemann2010role} or the regulation of the leading edge \cite{zimmermann2010leading}. Ziebert et al. \cite{ziebert2013effects} coupled a vector-field representation of the actin filament network with cell shape. Zhang, Levine, and Zhao \cite{zhang2025phase} combined the phase field method with Meinhardt's reaction-diffusion system on the membrane to study chemotaxis in Dictyostelium discoideum. Besides, recent progress has been made in applying phase field methods to solve PDEs on deformable surfaces \cite{teigen2009diffuse, li2009solving}, with applications to cell migration \cite{tao2020tuning}. A more comprehensive model has also been developed by Shao et al. \cite{shao2012coupling}, which integrated actin flow, discrete adhesion sites, and deformable boundaries to simulate cell migration in greater detail. Furthermore, the phase field approach has been applied to characterize specialized motility patterns, such as periodic migration \cite{camley2013periodic} and circular motion \cite{camley2014polarity,camley2017crawling}. Together, these studies demonstrate the effectiveness of the phase field model for the study of diverse cell motility by incorporating different biological, mechanical, and chemical components in the framework. For a comprehensive review of the phase field framework and its applications to cellular systems, we refer interested readers to the monographs \cite{du2011phase,du2020phase}. 

In this work, we develop a new computational modeling framework that couples a phase field formulation for membrane mechanics with the membrane-associated OK model to investigate vesicle phase separation and protein-driven membrane dynamics. In this framework, a phase field variable $\phi$ is introduced to distinguish the cell interior ($\phi\approx1$) from the exterior ($\phi\approx0$), with a thin diffuse interface smoothly transitioning between the two regions and represented by the $1/2$ level set. This phase field formulation captures complex morphology evolutions on a fixed computational grid without explicitly tracking the interface and governs the membrane deformation through a force balance in the new model. Protein segregation on the membrane surface is modeled by the OK dynamics, which is coupled to the membrane mechanics. This computational framework successfully reproduces experimentally observed multicomponent membrane patterns \cite{baumgart2003imaging} and is applicable to both two- and three-dimensional geometries, enabling comprehensive simulations of cell dynamics. Compared to earlier diffuse-interface models \cite{wang2008modelling}, this framework offers a more physically consistent representation of biochemical–mechanical coupling and provides mechanistic insight into protein-mediated membrane dynamics. Overall, it establishes a powerful and predictive platform for studying multicomponent membrane dynamics and advancing biomembrane modeling.

The rest of the paper is organized as follows. Section \ref{sec:model_dis} introduces our mechanochemical modeling framework, which uses two phase field functions: $\phi$ to describe vesicle shape and $u$ to capture microphase separation of membrane-associated proteins. Section \ref{sec:Num_Method} presents the numerical methods for solving the model. Section \ref{sec:Num} reports numerical simulations and compares them with the experimental observations in \cite{baumgart2003imaging}, demonstrating the accuracy and effectiveness of the framework. Finally, Section \ref{sec:Remark} concludes with several remarks.

% \newpage

\section{Phase Field Approach coupling with OK Model on the Membrane}\label{sec:model_dis}

In this section, we present in detail our new mechanochemical model for the study of patterns of multicomponent vesicle membranes.

\subsection{Phase Separation on a Phase Field Membrane}\label{subsec:OK_mem}

Since the OK model in Section \ref{subsec:OK} captures microphase separation, our first challenge is to place it on a dynamically evolving vesicle membrane. Following \cite{shao2010computational,zhang2025phase,painter2024biological}, we couple a surface phase separation equation to a phase field description of membrane geometry.

To this end, we introduce a phase field function $\phi$ to represent the vesicle membrane. The interior of the membrane is labeled by $\phi = 1$, and the exterior by $\phi = 0$. In the interfacial region near the membrane, $\phi$ transitions smoothly but rapidly across a diffuse interface of width $\epsilon_{\phi}$. To localize quantities to the membrane, we use
\begin{align}
    g(\phi) =  \frac{18}{\epsilon_{\phi}}(\phi^2-\phi)^2, \quad \mathrm{or} \quad \tilde{g}(\phi) =  \frac{\epsilon_{\phi}}{2}|\nabla \phi|^2 \label{eqn:gphi}
\end{align}
to indicate the interfacial region of $\phi$. Note that in the phase field formulation, $g(\phi)$ and $\tilde{g}(\phi)$ defined in \eqref{eqn:gphi} are asymptotically equivalent at equilibrium \cite{modica1987gradient,li2013variational}. Accordingly, we use whichever choice gives better accuracy or stability for later computations, see Section \ref{sec:Num_Method}.

We then introduce a second phase field $u$ with an interfacial width $\epsilon_u$ for the density of membrane-associated proteins, which takes values near $1$ in protein-rich regions and $0$ in protein-poor regions when microphase separated. Since $u$ is confined to the membrane, its off-membrane values are immaterial. In practice, the $u$-dependent terms are weighted by $g(\phi)$ or $\tilde g(\phi)$ so that they act only in the interfacial region. See below the equation \reff{eqn:pACOK_mem}.

Now given the two phase fields $\phi$ and $u$, we introduce an advection-reaction-diffusion equation to describe the dynamical evolution of $u$ near the interfacial region of $\phi$:
\begin{align}\label{eqn:pACOK_mem}
    \frac{\partial(g(\phi)u)}{\partial t} + \nabla\cdot(g(\phi)u\mathbf{v}) = \ & \epsilon_u \Delta_{\mathrm{S}}u  - \frac{1}{\epsilon_u} g(\phi)W'(u) + \gamma g(\phi)\Delta_{\mathrm{S}}^{-1}\Big(g(\phi)(u-\bar{u})\Big) \nonumber \\
    &- M g(\phi) \left(\int_{\Omega}g(\phi)\left(u-\bar{u}   \right)\ \text{d}x \right).
\end{align}
Here $\mathbf{v}$ is the interface velocity defined as 
\begin{align}\label{eqn:velocity}
    \mathbf{v}: = -\partial_t\phi\frac{\nabla\phi}{|\nabla\phi|^2},
\end{align}
and $\bar{u}$ is a fixed constant representing the volume fraction of $u$. The $\phi$-bounded surface Laplacian $\Delta_{\mathrm{S}}$ is defined as
\begin{align}\label{eqn:surface_laplacian}
    \Delta_{\mathrm{S}}u:= D_{\parallel}\nabla_{\parallel}\cdot(g(\phi)\nabla_{\parallel}u)+D_{\perp}\nabla_{\perp}\cdot(g(\phi)\nabla_{\perp}u),
\end{align}
in which the operators $\nabla_{\parallel}$ and $\nabla_{\perp}$ represent the tangential and normal diffusion, respectively. In the 2D case, they are defined as
\begin{align}
    \nabla_{\parallel}  = \begin{bmatrix}
        n_{y}^2 & -n_{x}n_{y} \\
        -n_{x}n_{y} & n_{x}^2
    \end{bmatrix}
    \begin{bmatrix}
        \partial_x \\
        \partial_y
    \end{bmatrix}, \quad
    \nabla_{\perp} = \begin{bmatrix}
        n_{x}^2 & n_{x}n_{y} \\
        n_{x}n_{y} & n_{y}^2
    \end{bmatrix}
    \begin{bmatrix}
        \partial_x \\
        \partial_y
    \end{bmatrix},
\end{align}
where the unit normal vector is $\mathbf{n} = [n_x,n_y]^\mathrm{T} = -\frac{\nabla\phi}{|\nabla\phi|}$. In the 3D case, they are given by 
\begin{align}\label{eqn:DeltaS_3D}
    \nabla_{\parallel} = \begin{bmatrix}
        n_y^2 + n_z^2 & -n_xn_y & -n_xn_z \\
        -n_xn_y & n_x^2+n_z^2 & -n_yn_z \\
        -n_xn_z & -n_yn_z & n_x^2+n_y^2
    \end{bmatrix}\begin{bmatrix}
    \partial_x \\
    \partial_y \\
    \partial_z
    \end{bmatrix}, \quad
    \nabla_{\perp} = \begin{bmatrix}
        n_x^2 & n_xn_y & n_xn_z \\
        n_xn_y & n_y^2 & n_yn_z \\
        n_xn_z & n_yn_z & n_z^2
      \end{bmatrix}\begin{bmatrix}
    \partial_x \\
    \partial_y \\
    \partial_z
    \end{bmatrix},
\end{align}
with the unit normal vector $\mathbf{n} = [n_x,n_y,n_z]^\mathrm{T} = -\frac{\nabla\phi}{|\nabla\phi|}$. $D_{\parallel}$ and $D_{\perp}$ in \reff{eqn:surface_laplacian} are the corresponding tangential and normal diffusion coefficients. Since $u$ evolves only on the membrane, one might expect purely tangential diffusion to be sufficient. In a phase field setting, however, the membrane is an interfacial layer of finite thickness, so values of $u$ on nearby level sets must remain consistent. We therefore include a small normal diffusion component to couple these layers, synchronize the reaction diffusion dynamics across the interface, and stabilize the tangential diffusion in the interfacial region \cite{zhang2025phase}.

The dynamics \reff{eqn:pACOK_mem} can be viewed as the $L^2$ gradient flow dynamics with respect to $u$ for the following membrane-bound (in other words, $\phi$-interface bound) OK free energy:
\begin{align}
    E_{\phi}[u]: = &\int_{\Omega} g(\phi)\bigg[ \frac{\epsilon_u}{2}\Big( D_{\parallel}|\nabla_{\parallel}u|^2 + D_{\perp} |\nabla_{\perp}u|^2\Big) + \frac{1}{\epsilon_u}W(u) \bigg] \mathrm{d}\mathbf{x}\nonumber\\
    & + \frac{\gamma}{2}\int_{\Omega} \bigg[(-\Delta_{\mathrm{S}})^{-1/2}\Big(g(\phi)(u-\bar{u})\Big)\bigg]^2 \mathrm{d}\mathbf{x} 
    + \frac{M}{2}\bigg[ \int_{\Omega} g(\phi)(u-\bar{u}) \mathrm{d}\mathbf{x} \bigg]^2, \label{eqn:mem_OK_energy}
\end{align}
except for an inclusion of the advection term $\nabla\cdot(g(\phi)u\mathbf{v})$. The first integral represents the short-range interaction of the OK model restricted to the membrane through the localization function $g(\phi)$. Here, the diffusion operator is defined on the membrane using a surface Laplacian that combines tangential and normal diffusion contributions, and $W(u)$ accounts for the local double-well potential. The second integral corresponds to the long-range interaction of the OK model, also restricted to the membrane by $g(\phi)$, where the inverse operator $(-\Delta_{\mathrm{S}})^{-1/2}$ captures long-range coupling effects of the protein distribution. The third term denotes a soft area constraint on the protein-rich area, again enforced through $g(\phi)$, which ensures that the protein concentration remains close to the prescribed average $\bar{u}$.

\begin{remark}
    The membrane-bound OK free energy \eqref{eqn:mem_OK_energy} is of the form similar to the original OK energy \eqref{eqn:OK}, except that it is constrained to the membrane of the phase field cell $\phi$ through the weighting function $g(\phi)$ (or $\tilde{g}(\phi)$). Therefore, this formulation allows us to conduct analysis such as the existence of the minimizers, $\Gamma$-limit of the energy, and other key properties of the membrane-bound OK model, by using analysis tools that we previously applied to the Euclidean domain \cite{LuoZhao_PhysicaD2024,Wang_Ren_Zhao2019}. We leave these directions for future investigation.
\end{remark}

\subsection{Deformation of the Membrane}\label{subsec:deform_main}

The dynamical change of the vesicle membrane is determined through a force balance. Several forces are involved: elastic bending force $\mathbf{F}_{\mathrm{ben}}$, surface tension force $\mathbf{F}_{\mathrm{surf}}$, area force $\mathbf{F}_{\mathrm{area}}$ that regulates the cell area, chemical force $\mathbf{F}_{\mathrm{chem}}$ proportional to the membrane-associated protein density $u$, line tension force $\mathbf{F}_{\mathrm{line}}$ between protein-rich and protein-poor phases, and an effective friction force $\mathbf{F}_{\mathrm{fr}}$ due to the interaction between the vesicle and the substrate. All these forces are formulated within the phase field framework, as described below.

The phase field elastic bending energy is defined as \cite{du2004phase,du2006simulating}
\begin{align}
    E_{\mathrm{ben}}(\phi) =  \int_{\Omega}\frac{\kappa}{2\epsilon_{\phi}}\left[\epsilon_{\phi}\Delta\phi - \frac{1}{\epsilon_{\phi}}W'(\phi) \right]^2dx \label{eqn:bending_eng_0sp}
\end{align}
where $\kappa$ is the bending rigidity, $\epsilon_{\phi}$ represents the phase field interfacial width. The surface tension energy in the phase field formulation is given by \cite{du2006simulating}
\begin{align}
    E_{\mathrm{surf}}(\phi) = \lambda_{\mathrm{surf}}\int_{\Omega}\left[\frac{\epsilon_{\phi}}{2}|\nabla\phi|^2+\frac{1}{\epsilon_{\phi}}W(\phi)\right]dx, \label{eqn:surf_area_onecomp}
\end{align}
in which $\lambda_{\mathrm{surf}}$ is the surface tension strength. Taking the variational derivative of the bending and tension energies, the elastic bending force and the surface tension force are defined by \cite{shao2010computational,zhang2025phase}
\begin{align}
    & \mathbf{F}_{\mathrm{ben}} = \frac{\delta E_{\mathrm{ben}}}{\delta\phi} \frac{\nabla\phi}{\epsilon_{\phi}|\nabla\phi|^2} = \frac{\kappa}{\epsilon_{\phi}^2}\left(\epsilon_{\phi}\Delta - \frac{1}{\epsilon_{\phi}}W''(\phi)\right)\left(\epsilon_{\phi}\Delta\phi - \frac{1}{\epsilon_{\phi}}W'(\phi)\right)\frac{\nabla\phi}{|\nabla\phi|^2}, \label{eqn:bend_force} \\
    & \mathbf{F}_{\mathrm{surf}} = \frac{\delta E_{\mathrm{surf}}}{\delta\phi} \frac{\nabla\phi}{\epsilon_{\phi}|\nabla\phi|^2} = \frac{\lambda_{\mathrm{surf}}}{\epsilon_{\phi}}\left(-\epsilon_{\phi}\Delta\phi+\frac{1}{\epsilon_{\phi}}W'(\phi)\right)\frac{\nabla\phi}{|\nabla\phi|^2}. \label{eqn:tension_force}
\end{align}
To constrain the cell area, the area force is formulated as a soft penalty as:
\begin{align}
    \mathbf{F}_{\mathrm{area}}  = M_{\mathrm{area}}\left(\int_{\Omega}\phi \ dx - A_0\right)\frac{\nabla\phi}{|\nabla\phi|}. \label{eqn:area_force}
\end{align}
where $M_{\mathrm{area}}$ is the penalty constant and $A_0$ is a prescribed area. Next, the line tension force is defined as 
\begin{align}
    \mathbf{F}_{\mathrm{line}} =  \lambda_{\mathrm{line}} \left(-\epsilon_u \Delta_{\mathrm{S}}u + \frac{1}{\epsilon_u}g(\phi)W'(u)\right) \frac{\nabla\phi}{|\nabla\phi|}, \label{eqn:line_ten_force}
\end{align}
where $\lambda_{\mathrm{line}}$ is the line tension coefficient. A multiplication of $g(\phi)$ confines the line tension between $\{u=1\}$ and $\{u=0\}$ only on the membrane $\phi$. The chemical force, due to the membrane-associated proteins, is given by
\begin{align}
    \mathbf{F}_{\mathrm{chem}} = - \alpha (u + u_0) \left(\epsilon_{\phi}\Delta\phi-\frac{1}{\epsilon_{\phi}}W'(\phi)\right) \frac{\nabla\phi}{|\nabla\phi|}. \label{eqn:chem_force}
\end{align}
Here $\alpha$ is the strength of the chemical force, the factor $(u+u_0)$ encodes the composition: protein-rich domains take $1+u_0$ and protein-poor domains take $u_0$. The term $\Big(\epsilon_{\phi}\Delta\phi-\frac{1}{\epsilon_{\phi}}W'(\phi)\Big)$ is a phase field formulation of the membrane curvature, so that $\mathbf{F}_{\mathrm{chem}}$ scales with the local curvature. More importantly, in regions of negative curvature, it will exert an inward force.

The force balance at quasi-steady state is 
\begin{align}\label{eqn:force_balance}
    \mathbf{F}_{\mathrm{ben}} + \mathbf{F}_{\mathrm{surf}} + \mathbf{F}_{\mathrm{area}} + \mathbf{F}_{\mathrm{line}} + \mathbf{F}_{\mathrm{chem}} + \mathbf{F}_{\mathrm{fr}} = 0.
\end{align}
Here $\mathbf{F}_{\mathrm{fr}}$ is the friction force due to the interaction between the cell and the surrounding environment, such as the adhesion between them, attachment and detachment of the cell from the substrate, which is assumed to be proportional to the local speed: $\mathbf{F}_{\mathrm{fr}} = -\mu\mathbf{v}$ in which $\mu$ is the friction constant. Therefore, the force balance equation \reff{eqn:force_balance} leads to 
\begin{align}
    \mathbf{v} = -\frac{1}{\mu}\mathbf{F}_{\mathrm{fr}} = \frac{1}{\mu}\left(\mathbf{F}_{\mathrm{ben}} + \mathbf{F}_{\mathrm{surf}} + \mathbf{F}_{\mathrm{area}} + \mathbf{F}_{\mathrm{line}} + \mathbf{F}_{\mathrm{chem}}\right).\nonumber 
\end{align}
Along with the transport equation of $\phi$ by the velocity field $\mathbf{v}$: 
$
    \frac{\partial\phi}{\partial t} = - \mathbf{v}\cdot\nabla\phi, 
$
we have the final equation of $\phi$ as:
\begin{align}
    \mu\frac{\partial\phi}{\partial t} & = \lambda_{\mathrm{surf}}\left(\Delta\phi-\frac{1}{\epsilon_{\phi}^2}W'(\phi)\right) - k\left(\Delta - \frac{W''(\phi)}{\epsilon_{\phi}^2}\right)\left(\Delta \phi - \frac{W'(\phi)}{\epsilon_{\phi}^2}\right) \nonumber\\
    & \quad + \lambda_{\mathrm{line}}\left(\epsilon_u \Delta_{\mathrm{S}}u - \frac{1}{\epsilon_u}g(\phi)W'(u)\right)|\nabla\phi| - M_{\mathrm{area}}\left(\int_{\Omega}\phi \ dx - A_0\right)|\nabla\phi| \nonumber \\ 
    & \quad + \alpha (u + u_0)\left(\epsilon_{\phi}\Delta\phi-\frac{1}{\epsilon_{\phi}}W'(\phi)\right) |\nabla\phi|. \label{eqn:final_mem3} 
\end{align}

In summary, equation \reff{eqn:final_mem3} governs the dynamical evolution of the membrane variable $\phi$, and equation \ref{eqn:pACOK_mem} captures the microphase separation of the membrane associated protein $u$. For completeness, we restate both equations below:
\begin{align}
    & \mu\frac{\partial\phi}{\partial t} = \lambda_{\mathrm{surf}}\left(\Delta\phi-\frac{1}{\epsilon_{\phi}^2}W'(\phi)\right) - \kappa\left(\Delta - \frac{W''(\phi)}{\epsilon_{\phi}^2}\right)\left(\Delta\phi - \frac{W'(\phi)}{\epsilon_{\phi}^2}\right) \nonumber\\
    & \quad\quad\quad + \lambda_{\mathrm{line}} \left(\epsilon_u \Delta_{\mathrm{S}}u - \frac{1}{\epsilon_u}g(\phi)W'(u)\right)|\nabla\phi| - M_{\mathrm{area}}\left(\int_{\Omega}\phi \ dx - A_0\right)|\nabla\phi| \nonumber \\ 
    & \quad\quad\quad + \alpha (u + u_0)\left(\epsilon_{\phi}\Delta\phi-\frac{1}{\epsilon_{\phi}}W'(\phi)\right)|\nabla\phi|, \label{eqn:model_force} \\
    & \frac{\partial(g(\phi)u)}{\partial t} + \nabla\cdot(g(\phi)u\mathbf{v}) = \ \epsilon_u \Delta_{\mathrm{S}}u  - \frac{1}{\epsilon_u} g(\phi)W'(u) + \gamma g(\phi)\Delta_{\mathrm{S}}^{-1}\Big(g(\phi)(u-\bar{u})\Big) \nonumber \\  &\quad\quad\quad\quad\quad\quad\quad\quad\quad\quad\quad\quad\quad - M g(\phi) \left(\int_{\Omega}g(\phi)\left(u-\bar{u}\right)\ \text{d}x \right).\label{eqn:model_OK}
\end{align}

\subsection{Model Discussion}

Building on the previous two-component membrane model \eqref{eqn:two_comp_ene}, Wang and Du \cite{wang2008modelling} reformulated it in a phase field framework and compared their simulations with the experimental findings in \cite{baumgart2003imaging}. Their model minimizes the total energy that combines phase field elastic bending energy with component-dependent bending rigidities and a line tension along component interfaces, subject to constraints on surface area, enclosed volume, and an orthogonality condition. In contrast, our model, which also reproduces the experimental findings \cite{baumgart2003imaging}, relies on a different mechanism: segregation of membrane-bound proteins generates biochemical forces that, together with other mechanical forces, drive the membrane deformation. 

Wang and Du’s model relies on the coupling of two vesicles through energies and geometric constraints, and then solves the constrained energy minimization problem. This procedure is less reflective of the actual cellular processes. Our approach is biologically more relevant and has two advantages. First, it avoids the orthogonality condition by directly confining the protein density $u$ to the membrane $\phi$, and evolving $\phi$ through a balance of mechanical and biochemical forces on the membrane. Second, it models protein dynamics on the membrane through a reaction–advection–diffusion equation within the OK framework. This unified framework captures the interplay between the membrane-associated proteins and membrane mechanics, and potentially allows us to investigate biological processes such as budding, endocytosis, and fission with greater flexibility.

Our framework is adaptive to incorporating additional components such as external chemical cues, cortical actin flow, or substrate adhesion. Adhesion is known to promote phase separation in multicomponent membranes \cite{gordon2008adhesion}.
In particular, Zhao et al. \cite{zhao2010adhesion} showed that coupling bending energy to adhesion potentials in a phase field model facilitates the separation of compositionally distinct domains on vesicle membranes. Motivated by these findings, we may consider our coupled system with an adhesion term and study how adhesion dynamically modulates protein phase separation on the membrane and the resulting membrane morphology. This extension increases biological relevance and broadens applications to problems such as vesicle docking \cite{schimmoller1998rab}, cell–substrate contact \cite{axelrod1981cell}, and adhesion-driven domain patterning \cite{sackmann2002cell}. 

Finally, our model can integrate hydrodynamic equations to explicitly determine the velocity field. By coupling our mechanochemical model to the hydrodynamic flow, we can obtain a self-consistent framework that simultaneously solves for cell velocity, cell shape dynamics, and membrane-bound protein density evolution. This extension provides a more realistic description of multicomponent cell motility when interacting with surrounding environments.

\section{Numerical Methods}\label{sec:Num_Method}

In this section, we present the numerical method for solving the coupled system \eqref{eqn:model_force}–\eqref{eqn:model_OK} under periodic boundary conditions. Let $N$ be a positive even integer, we define the domain as $\Omega = [-L,L]^d$ with $d=2,3$, and set the uniform spatial size $h = \frac{2L}{N}$. The corresponding discrete domain is then given by $\Omega_h = \Omega\cap (\overset{d}{\underset{i = 1}{\otimes}} h\mathbb{Z})$. For the temporal discretization, given a time interval $[0,T]$ and an integer $M>0$, we take the uniform time step size $\tau = \frac{T}{M}$ and $t_n = n\tau$ for $n=0,1,\cdots,M$. The approximate solutions of the phase field cell $\phi$ and the membrane-bound protein density $u$ at grid points in $\Omega_h$ and time $t_n$ are denoted by $(\phi^n, u^n)$. Given the initial conditions $\phi^0$ and $u^0$, the following numerical algorithm computes $(\phi^n, u^n)$ for $n = 1, 2, \dots, M$.

At first, we use a semi-implicit Fourier spectral method to discretize \eqref{eqn:model_force} for the phase field cell $\phi$ as follows:
\begin{align}
    \mu\frac{\phi^{n+1}-\phi^n}{\tau} = \ & \lambda_{\mathrm{surf}}\Delta \phi^{n+1} - \frac{\lambda_{\mathrm{surf}}}{\epsilon_{\phi}^2}W'(\phi^{n}) - \kappa\Delta^2\phi^{n+1} + \frac{\kappa}{\epsilon_{\phi}^2}\Delta\left(W'(\phi^{n})\right) \nonumber\\
    & + \frac{\kappa}{\epsilon_{\phi}^2}W''(\phi^n)\left(\Delta\phi^n - \frac{1}{\epsilon_{\phi}^2}W'(\phi^n)\right)  - \lambda_{\mathrm{line}} \left(\epsilon_u\Delta_{\mathrm{S}}u^n - \frac{1}{\epsilon_u}g(\phi^n)W'(u^n)\right)|\nabla\phi^n| \nonumber\\
    & - M_{\mathrm{area}}\left(\int\phi^n \mathrm{d}x - A_0\right)|\nabla\phi^n|  + \alpha(u^n + u_0)\left(\epsilon_{\phi}\Delta\phi^n - \frac{1}{\epsilon_{\phi}}W'(\phi^n)\right)|\nabla\phi^n|.\label{eqn:num_force}
\end{align}
This can be rewritten in the compact form:
\begin{align}
    \left(\frac{\mu}{\tau} - \lambda_{\mathrm{surf}}\Delta + \kappa\Delta^2\right)\phi^{n+1} = \mathrm{RHS}, \label{eqn:num_force_simp}
\end{align}
where 
\begin{align*}
    \mathrm{RHS} = \ & \frac{\mu}{\tau}\phi^n - \frac{\lambda_{\mathrm{surf}}}{\epsilon_{\phi}^2}W'(\phi^{n}) + \frac{\kappa}{\epsilon_{\phi}^2}\Delta\left(W'(\phi^{n})\right) + \frac{\kappa}{\epsilon_{\phi}^2}W''(\phi^n)\left(\Delta\phi^n - \frac{1}{\epsilon_{\phi}^2}W'(\phi^n)\right) \\
    & - \lambda_{\mathrm{line}} \left(\epsilon_u\Delta_{\mathrm{S}}u^n - \frac{1}{\epsilon_u}g(\phi^n)W'(u^n)\right)|\nabla\phi^n| - M_{\mathrm{area}}\left(\int\phi^n \mathrm{d}x - A_0\right)|\nabla\phi^n|  \\
    & + \alpha(u^n + u_0)\left(\epsilon_{\phi}\Delta\phi^n - \frac{1}{\epsilon_{\phi}}W'(\phi^n)\right)|\nabla\phi^n|. 
\end{align*}
Thus, equation \eqref{eqn:num_force_simp} can be efficiently solved for $\phi^{n+1}$ using the Fourier spectral method.

Next, we present the numerical method for the OK model on the membrane \eqref{eqn:model_OK}. In our algorithm, we note that $g(\phi)$ and $\tilde{g}(\phi)$ defined in \eqref{eqn:gphi} are equivalent at equilibrium in the phase field formulation \cite{modica1987gradient,li2013variational}. Accordingly, we use $g(\phi)$ in all terms of the system, except for the diffusion operator $\Delta_{\mathrm{S}} = D_{\parallel}\nabla_{\parallel}\cdot(\tilde{g}(\phi)\nabla_{\parallel})+D_{\perp}\nabla_{\perp}\cdot(\tilde{g}(\phi)\nabla_{\perp})$ and its inverse $\Delta_{\mathrm{S}}^{-1}$.  This is because $\tilde{g}(\phi)$ can simplify the calculation of $\Delta_{\mathrm{S}}$. The full expansion for $\Delta_{\mathrm{S}}$ in the 2D case is presented below. The calculation of the 3D case in \reff{eqn:DeltaS_3D} is similar and straightforward, so we omit it here.  
\begin{align*}
    \nabla_{\parallel}
    \cdot(\tilde{g}(\phi)\nabla_{\parallel}u) 
    & = \begin{bmatrix}
        n_{y}^2 & -n_{x}n_{y} \\
        -n_{x}n_{y} & n_{x}^2
    \end{bmatrix}
    \begin{bmatrix}
        \partial_x \\
        \partial_y
    \end{bmatrix}
    \cdot\left(\frac{\epsilon_{\phi}}{2}|\nabla\phi|^2\begin{bmatrix}
        n_{y}^2 & -n_{x}n_{y} \\
        -n_{x}n_{y} & n_{x}^2
    \end{bmatrix}
    \begin{bmatrix}
        \partial_x \\
        \partial_y
    \end{bmatrix}u
    \right)\\
    & = \frac{\epsilon_{\phi}}{2}\left[ (n_y^2\partial_x - n_{x}n_{y}\partial_y)(\phi_{y}^2u_x  -\phi_{x}\phi_{y}u_y) + (n_{x}^2\partial_y-n_{x}n_{y}\partial_x)( \phi_{x}^2u_y-\phi_{x}\phi_{y}u_x)\right], \\
    \nabla_{\perp}\cdot(\tilde{g}(\phi)\nabla_{\perp}u) 
    & = 
    \begin{bmatrix}
        n_{x}^2 & n_{x}n_{y} \\
        n_{x}n_{y} & n_{y}^2
    \end{bmatrix}
    \begin{bmatrix}
        \partial_x \\
        \partial_y
    \end{bmatrix}
    \cdot\left(\frac{\epsilon_{\phi}}{2}|\nabla\phi|^2\begin{bmatrix}
        n_{x}^2 & n_{x}n_{y} \\
        n_{x}n_{y} & n_{y}^2
    \end{bmatrix}
    \begin{bmatrix}
        \partial_x \\
        \partial_y
    \end{bmatrix}u
    \right)\\
    & = \frac{\epsilon_{\phi}}{2}\left[ (n_x^2\partial_x + n_{x}n_{y}\partial_y)(\phi_{x}^2u_x  + \phi_{x}\phi_{y}u_y) + (n_{y}^2\partial_y + n_{x}n_{y}\partial_x)(\phi_{y}^2u_y + \phi_{x}\phi_{y}u_x)\right].
\end{align*}

Now we present the numerical treatments for $\Delta_{\mathrm{S}}$. At first, we apply the Fourier spectral method to evaluate $(\phi_x, \phi_y)$, $((\phi_x)^2_x , (\phi_x)^2_y)$, $((\phi_y)^2_x , (\phi_y)^2_y)$, and $((\phi_x\phi_y)_x , (\phi_x\phi_y)_y)$.
Secondly, the unit norm vector $\mathbf{n} = -\frac{\nabla\phi}{|\nabla\phi|}= [n_x,n_y]^{\mathrm{T}}$ is calculated as
\begin{align*}
  n_x = -\frac{\phi_x}{\sqrt{\phi_x^2+\phi_y^2+\epsilon_0}} , \quad  n_y = -\frac{\phi_y}{\sqrt{\phi_x^2+\phi_y^2+\epsilon_0}},
\end{align*}
where $\epsilon_0$  is a regularization constant to prevent division by zero. Thirdly, we use the central difference scheme to compute the 
derivatives of $u$ as:
\begin{align*}
    & (u_x)_{i,j} \approx \frac{u_{i+1,j} - u_{i-1,j}}{2h_x}, \quad (u_y)_{i,j} \approx\frac{u_{i,j+1} - u_{i,j-1}}{2h_y}, \quad
    (u_{xx})_{i,j} \approx \frac{u_{i+1,j} - 2u_{i,j} + u_{i-1,j}}{h_x^2}, \\
    & (u_{xy} = u_{yx})_{i,j}\approx\frac{u_{i+1,j+1} - u_{i-1,j+1} - u_{i+1,j-1} + u_{i-1,j-1}}{4h_xh_y}, \quad (u_{yy})_{i,j} \approx \frac{u_{i,j+1} - 2u_{i,j} + u_{i,j-1}}{h_y^2}.
\end{align*}
Then using the above approximations for $\mathbf{n}$ and derivatives of $u$, we obtain an approximation of the $\phi$-surface Laplacian $\Delta_{\mathrm{S},h}$.

With the above discretizations in place, we compute $f = \Delta_{\mathrm{S}}^{-1}(g(\phi)(u-\bar{u}))$ by solving the discrete membrane-bound Poisson equation:
\begin{align*}   
\Delta_{\mathrm{S},h} f_h = g(\phi{^n})(u{^n} - \bar{u}).
\end{align*}
Then we obtain the required discrete approximations $\Delta_{\mathrm{S},h}u{^n}$
and $\Delta_{\mathrm{S},h}^{-1}(g(\phi{^n})(u{^n} - \bar{u}))$ for the use in the time-stepping scheme.

The advection term in \reff{eqn:model_OK} is approximated by a central difference method as
\begin{align*}
    \nabla\cdot(g(\phi)u\mathbf{v})  \approx (\nabla\cdot(g(\phi)u\mathbf{v}))^{n} := \ & \frac{g(\phi{^n}_{i+1,j})u^n_{i+1,j}v^{x,n}_{i+1,j} - g(\phi^n_{i-1,j})u{^n}_{i-1,j}v^{x,n}_{i-1,j}}{2h_x} \\
    & + \frac{g(\phi{^n}_{i,j+1})u{^n}_{i,j+1}v^{y,n}_{i,j+1} - g(\phi{^n}_{i,j-1})u{^n}_{i,j-1}v^{y,n}_{i,j-1}}{2h_y}.
\end{align*}

With these approximations and the forward Euler scheme for time evolution, the numerical scheme for solving \eqref{eqn:model_OK} is given by
\begin{align}
    &\frac{g(\phi^{n+1})u^{n+1}-g(\phi^n)u^n}{\tau} + {(\nabla\cdot(g(\phi)u\mathbf{v}))^{n}}  \nonumber\\
    = \ &  \epsilon_u\Delta_{\mathrm{S},{h}}u^n - \frac{1}{\epsilon_u}g(\phi^n)W'(u^n) + \gamma g(\phi{^n})\Delta_{\mathrm{S},{h}}^{-1}\Big(g(\phi^n)(u^n-\bar{u})\Big) \nonumber \\  &- M g(\phi^n) \left(\int_{\Omega}g(\phi^n)\left(u^n-\bar{u}\right)\ \text{d}x \right), \label{eqn:num_OK}
\end{align}
which updates $u^n \to u^{n+1}$ at $n$-th step. Here, the update is obtained in terms of the product $g(\phi^{n+1})u^{n+1}$, and an additional step is required to recover $u^{n+1}$. In practice, we introduce a computational box that includes all grid points where $g(\phi^{n+1})>10^{-3}$. Inside this box, $u^{n+1}$ is computed explicitly as
\begin{align*}
    u^{n+1} \leftarrow \frac{g(\phi^{n+1})u^{n+1}}{g(\phi^{n+1})}. 
\end{align*}
Outside this box, where $g(\phi^{n+1})$ is negligible, we simply assign $u^{n+1} = g(\phi^{n+1})u^{n+1}$.

\section{Numerical Experiments}\label{sec:Num}

In this section, we numerically simulate the proposed model in both 2D and 3D, and compare the results with the biological experiments on multicomponent membranes reported in \cite{baumgart2003imaging}.

For the 2D simulations, we consider the computational domain $\Omega = [-L_x,L_x]\times[-L_y,L_y]$ with $L_x = L_y = 10$. The number of mesh grid points is $N_x = N_y = 2^8$ with the mesh grid size $h_x = h_y = \frac{2L_x}{N_x} = \frac{2L_y}{N_y}$. The time step size $\tau $ is $5\times 10^{-4}$, and the membrane width is set to be $\epsilon_{\phi} = 10h_x$. For the 3D simulations, we take the domain $\Omega = [-L_x,L_x]\times[-L_y,L_y]\times[-L_z,L_z]$ with $L_x = L_y = L_z = 10$, using smaller mesh grid points $N_x = N_y = N_z = 2^7$ and membrane width $\epsilon_{\phi} = 10h_x$.
% using the same mesh grid size and membrane width as in the 2D case. 
A smaller time step $\tau = 1e-4$ is used to maintain numerical stability. In the following numerical simulations, the white circle represents the cell membrane, while the yellow dashed box denotes the subdomain $\tilde{\Omega} = [-0.875L_x,0.875L_x]\times[-0.875L_y,0.875L_y]$. This subdomain is introduced to prevent the phase-field cell $\phi$ from too closely approaching the boundary of the full computational domain $\Omega$. To achieve this, the cell is periodically shifted to remain within $\tilde{\Omega}$, and the subdomain is recentered whenever $\phi$ reaches the boundary of the yellow dashed region. Other parameters for each example are provided in the captions of the corresponding figures.

Although the OK model can theoretically start from random initials on the membrane and lead to the microphase separation of $u$, appropriate initial conditions can often accelerate the dynamic evolution and help the system reach equilibrium more efficiently. Therefore, different initial conditions are selected for various numerical experiments presented in this section.

\subsection{OK Model on the Deformable Membrane in 2D case}

In this subsection, we use the proposed model in the 2D setting and compare the results in \cite{baumgart2003imaging}. We set $\phi^0$ as a disk centered at the origin with radius $r_0 = 4$:
\begin{align}
    \phi^0(x) = 0.5 + 0.5\tanh{\left(\frac{r_0 - \mathrm{dist}(x,0)}{\epsilon_{\phi}/3}\right)}, \label{eqn:phi_initial}
\end{align} 
where $\mathrm{dist}(x,0)$ denotes the Euclidean distance between $x$ and the origin. 

\begin{figure}[b!]
    \begin{center}
    \includegraphics[width=0.16\textwidth]{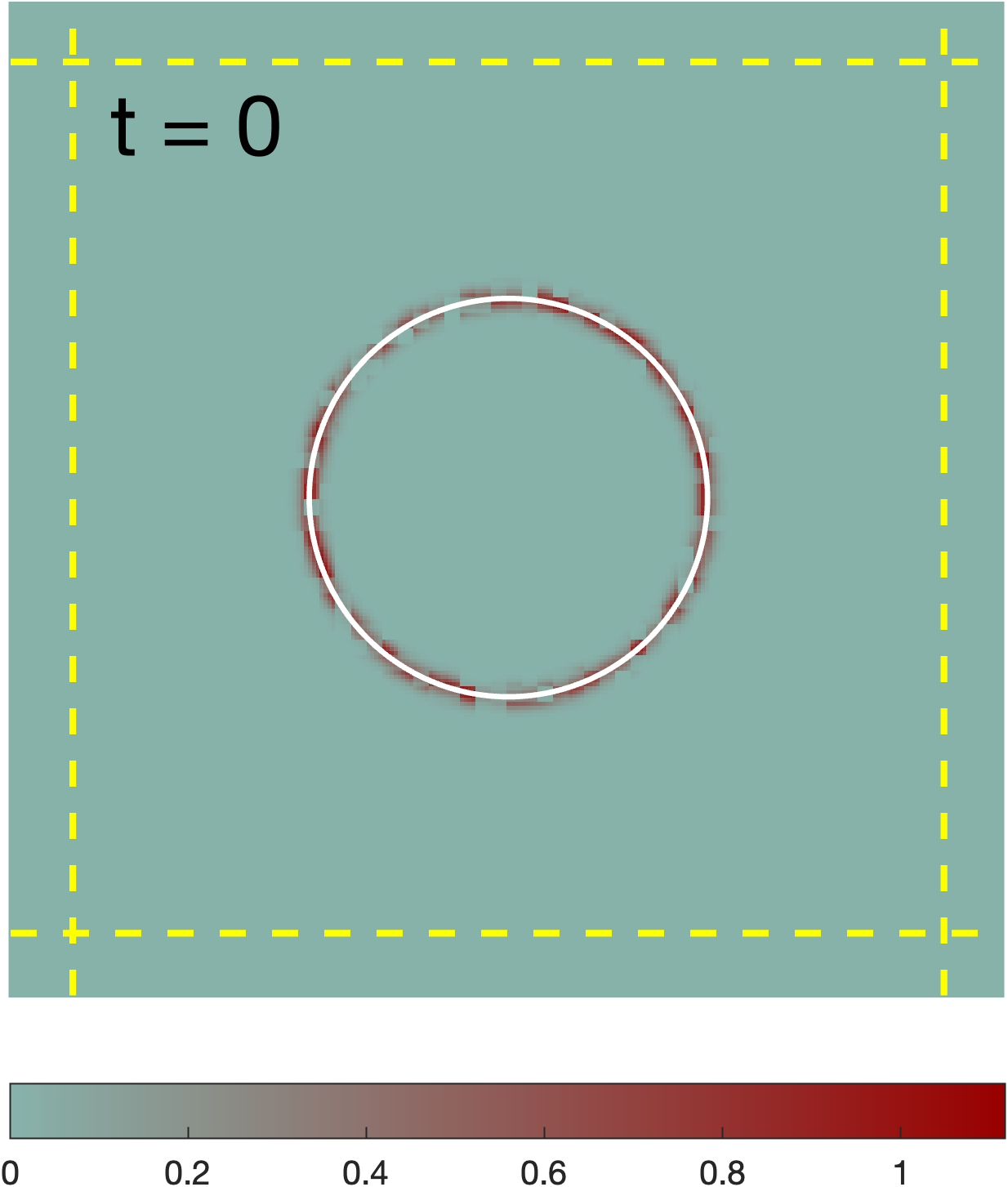}
    \includegraphics[width=0.16\textwidth]{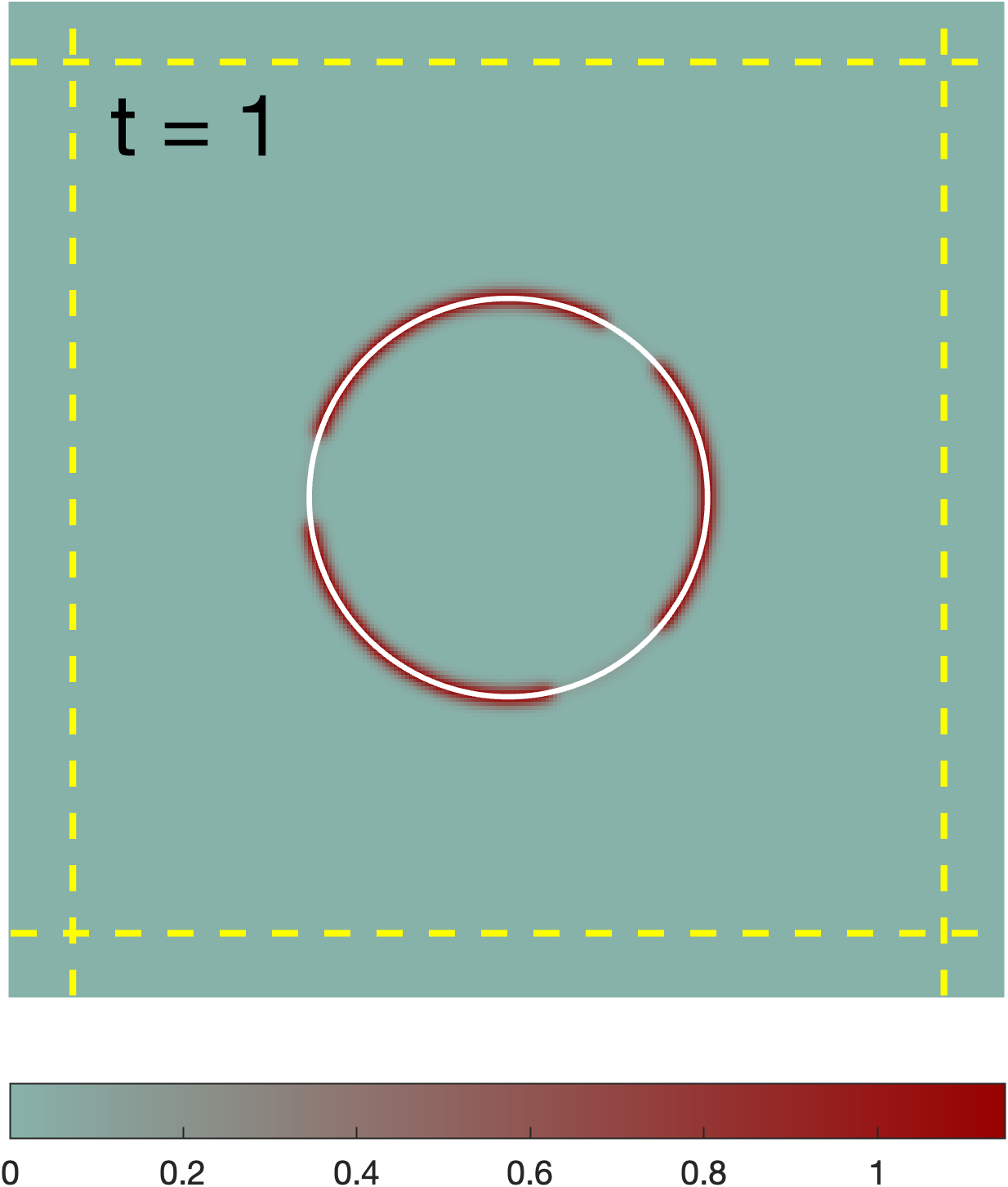}
    \includegraphics[width=0.16\textwidth]{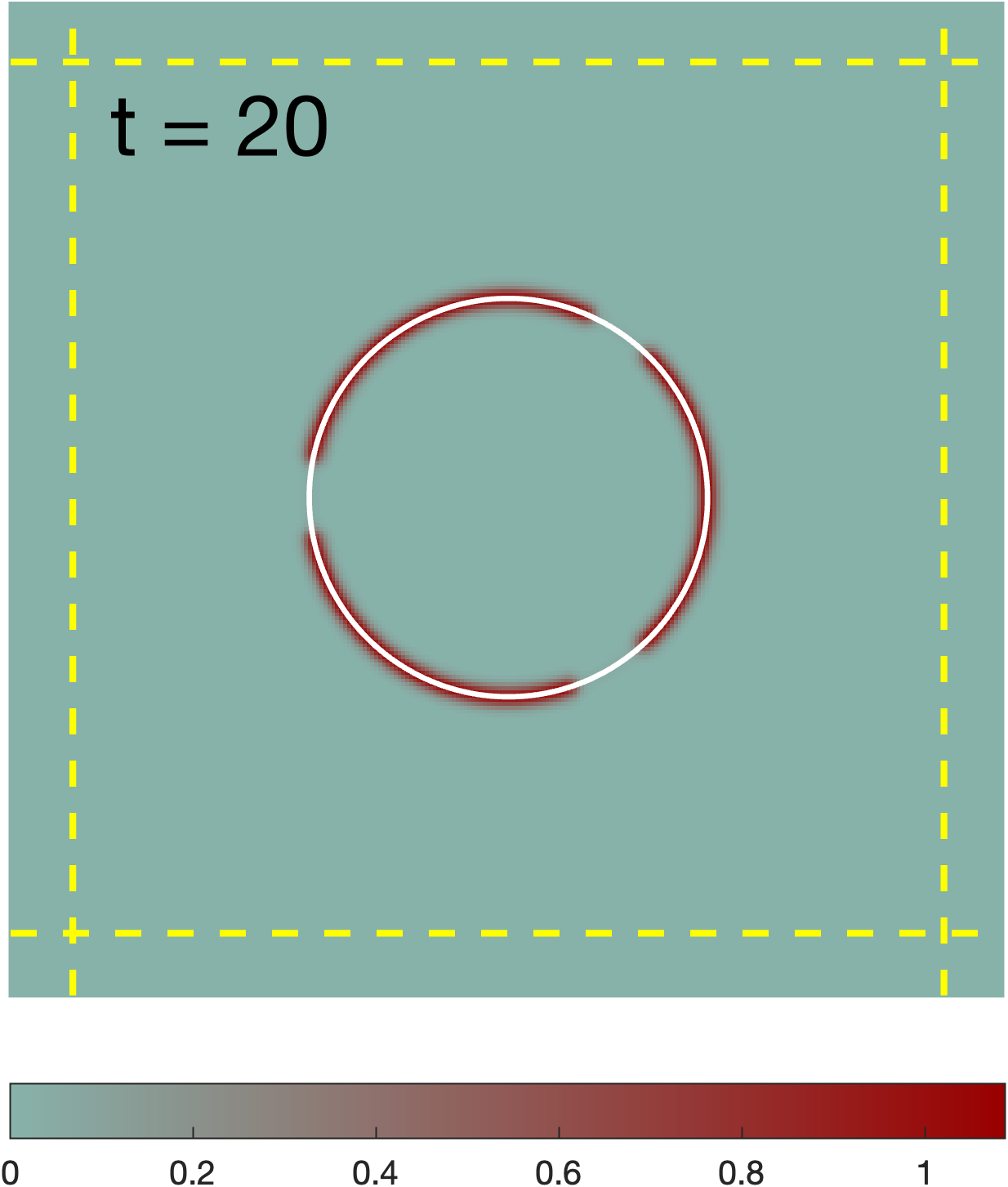}
    \includegraphics[width=0.16\textwidth]{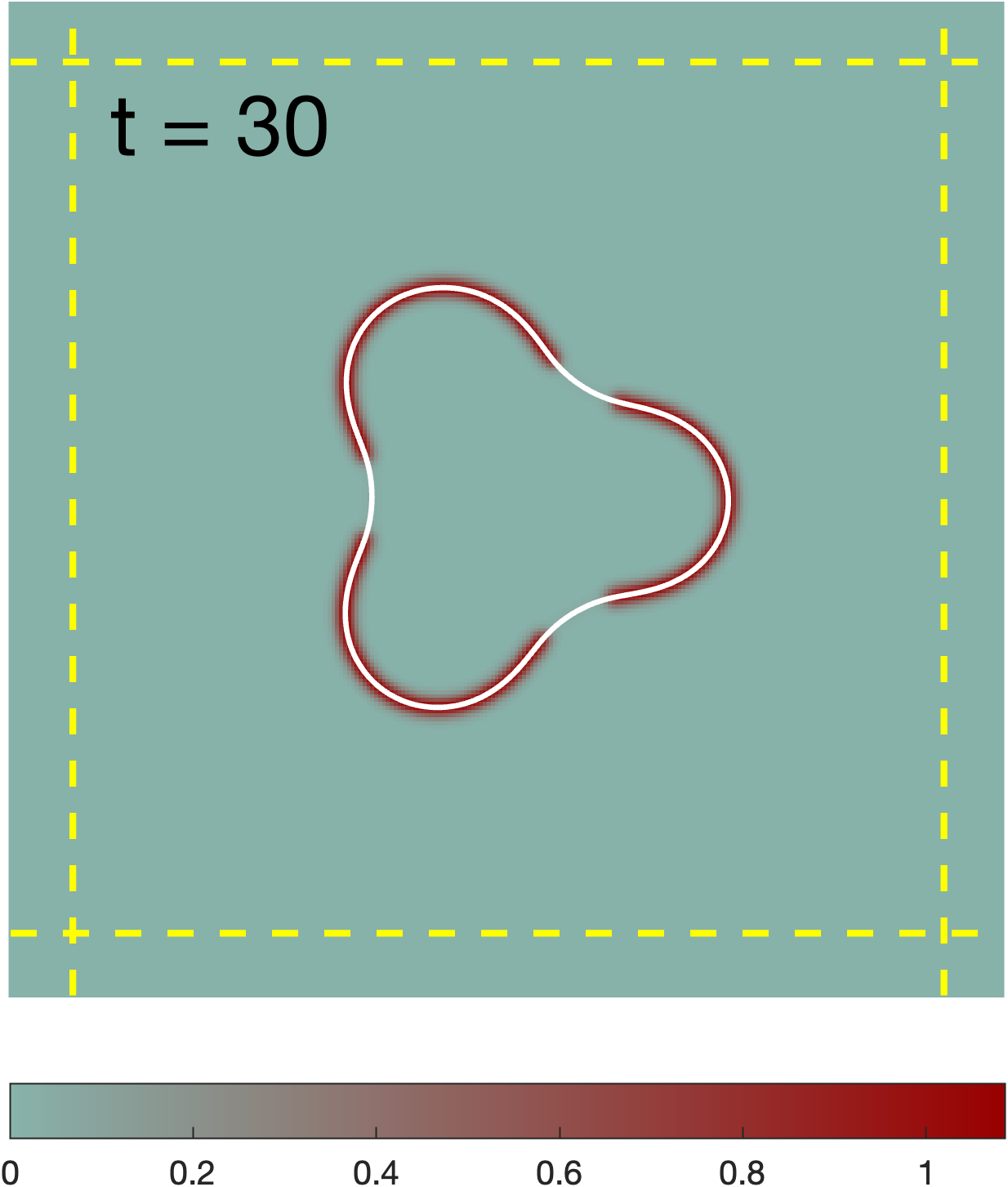}
    \includegraphics[width=0.16\textwidth]{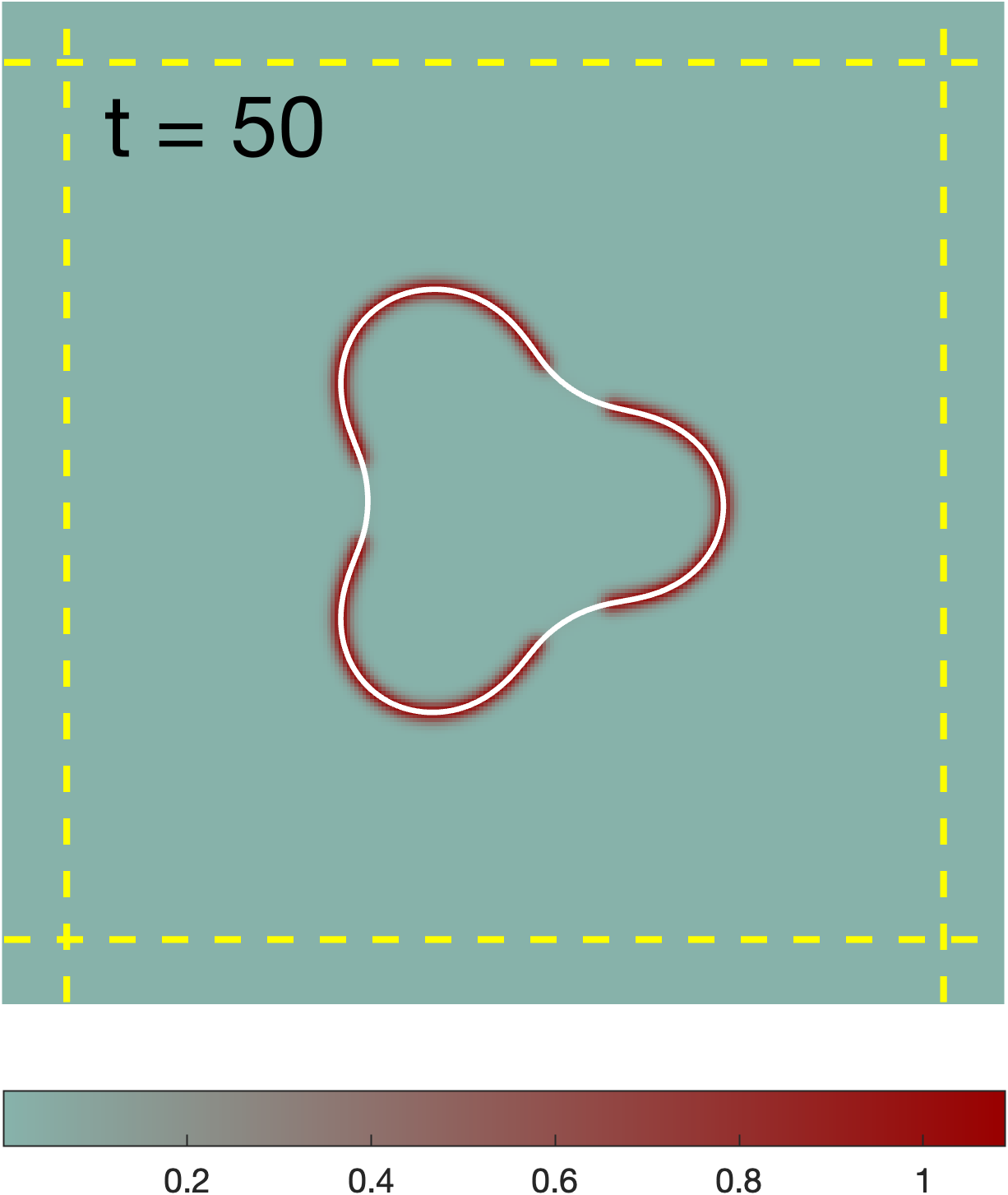}
     \raisebox{5.2mm}{\includegraphics[width=0.16\textwidth]{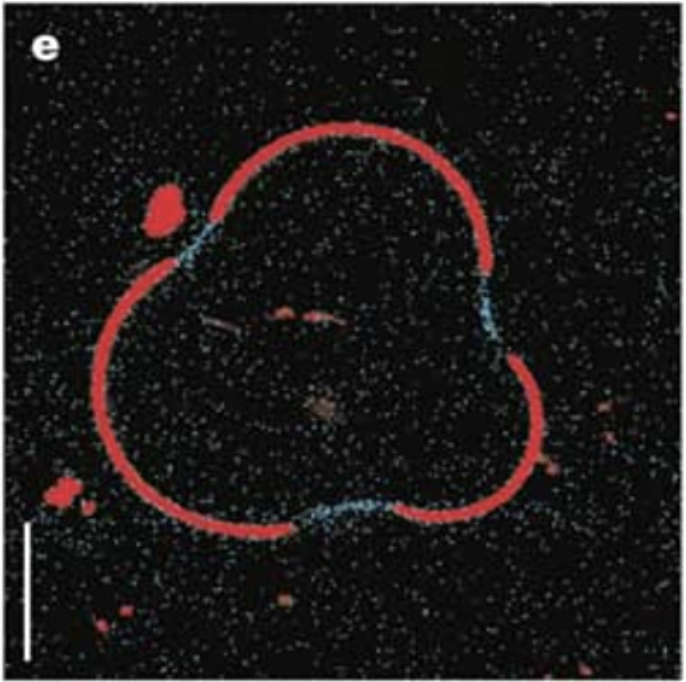}}
    \end{center}
    \caption{ Example illustrating phase separation and membrane deformation with three protein-rich subdomains. First three columns (from left): the evolution of membrane-bound protein domains starting from a random initial distribution under the OK model on a fixed membrane. Next two columns: the subsequent membrane deformation governed by the coupled system. Final column: the corresponding biological experiment in Figure 1(e) of \cite{baumgart2003imaging}. Parameters are given as: $\lambda_{\mathrm{surf}} = 2$, $\lambda_{\mathrm{line}} = 6$, $\kappa=1$, $\alpha = 1.5$, $u_0 = 0.2$, $\bar{u} = 0.8$, $\gamma = 80$, and $\epsilon_u = 20h_x$. }
    \label{fig:OK_3b}
\end{figure}

\begin{figure}[t!]
    \begin{center}
    \includegraphics[width=0.16\textwidth]{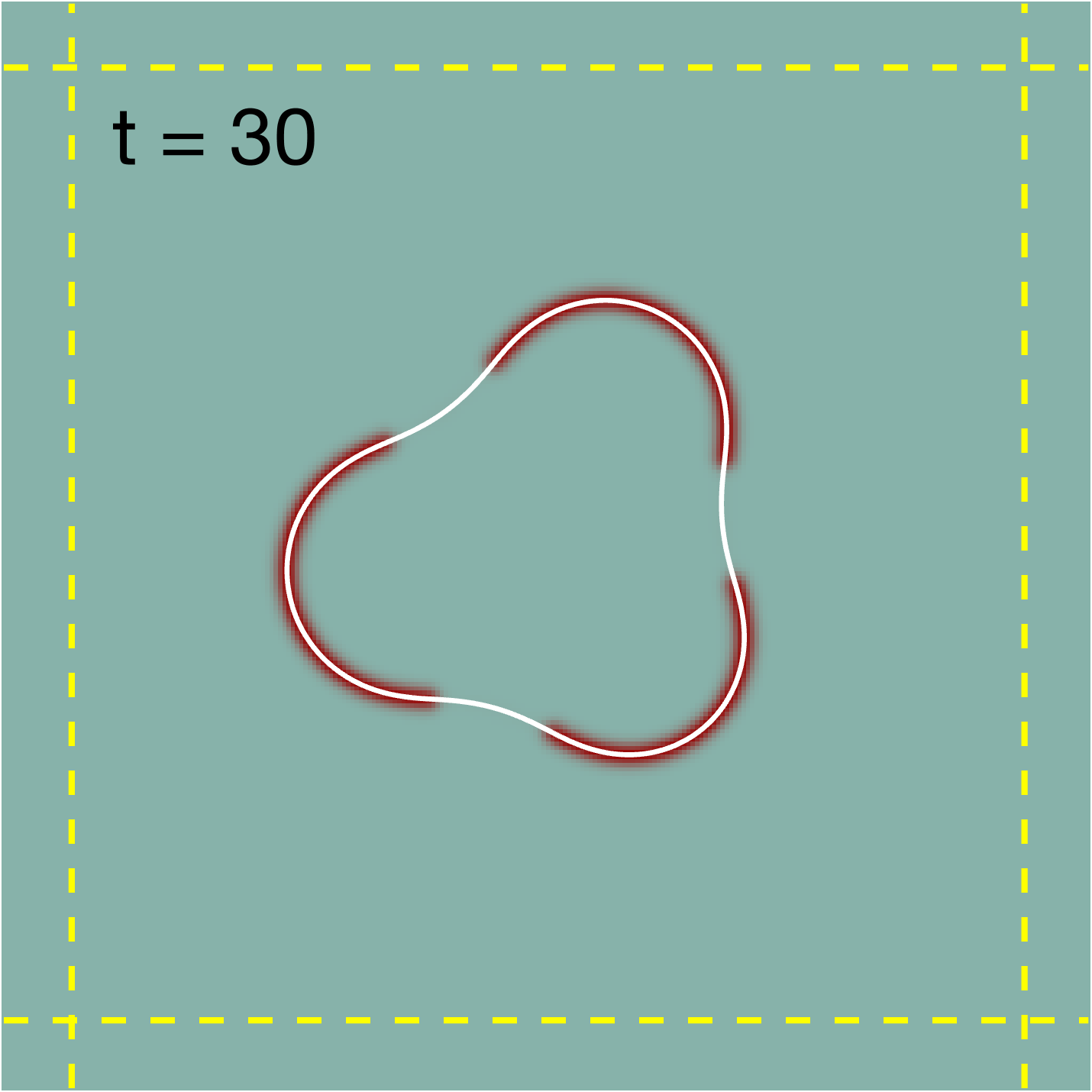}
    \includegraphics[width=0.16\textwidth]{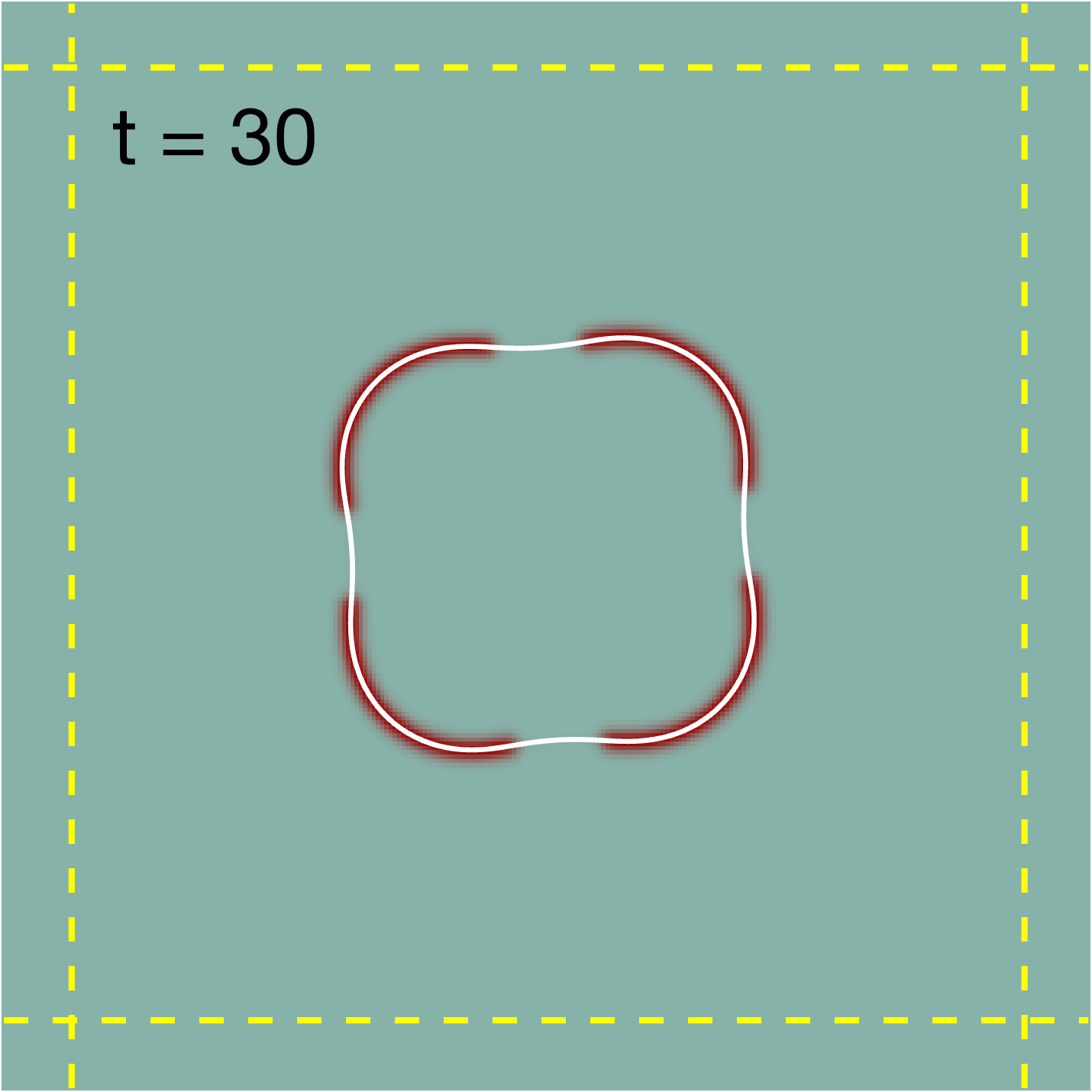}
    \includegraphics[width=0.16\textwidth]{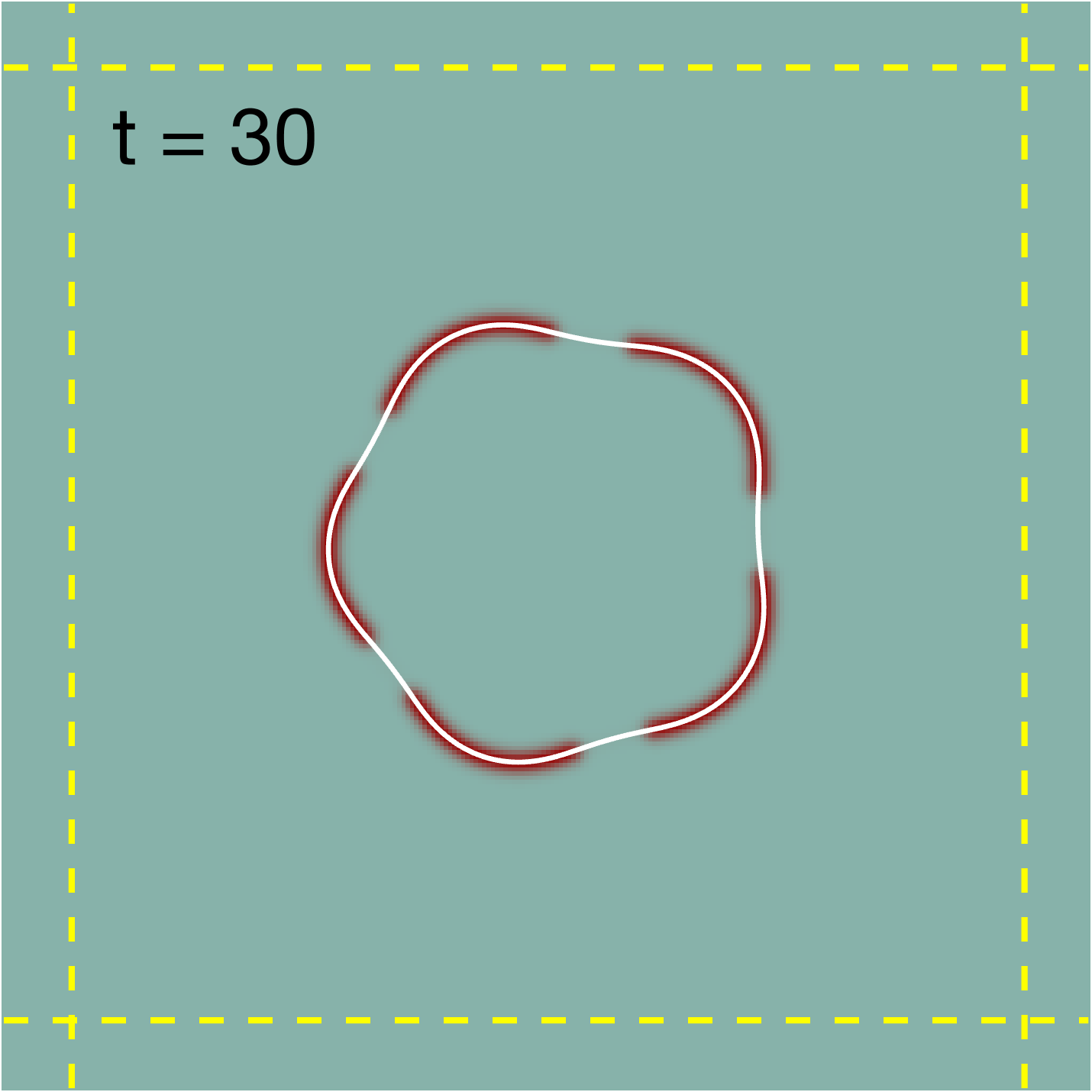}
    \includegraphics[width=0.16\textwidth]{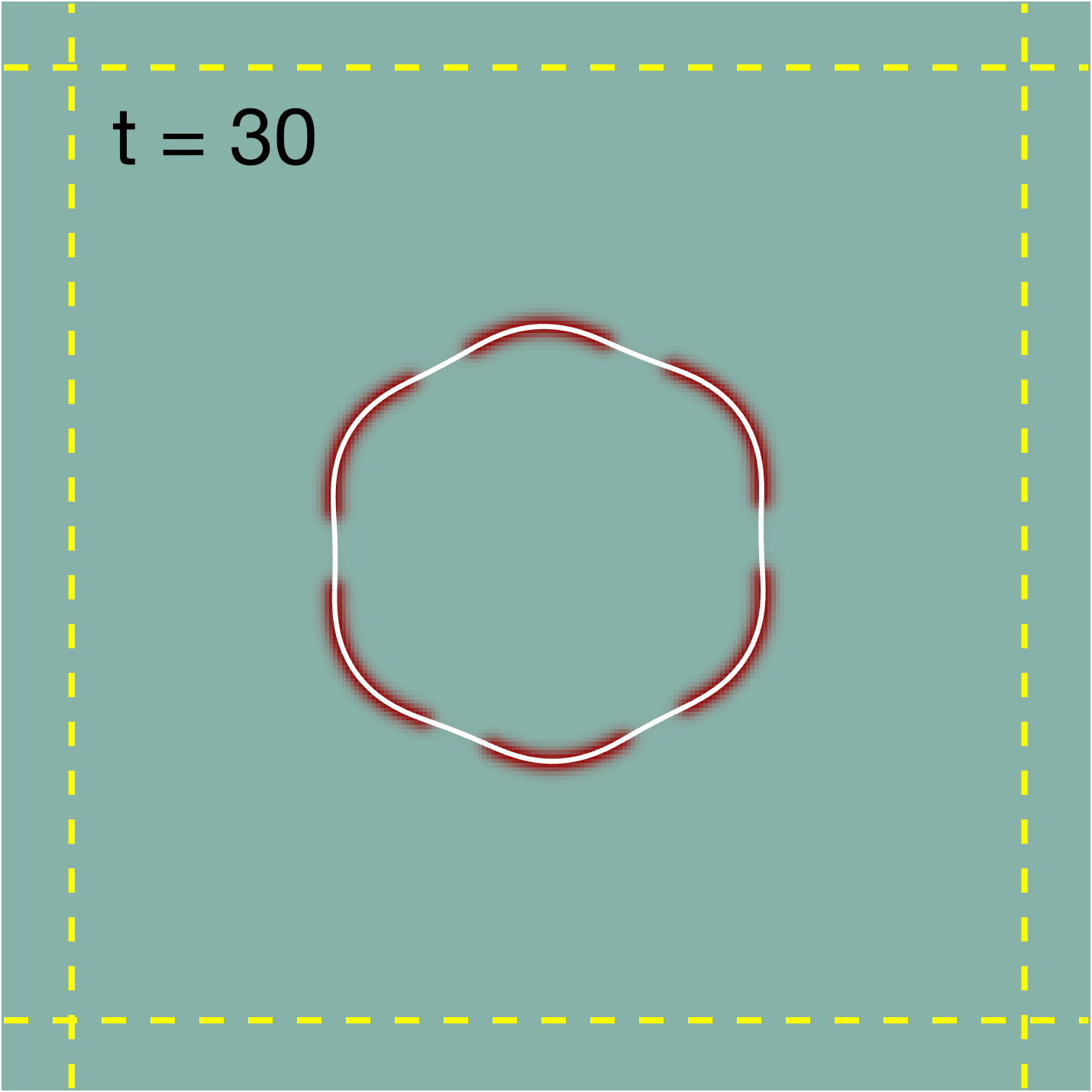}
    \includegraphics[width=0.16\textwidth]{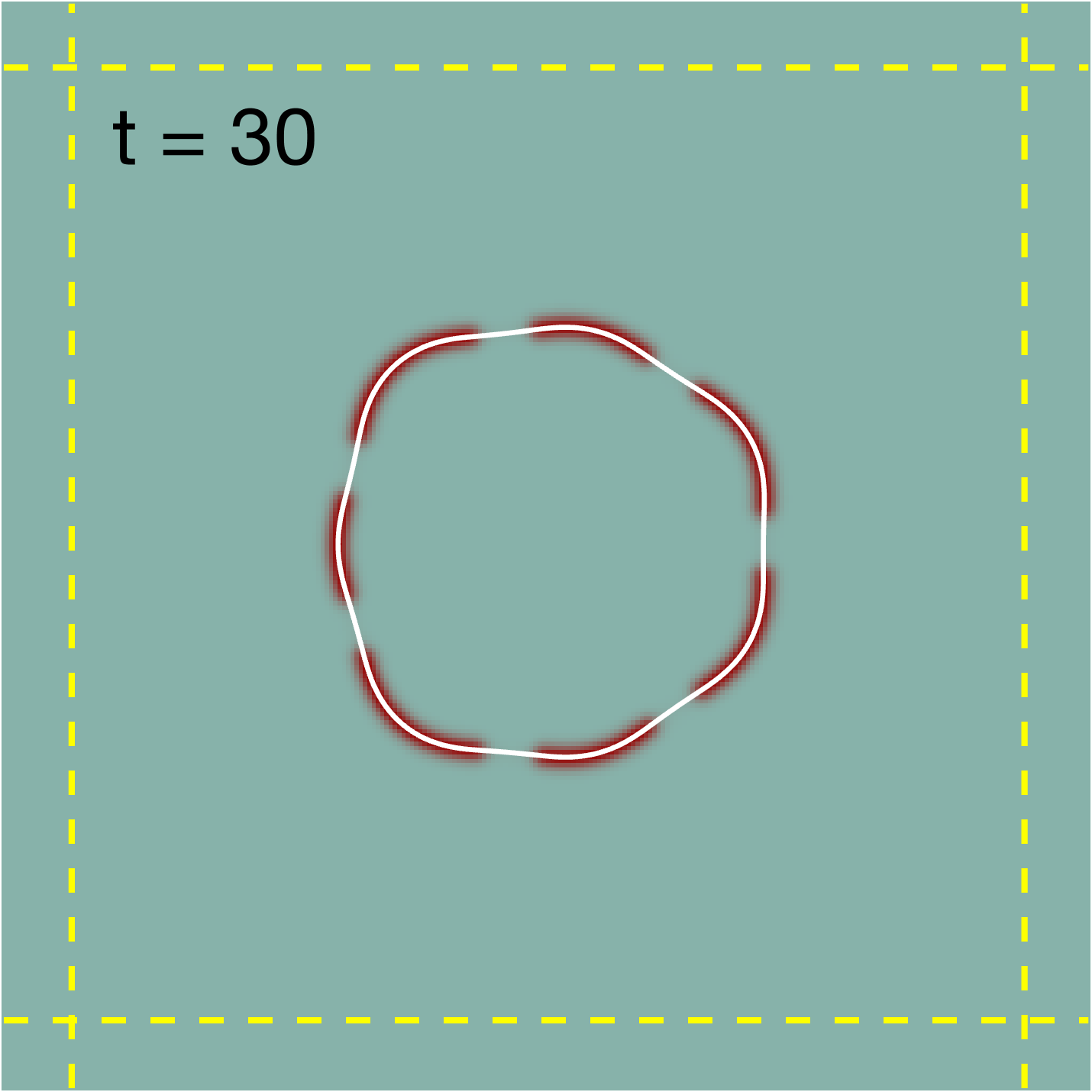}
    \includegraphics[width=0.16\textwidth]{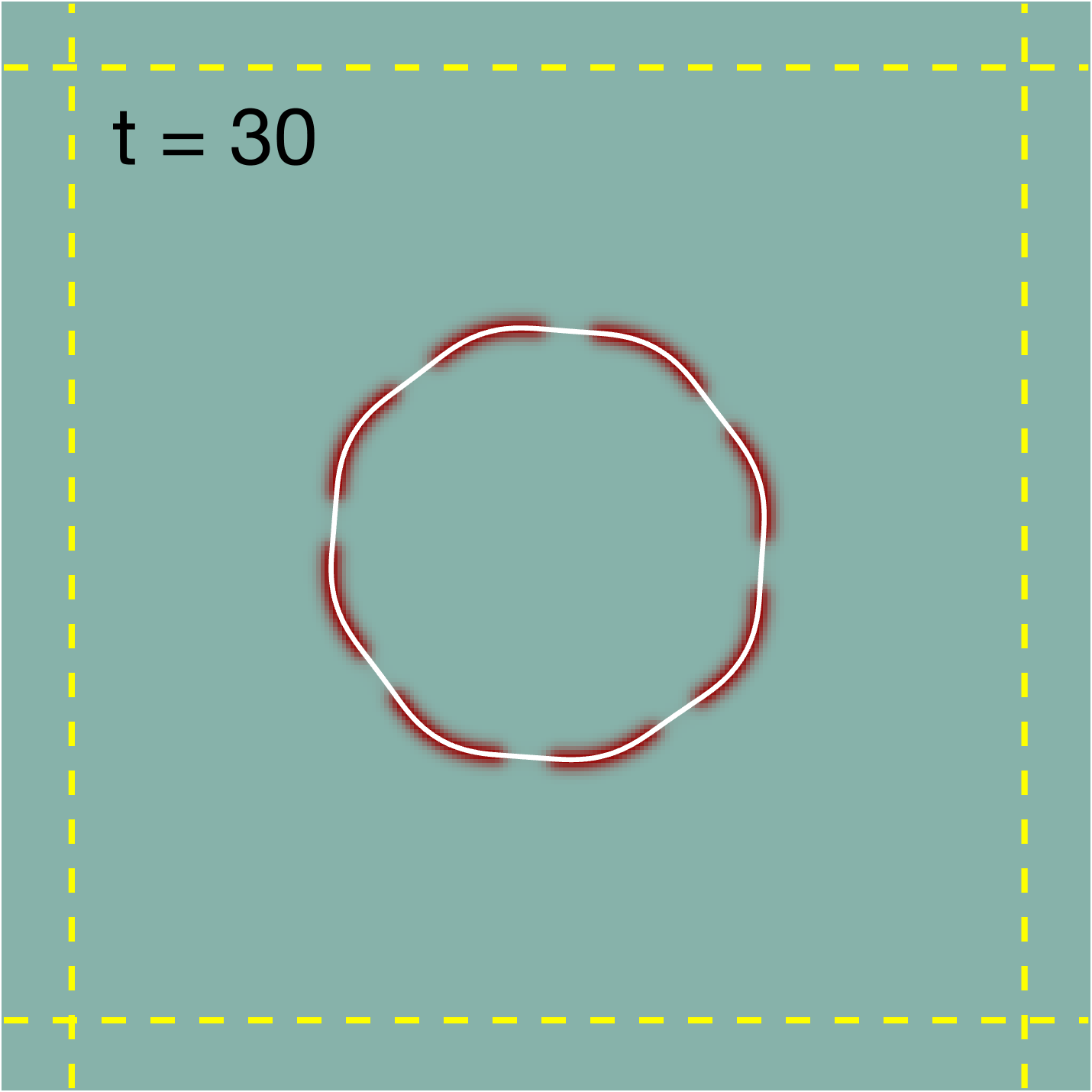}
    \end{center}
    \caption{Effect of long-range repulsive strength $\gamma$ for the phase separation on membrane surface, $\gamma = 60, 90, 125, 155, 190, 210$ for three, four, five, six, seven, and eight protein-rich regions, respectively. Other parameters are given as: $\lambda_{\mathrm{surf}} = 3$, $\lambda_{\mathrm{line}} = 3$, $\kappa=1$, $\alpha = 2$, $u_0 = 0$, $\bar{u} = 0.75$, and $\epsilon_u = 20h_x$. }
    \label{fig:OK_gammaeff}
\end{figure}
\begin{figure}[b!]
    \begin{center}
    \includegraphics[width=0.16\textwidth]{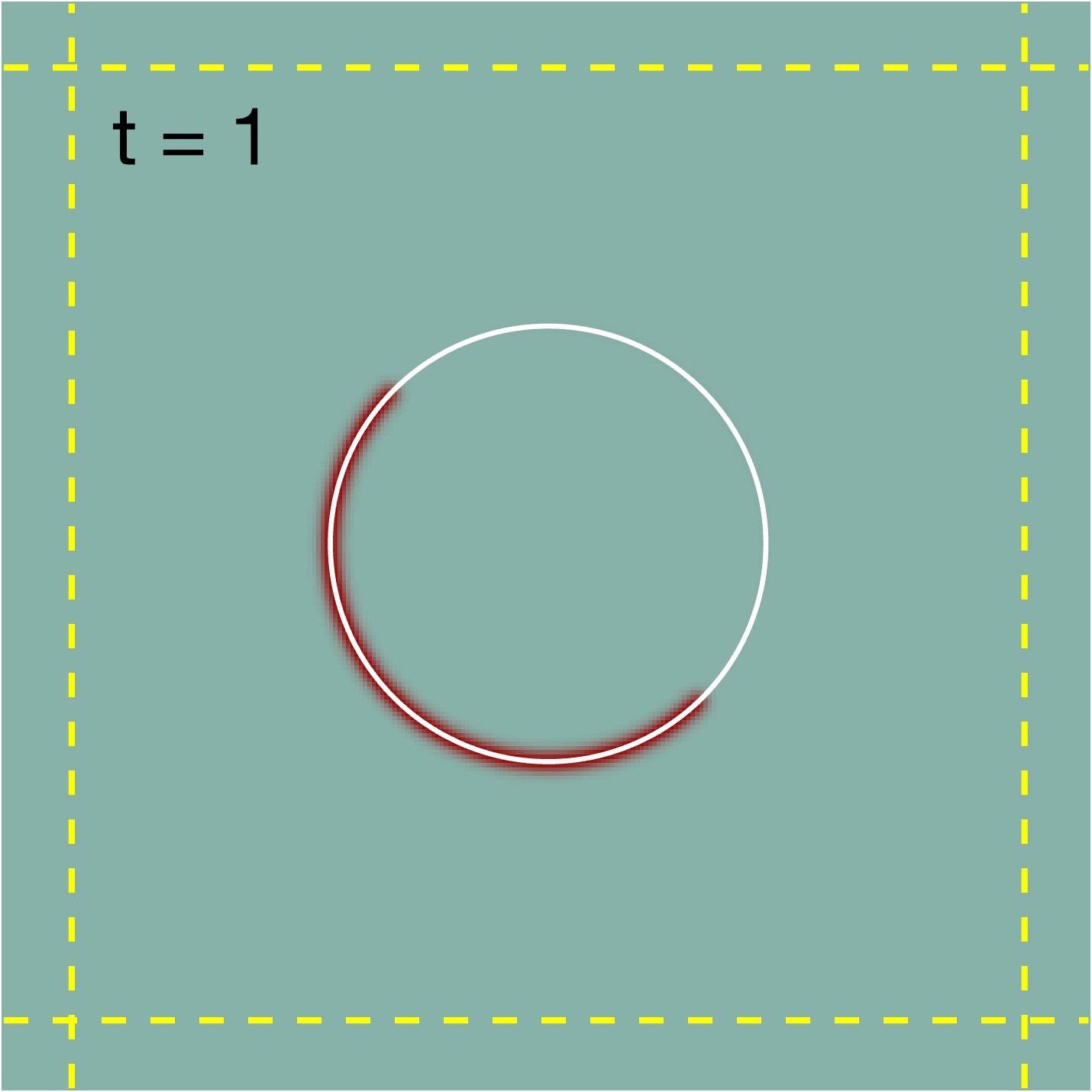}
    \includegraphics[width=0.16\textwidth]{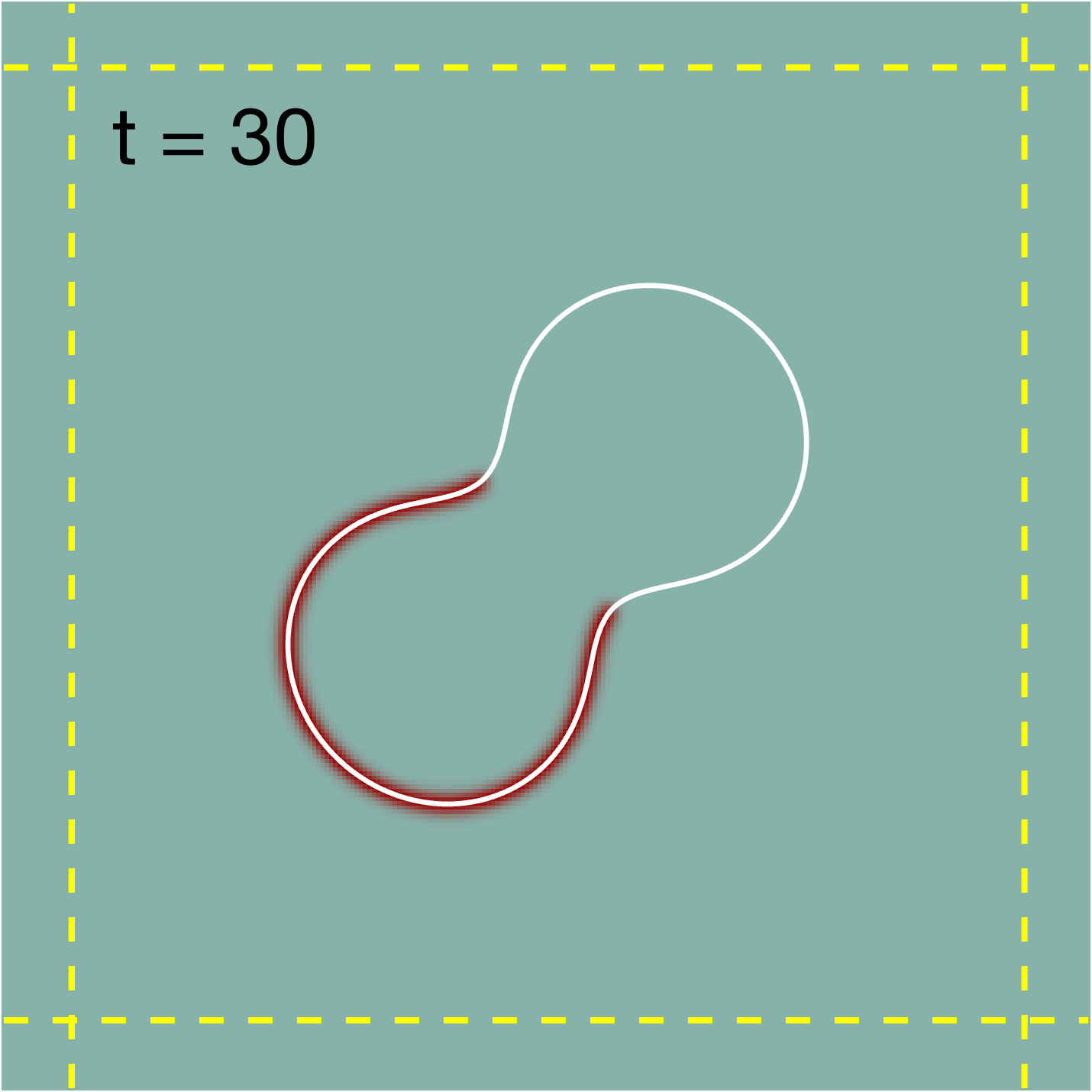}
    \includegraphics[width=0.16\textwidth]{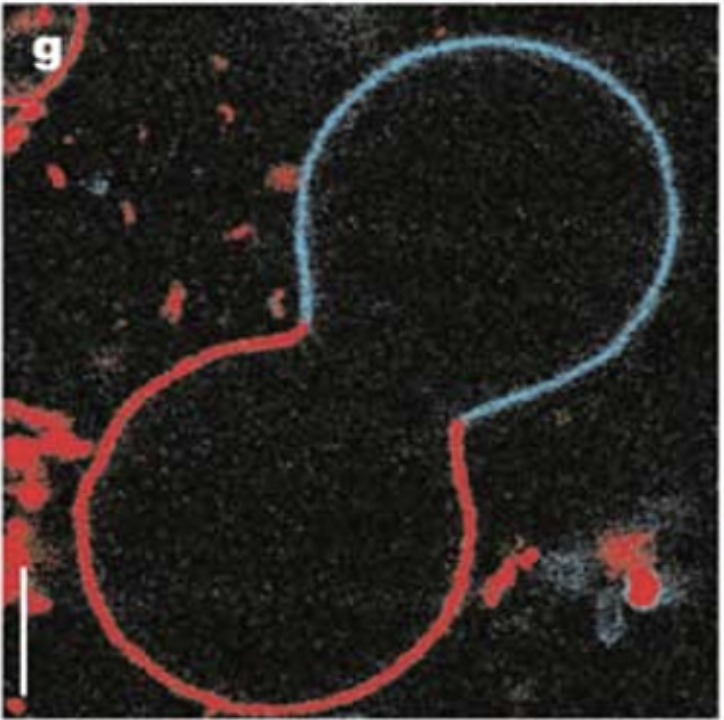} 
    \includegraphics[width=0.16\textwidth]{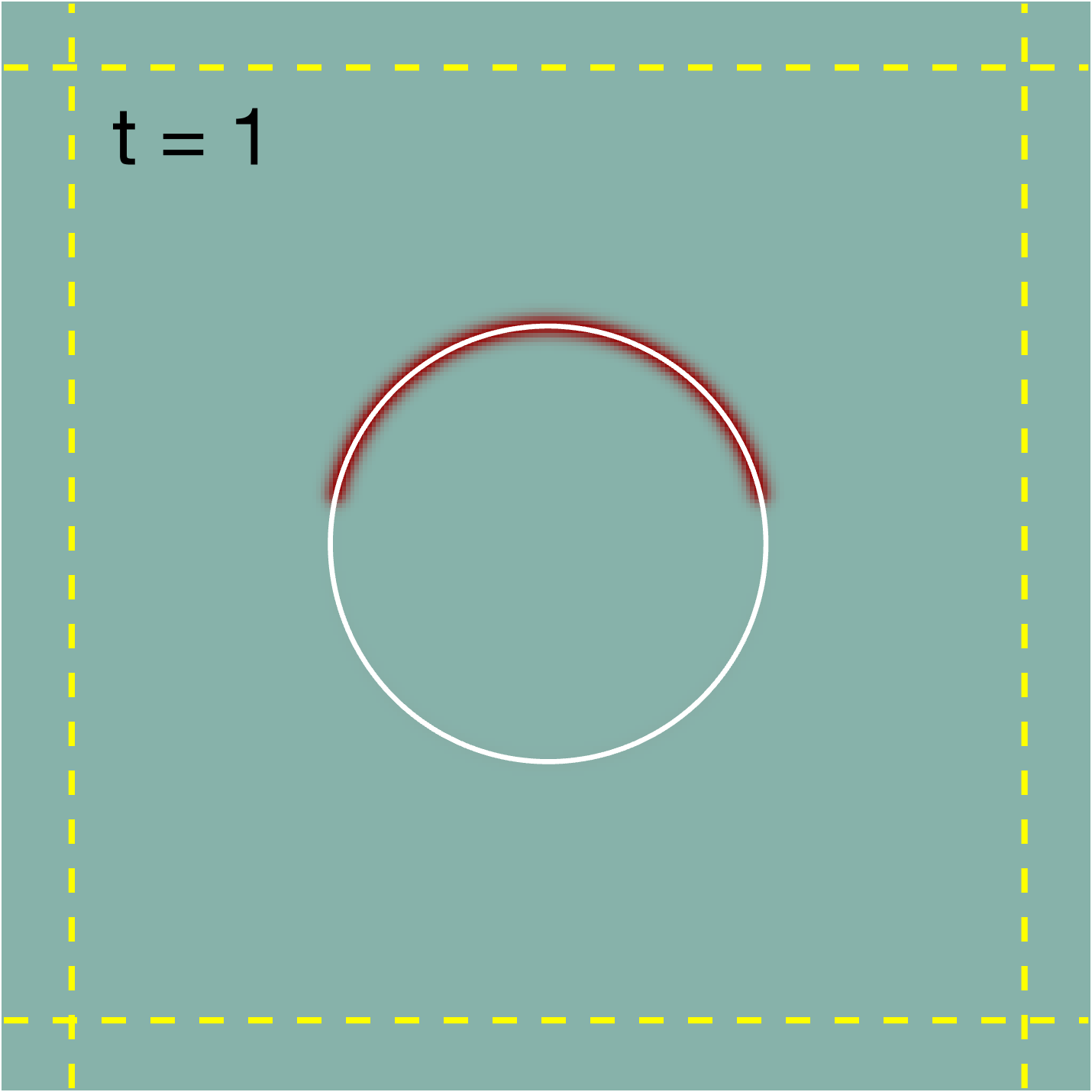}
    \includegraphics[width=0.16\textwidth]{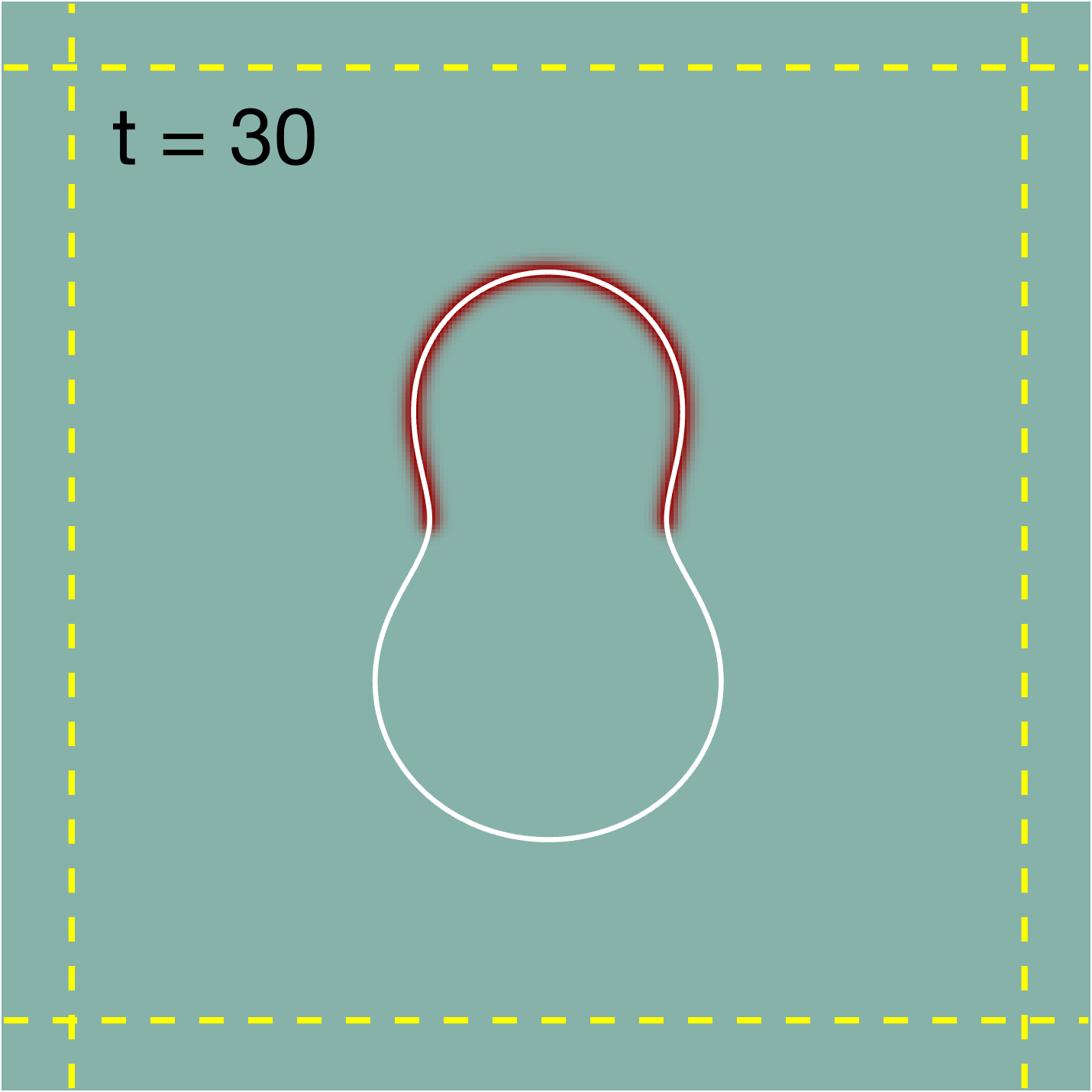}
    \includegraphics[width=0.16\textwidth]{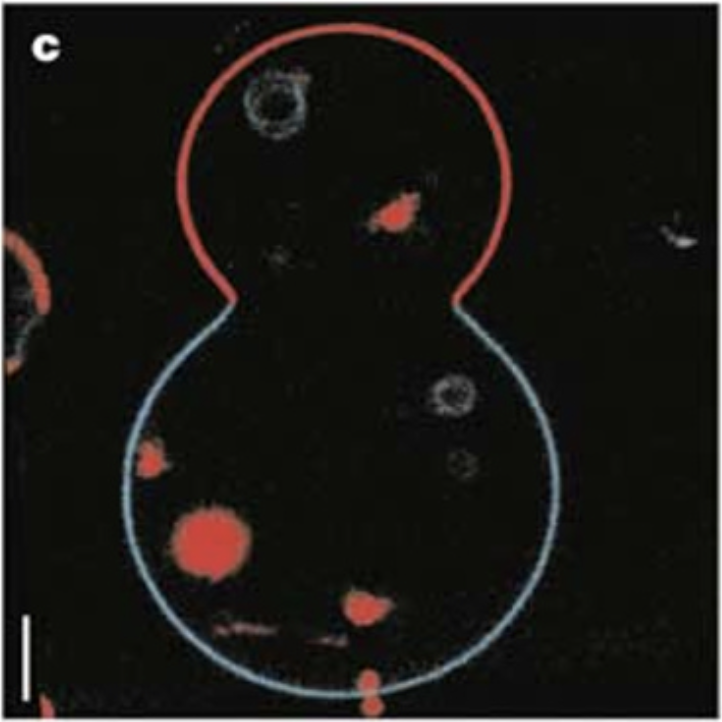} \\
    \includegraphics[width=0.16\textwidth]{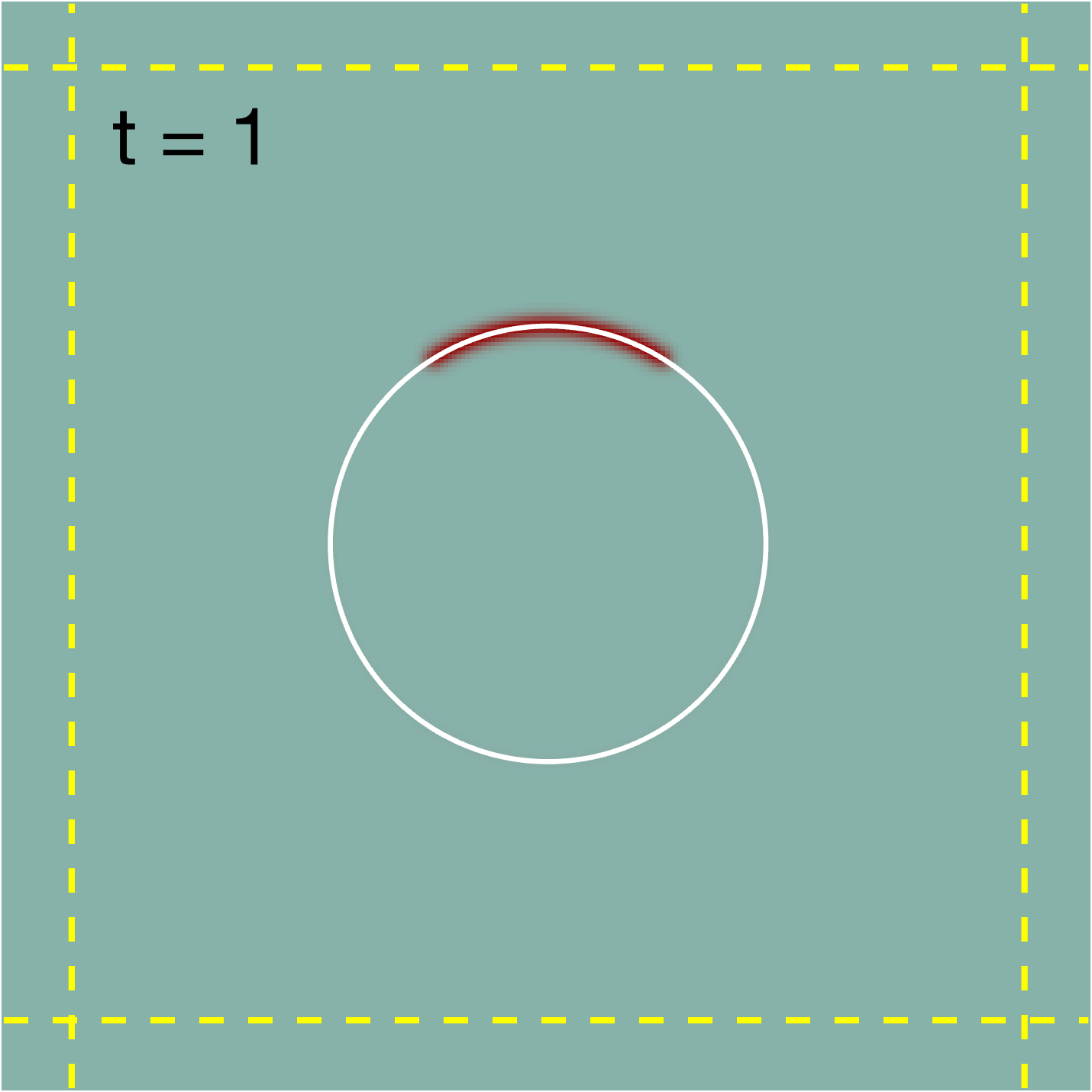}
    \includegraphics[width=0.16\textwidth]{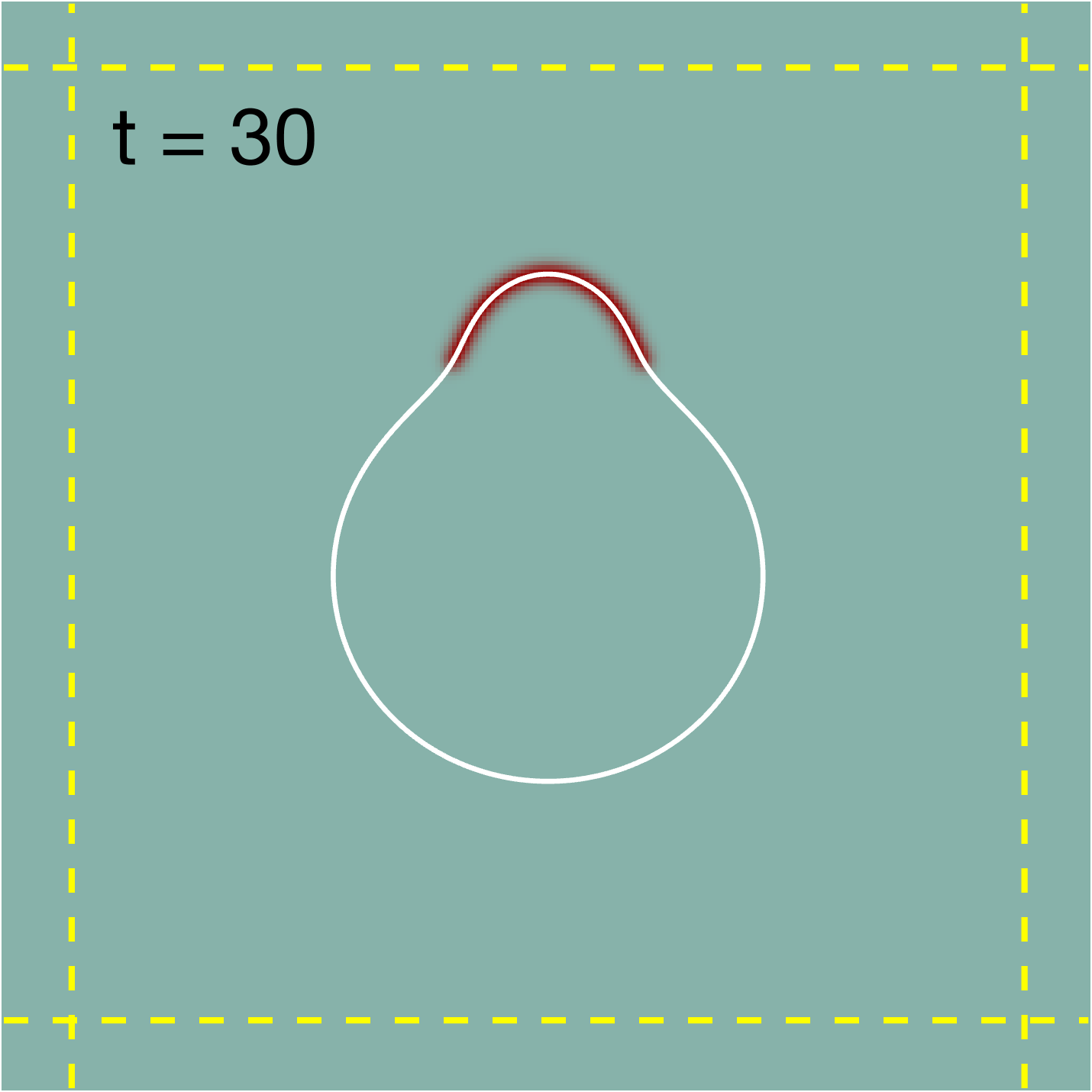}
    \includegraphics[width=0.16\textwidth]{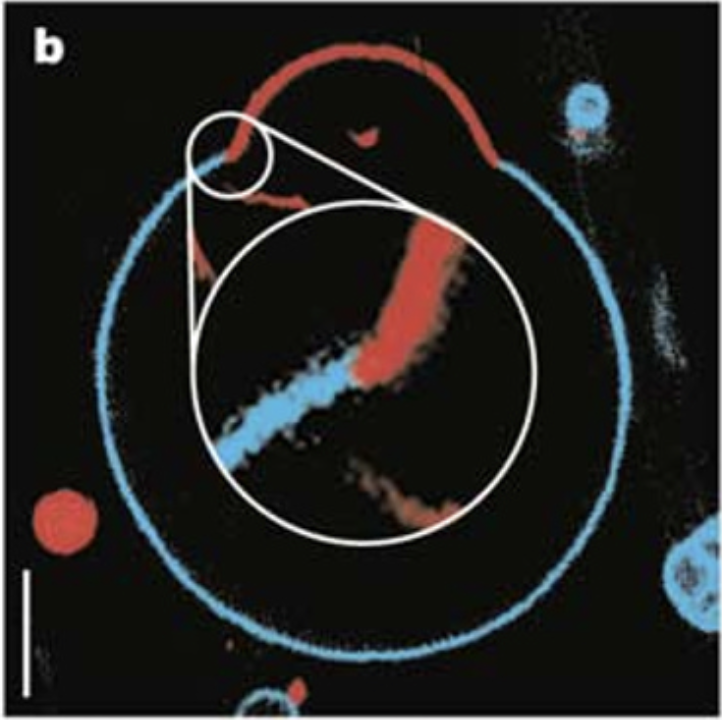} 
    \includegraphics[width=0.16\textwidth]{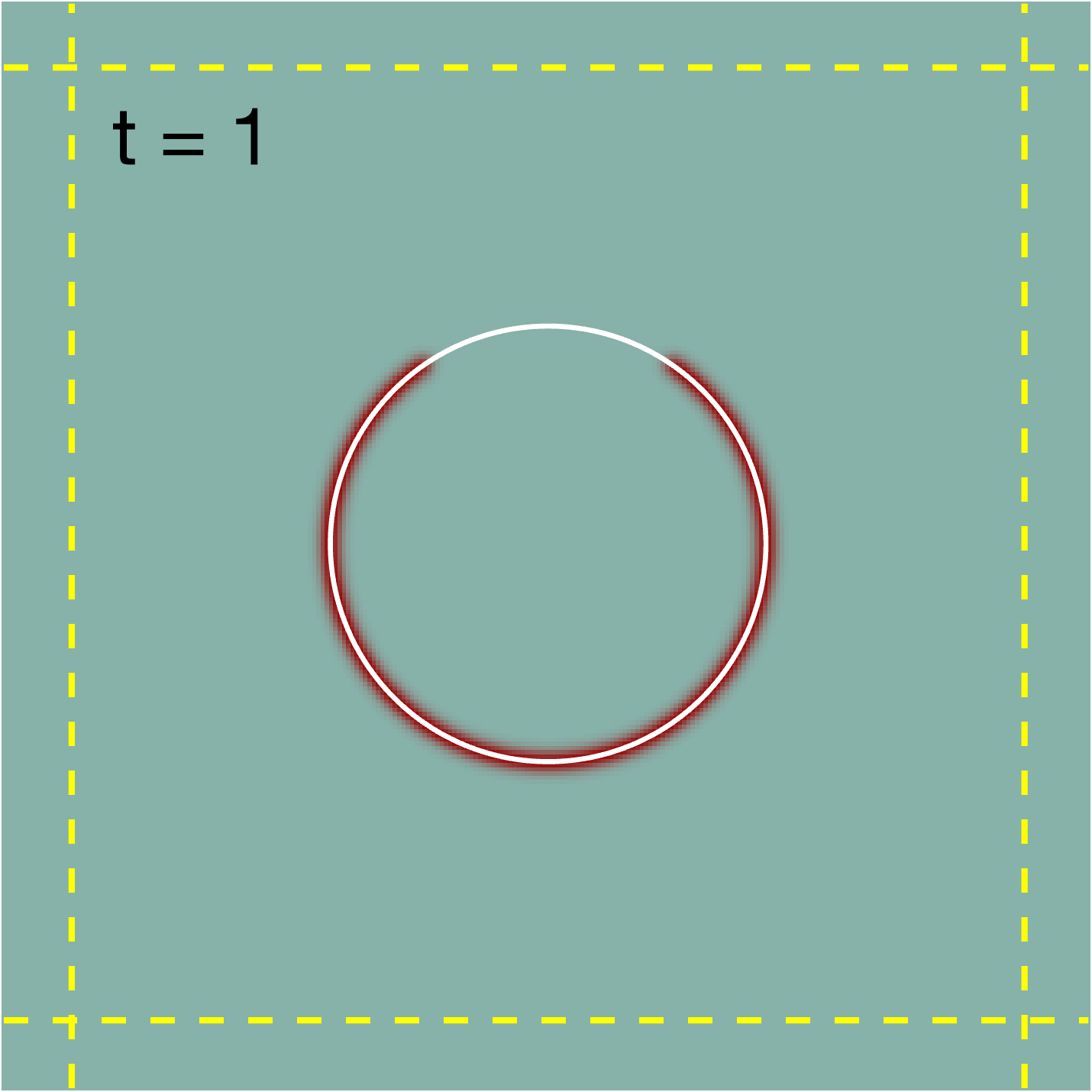}
    \includegraphics[width=0.16\textwidth]{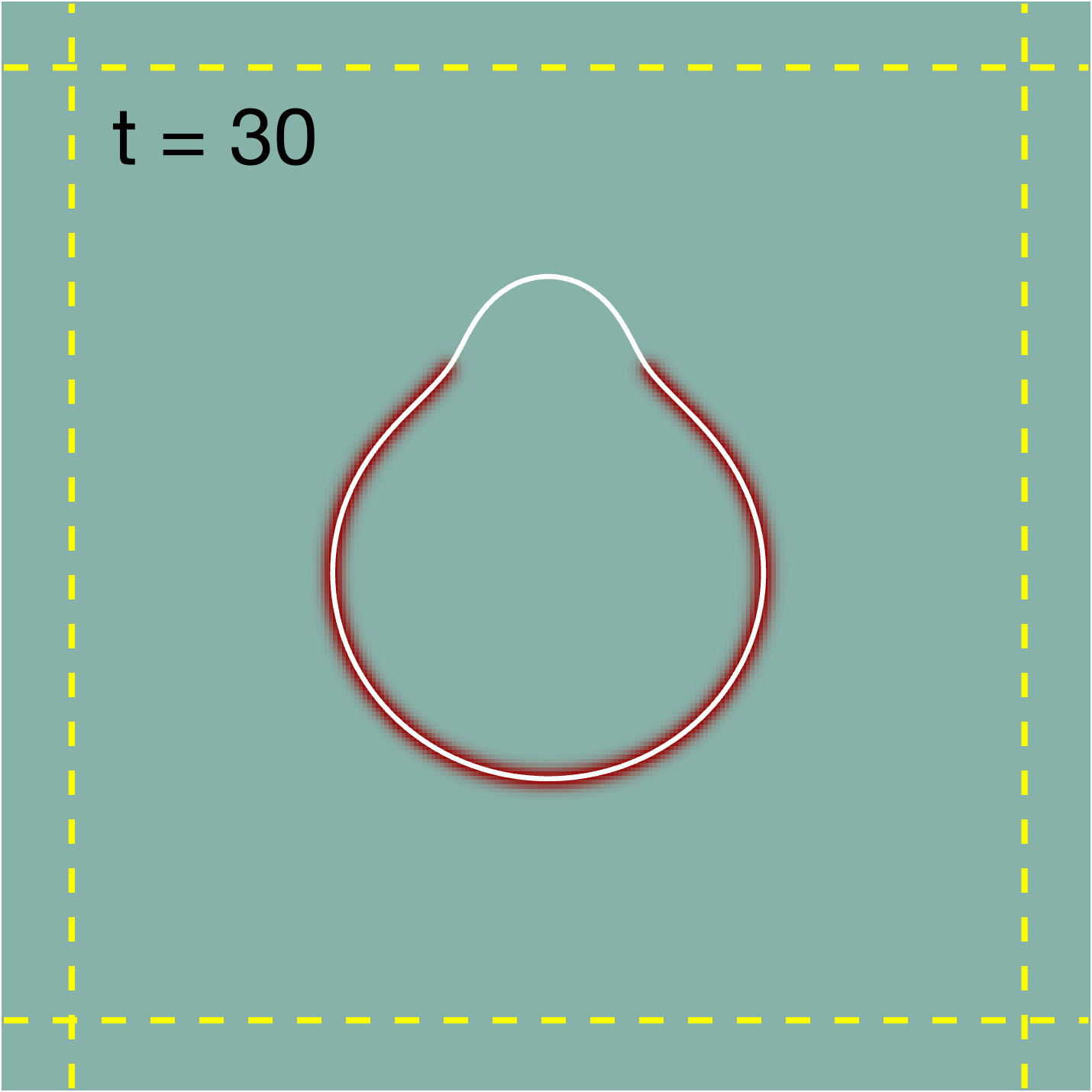}
    \includegraphics[width=0.16\textwidth]{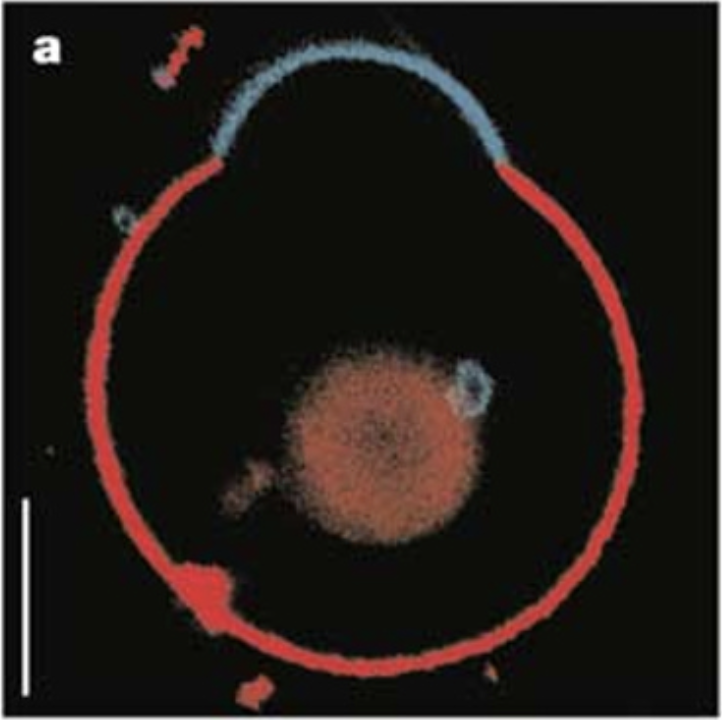}
    \end{center}
    \caption{Four examples with a single protein-rich subdomain on membrane at the dynamical steady state. Top left three: Volumes of the protein-rich and protein-poor domains are equal. Parameter values are taken as: $\lambda_{\mathrm{surf}} = 1$, $\lambda_{\mathrm{line}} = 25$, $\kappa=1$, $\alpha = 0.1$, $u_0 = 9$, $\bar{u} = 0.5$, $\gamma = 5$, and $\epsilon_u = 20h_x$; Top right three: volume of the red region is slightly lower than that of the white region. Parameter values are: $\lambda_{\mathrm{surf}} = 2$, $\lambda_{\mathrm{line}} = 35$, $\kappa=1$, $\alpha = 0.15$, $u_0 = 9$, $\bar{u} = 0.4$, $\gamma = 5$, and $\epsilon_u = 20h_x$; Bottom left three: membrane with a small protein-rich domain. Parameter values are: $\lambda_{\mathrm{surf}} = 1$, $\lambda_{\mathrm{line}} = 15$, $\kappa=1$, $\alpha = 0.4$, $u_0 = 9$, $\bar{u} = 0.2$, $\gamma = 5$, and $\epsilon_u = 20h_x$; Bottom right three: A membrane with a small protein-poor domain. Parameter values are: $\lambda_{\mathrm{surf}} = 1$, $\lambda_{\mathrm{line}} = 15$, $\kappa=1$, $\alpha = 0.1$, $u_0 = 7$, $\bar{u} = 0.8$, $\gamma = 5$, and $\epsilon_u = 20h_x$.} 
    \label{fig:OK_1red}
\end{figure}
For the numerical experiment presented in Figure \ref{fig:OK_3b}, we show snapshots at several time points to illustrate the microphase separation of $u$ on the membrane and the resulting membrane shape evolution. 
Starting from a random initial $u^0$ at $t=0$, we first test the OK model on a fixed membrane defined in \eqref{eqn:phi_initial} up to $t = 20$.
Phase separation occurs rapidly, where red regions indicate protein-rich subdomains and white regions denote protein-poor subdomains. At $t = 1$, {three} protein-rich subdomains appear with different sizes and distances. 
By $t = 20$, these red regions evolve into nearly equal-sized domains that are uniformly spaced. We then allow the cell membrane to deform by solving the coupled system \eqref{eqn:model_force},\eqref{eqn:model_OK}, yielding the results at $t = 30$ and $t = 50$. The simulation is consistent with the experimental observation in Figure 1(e) of \cite{baumgart2003imaging}. 

Next, we investigate the influence of the repulsive strength $\gamma$ in the membrane-associated OK model on protein segregation along the membrane surface, as illustrated in Figure \ref{fig:OK_gammaeff}. Starting from a random initial, the equilibrium states of the coupled system from left to right display three, four, five, six, seven, and eight protein-rich domains (red regions) corresponding to $\gamma = 60, 90, 125, 155, 190, 210$, respectively. These results indicate that larger values of $\gamma$ promote protein segregation on the cell membrane, leading to more protein-rich subdomains. This trend is consistent with the behavior of the OK model on different geometries in our previous work \cite{LuoZhao_NumPDE2024,luo2025fourier,LuoZhao_AAMM2024}.
% {\color{red} Wangbo, remember to make the font of $t= 30$ larger. Meanwhile, it looks to me that the seven bump case has not reach to the equilibrium yet. }

Figure \ref{fig:OK_1red} examines four cases in which the initial condition $u^0$ consists of a single protein-rich region of varying size and location on the fixed membrane of $\phi^0$. At $t = 1$, the protein-rich area evolves with the prescribed {volume} $\bar{u}$, and by $t = 30$, the membrane deforms according to the coupled system. At this stage, the system reaches a dynamical steady state in which the cell shape remains unchanged, but the cell continues to move slowly due to the chemical force generated by the protein-rich domain. These simulations closely match the experimental results in \cite{baumgart2003imaging}, see the last column. 
% {\color{red} so here all the figures at t = 30 is a dynamical steady state, the cell still solwly moving }

\begin{figure}[t]
    \begin{center}
    \includegraphics[width=0.8\textwidth]{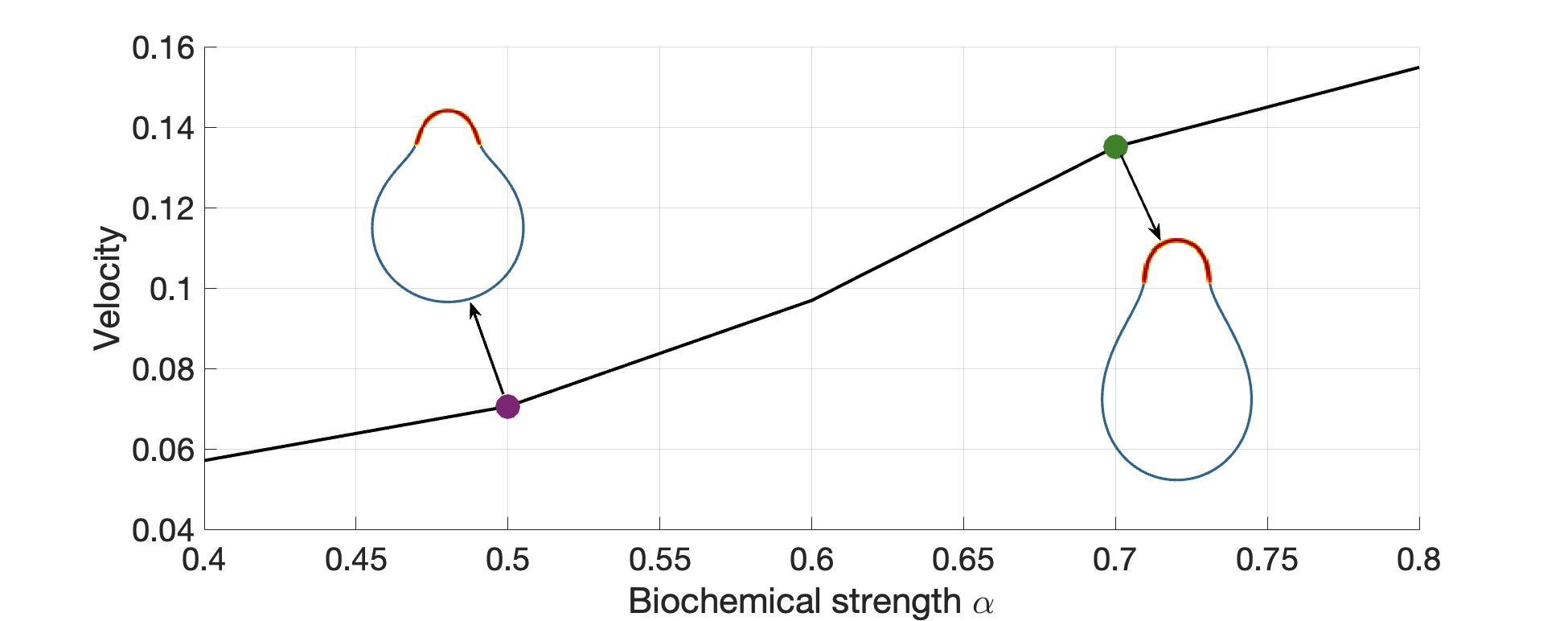}
    \end{center}
    \vspace{-4mm}
    \caption{Cell velocity versus biochemical strength $\alpha$ for a cell with a small protein-rich domain (red region). Left insert subfigure is for $\alpha=0.5$, and right insert subfigure is for $\alpha=0.7$. Other parameter values are: $\lambda_{\mathrm{surf}} = 1$, $\lambda_{\mathrm{line}} = 15$, $\kappa=1$, $u_0 = 9$, $\bar{u} = 0.2$, $\gamma = 5$, and $\epsilon_u = 20h_x$.} 
    \label{fig:OK_Cellmove}
\end{figure}

Since a cell with a single protein-rich subdomain reaches the steady state, we further explore how the cell's velocity at steady state is influenced by the system parameters, specifically the biochemical strength $\alpha$, as shown in Figure \ref{fig:OK_Cellmove}.
% \sout{This figure provides a quantitative study of the relationship between $\alpha$ and cell velocity, where velocity is defined as the maximum speed of cell movement.} 
The two inset panels display the steady states for $\alpha = 0.5$ and $\alpha = 0.7$. Increasing $\alpha$ strengthens the biochemical activity, leading to higher cell speeds and greater elongation of the steady state shape. This numerical result shows how $\alpha$ modulates both cell motility and membrane morphology through the interaction of protein-induced forces with membrane mechanics.

\begin{figure}[t]
    \begin{center}
    \includegraphics[width=0.18\textwidth]{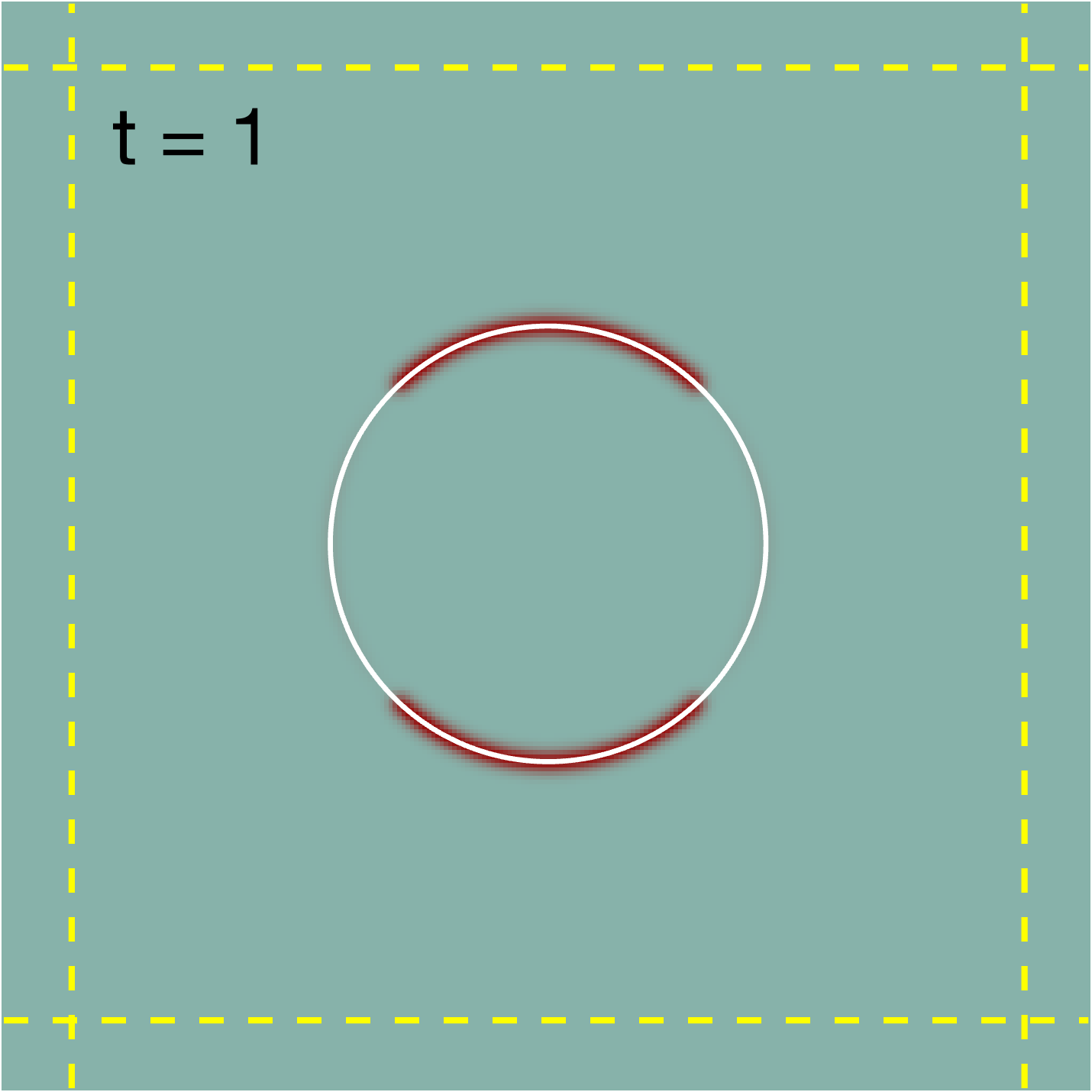}
    \includegraphics[width=0.18\textwidth]{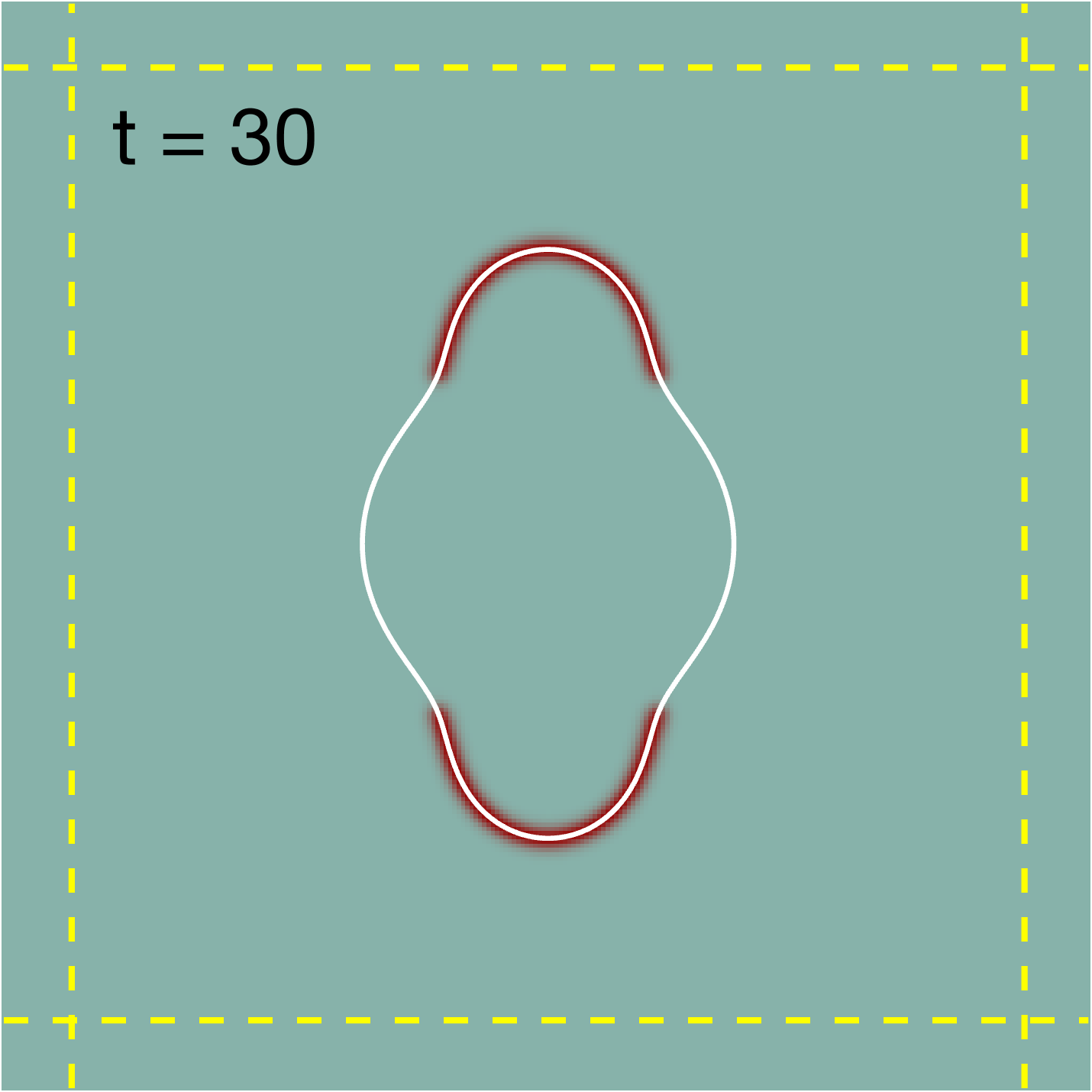}
    \includegraphics[width=0.18\textwidth]{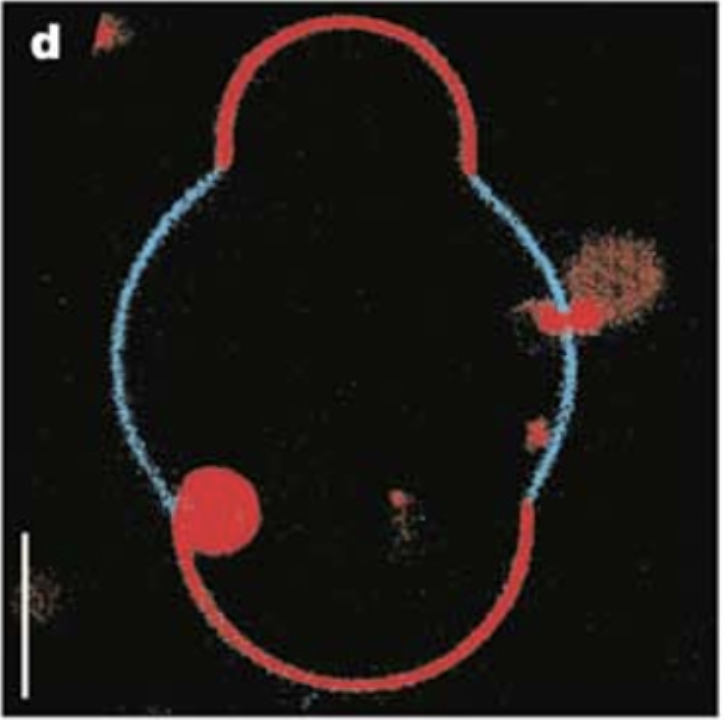} \\
    \includegraphics[width=0.18\textwidth]{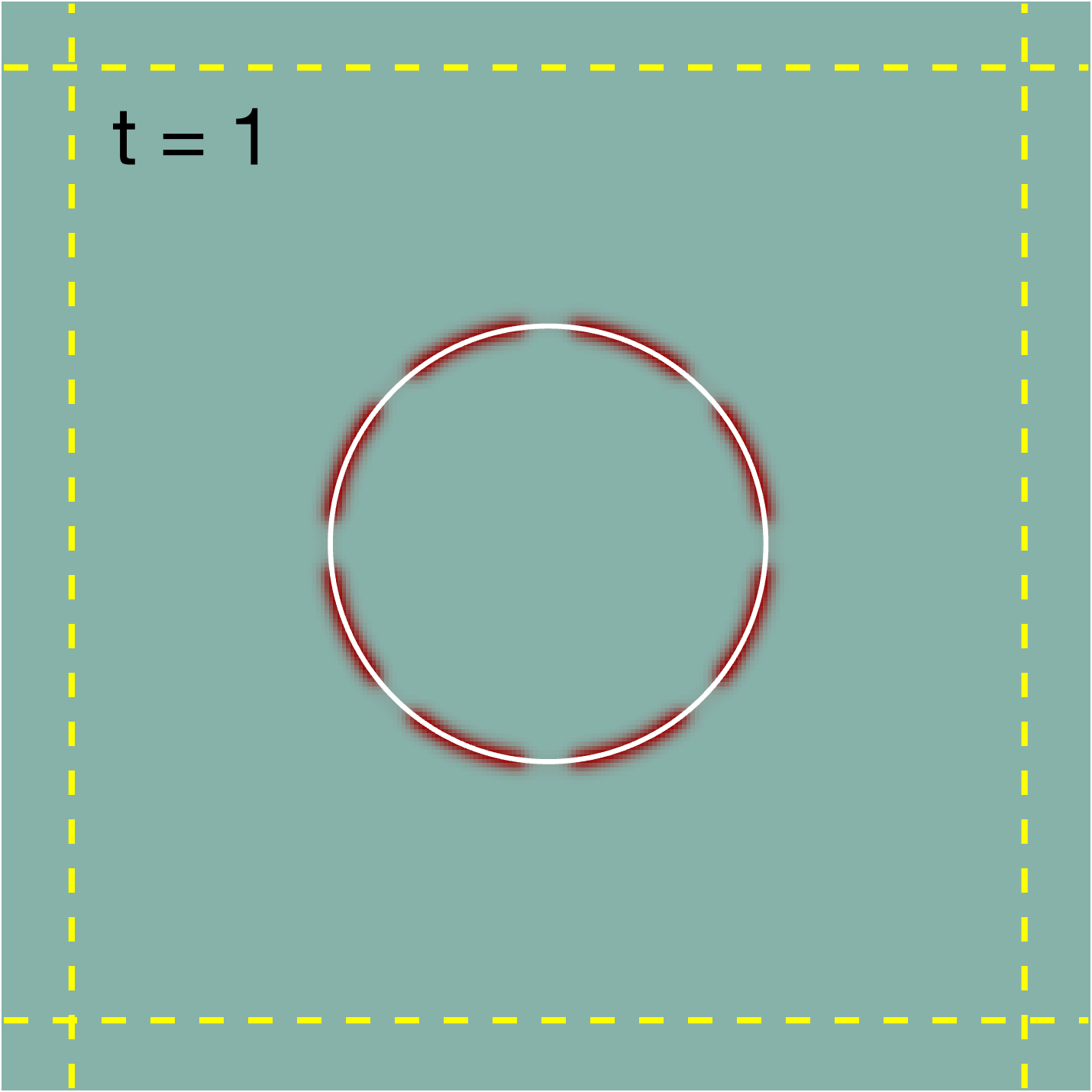}
    \includegraphics[width=0.18\textwidth]{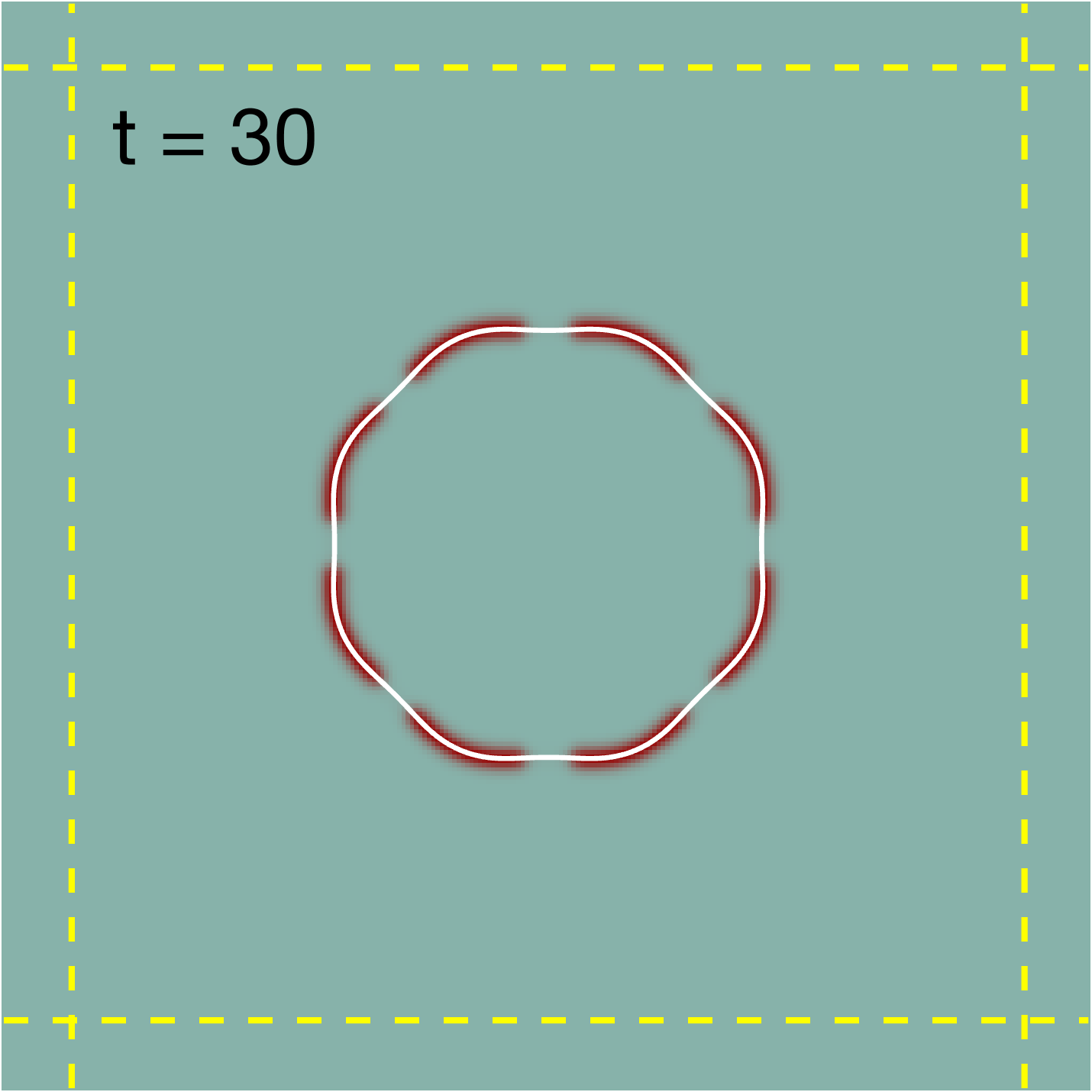}
    \includegraphics[width=0.18\textwidth]{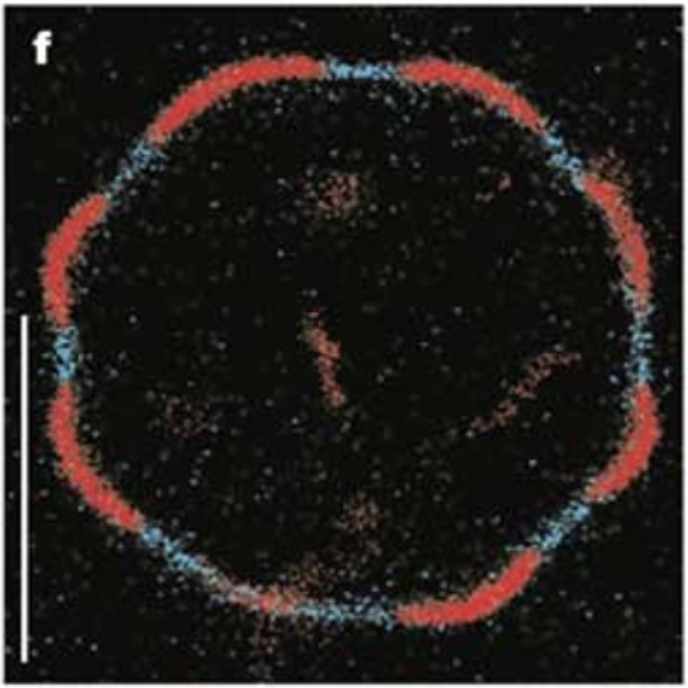} \\
    \includegraphics[width=0.18\textwidth]{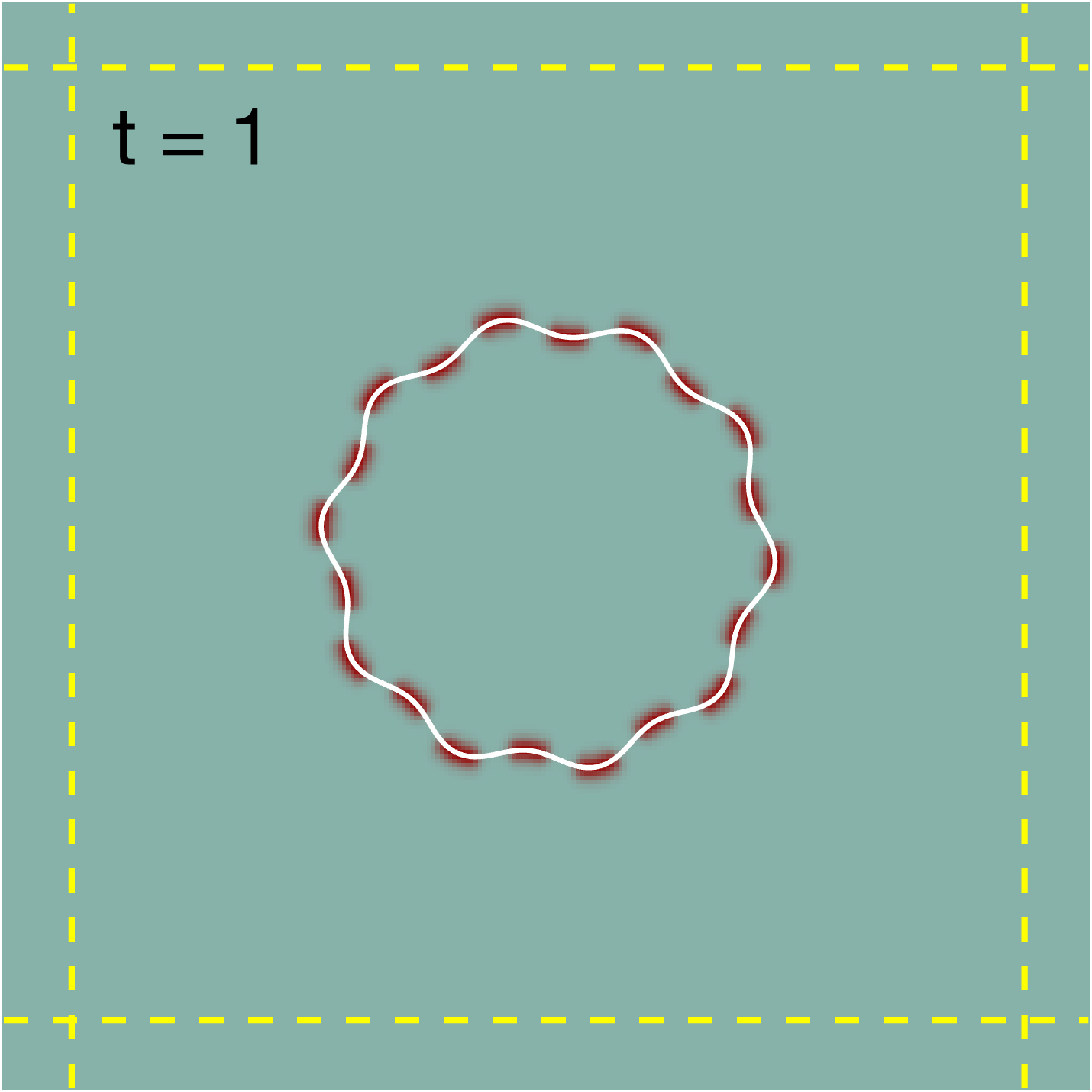}
    \includegraphics[width=0.18\textwidth]{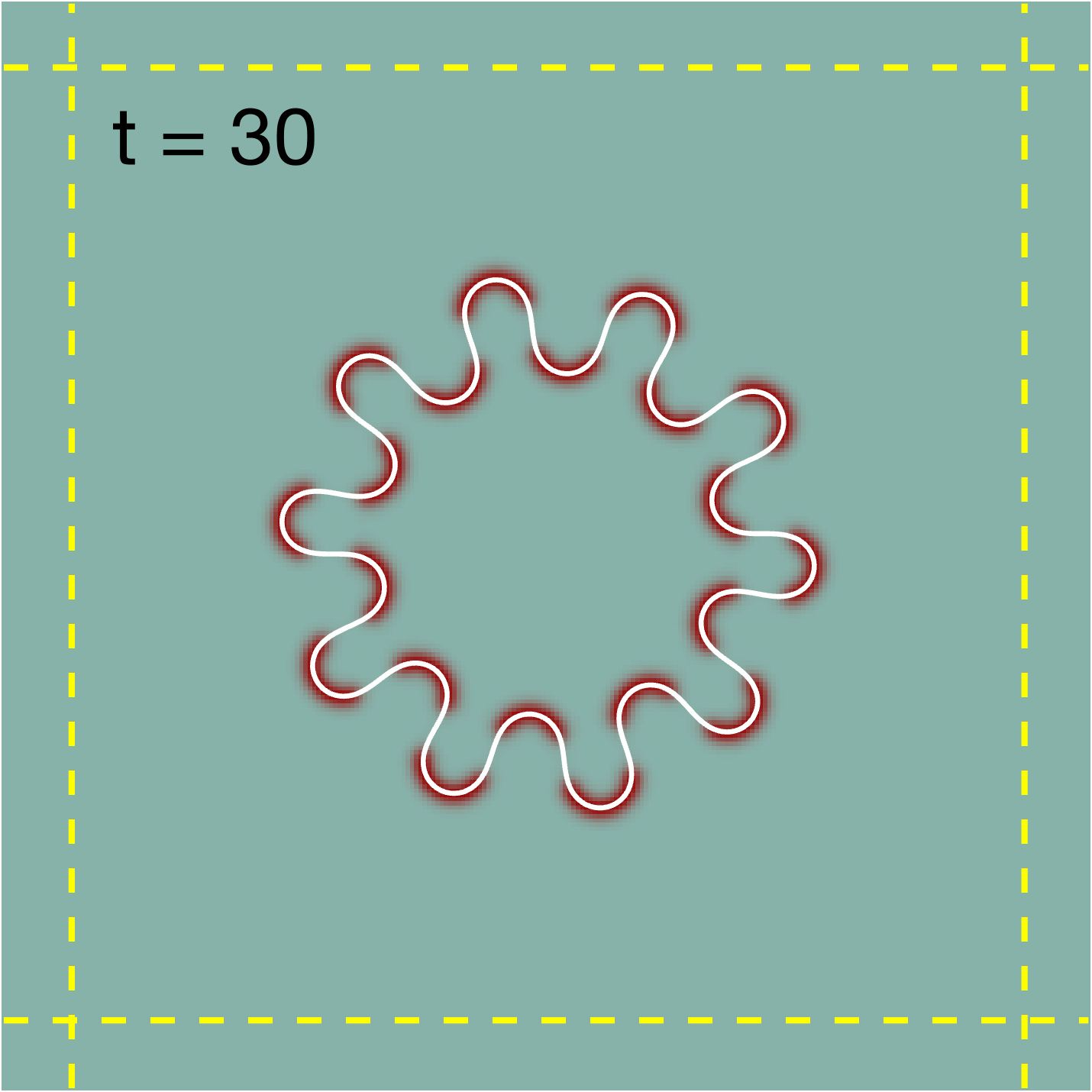} 
    \includegraphics[width=0.18\textwidth]{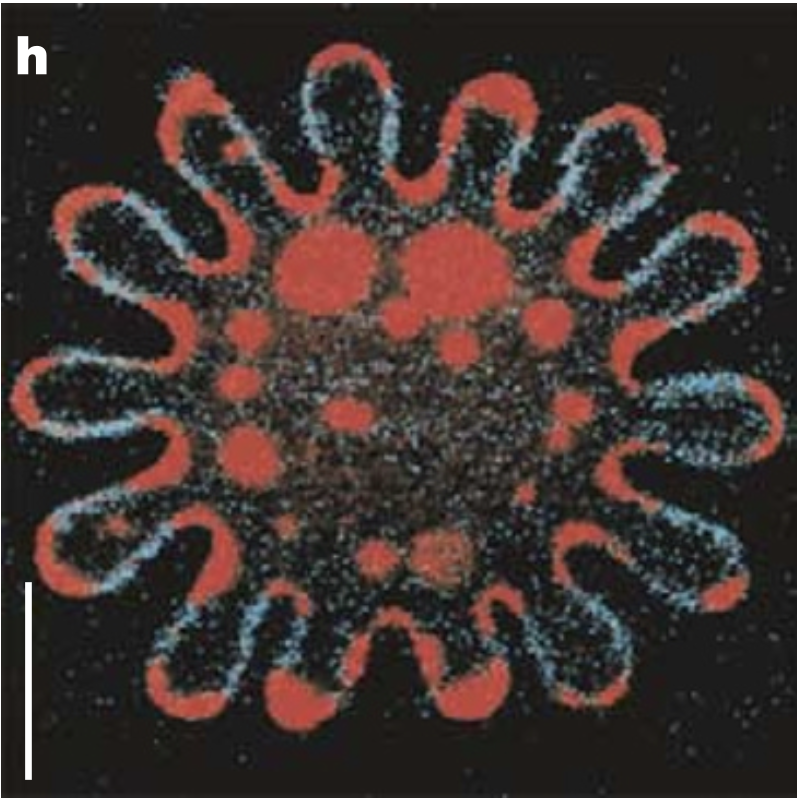}
    \end{center}
    \caption{Three examples with multiple protein-rich subdomains. Top row: membrane with two protein-rich subdomains. Parameter values are given as: $\lambda_{\mathrm{surf}} = 1$, $\lambda_{\mathrm{line}} = 17$, $\kappa=1$, $\alpha = 0.5$, $u_0 = 0.2$, $\bar{u} = 0.5$, $\gamma = 30$, and $\epsilon_u = 20h_x$. Middle row: eight-bump membrane. Parameter values are $\lambda_{\mathrm{surf}} = 3$, $\lambda_{\mathrm{line}} = 3$, $\kappa=1$, $\alpha = 3$, $u_0 = 0.2$, $\bar{u} = 0.75$, $\gamma = 200$, and $\epsilon_u = 20h_x$. Bottom row: ten-finger membrane. Parameter values are: $\lambda_{\mathrm{surf}} = 0.7$, $\lambda_{\mathrm{line}} = 3$, $\kappa=1$, $\alpha = 10$, $u_0 = 0$, $\bar{u} = 0.6$, $\gamma = 400$, and $\epsilon_u = 20h_x$.}
    \label{fig:OK_multiplered}
\end{figure}

In Figure \ref{fig:OK_multiplered}, we present several examples, each with multiple protein-rich (red) subdomains. The top row starts an initial $u^0$ consisting of $2$ protein-rich subdomains located at the top and bottom of the membrane of $\phi^0$. At $t = 1$, the protein-rich subdomains evolve to equal size and are equally spaced. By $t = 30$, the membrane deforms while maintaining the two protein-rich subdomains, reproducing the experimental observation in Figure 1(d) of \cite{baumgart2003imaging}. The middle row takes the initial $u^0$ with eight protein-rich regions distributed across the fixed membrane of $\phi^0$, and eventually reaches equilibrium with eight bumps. The last row is a ten-finger case in which all the inward red subdomains exert inward biochemical forces.

\begin{figure}[t]
    \begin{center}
    \includegraphics[width=0.18\textwidth]{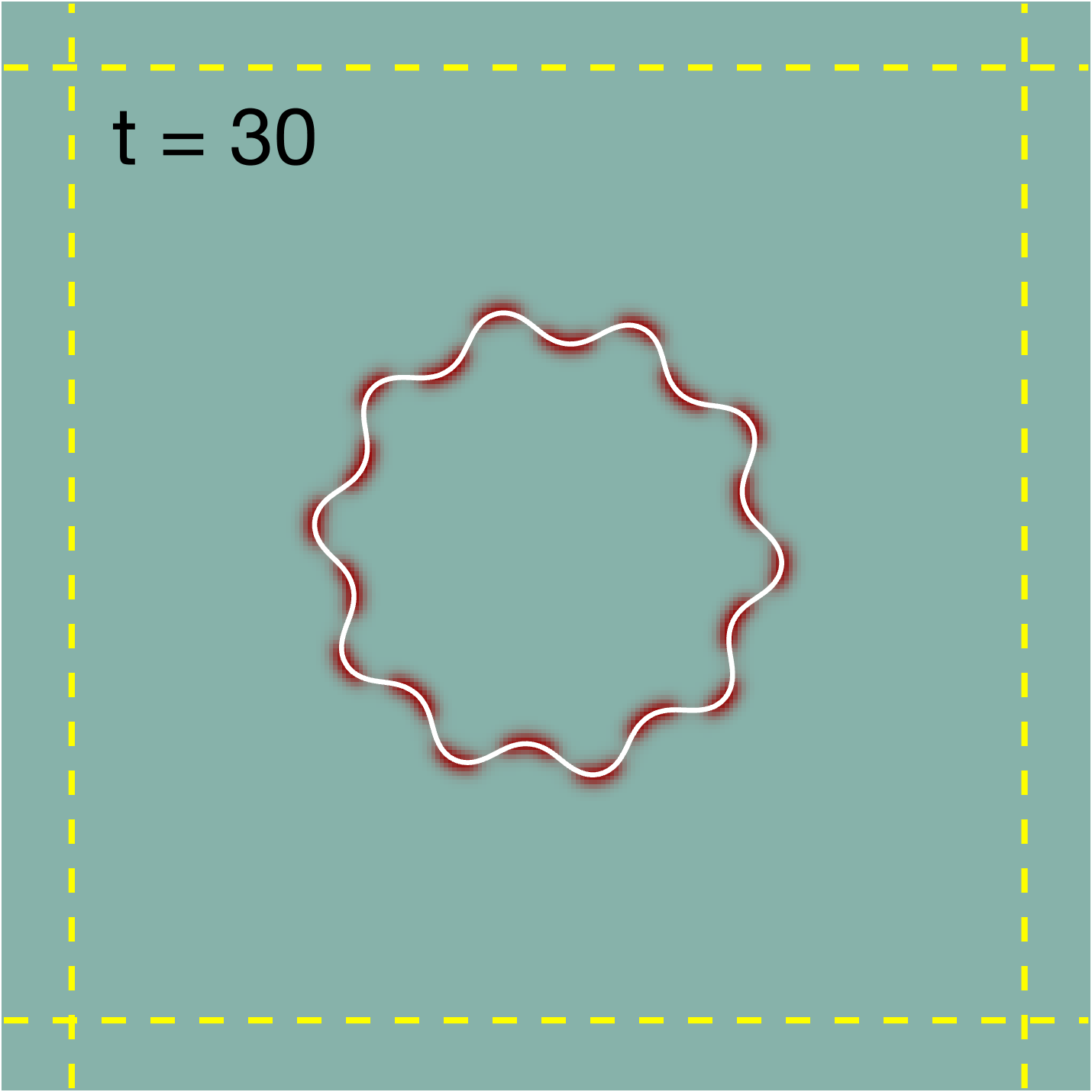}
    \includegraphics[width=0.18\textwidth]{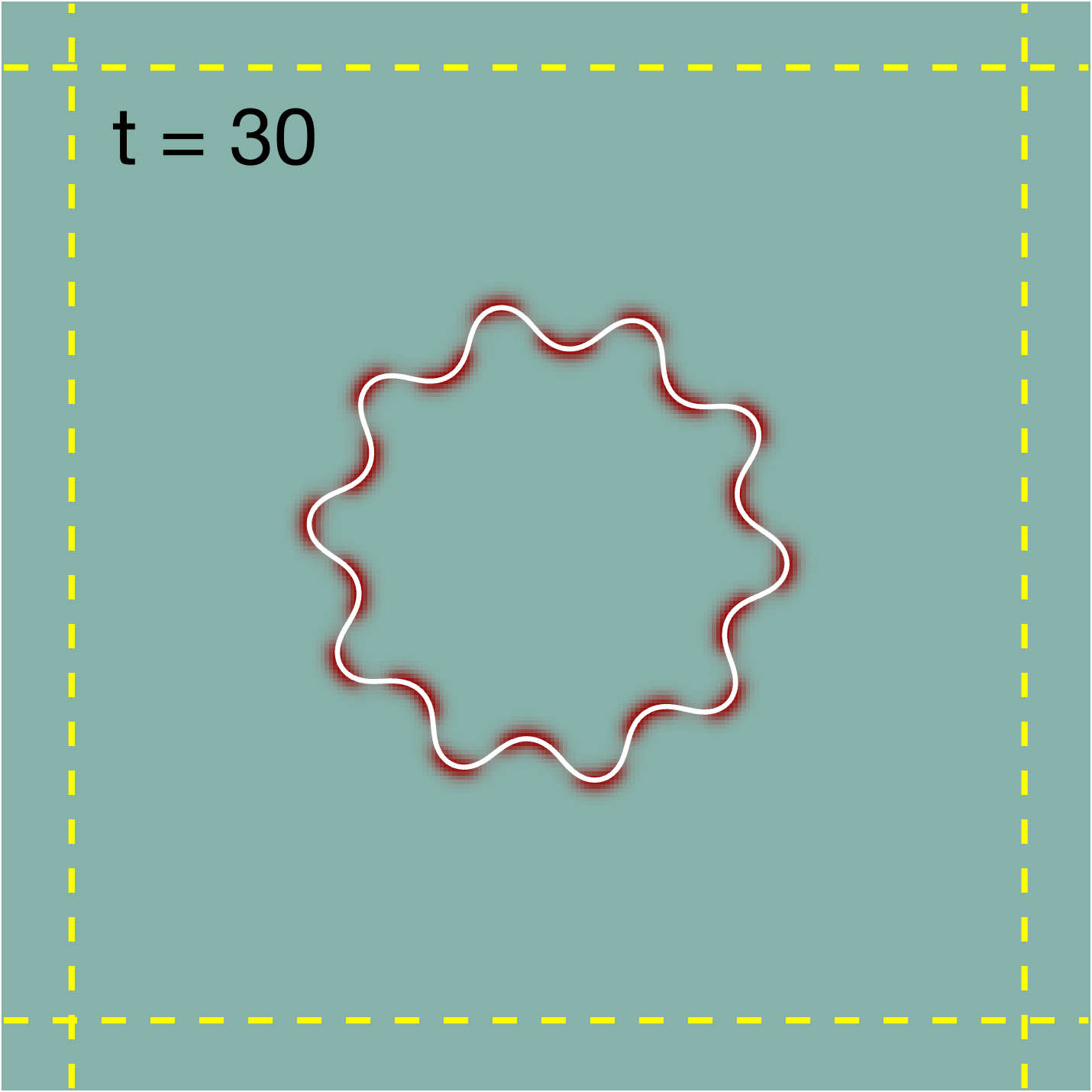}
    \includegraphics[width=0.18\textwidth]{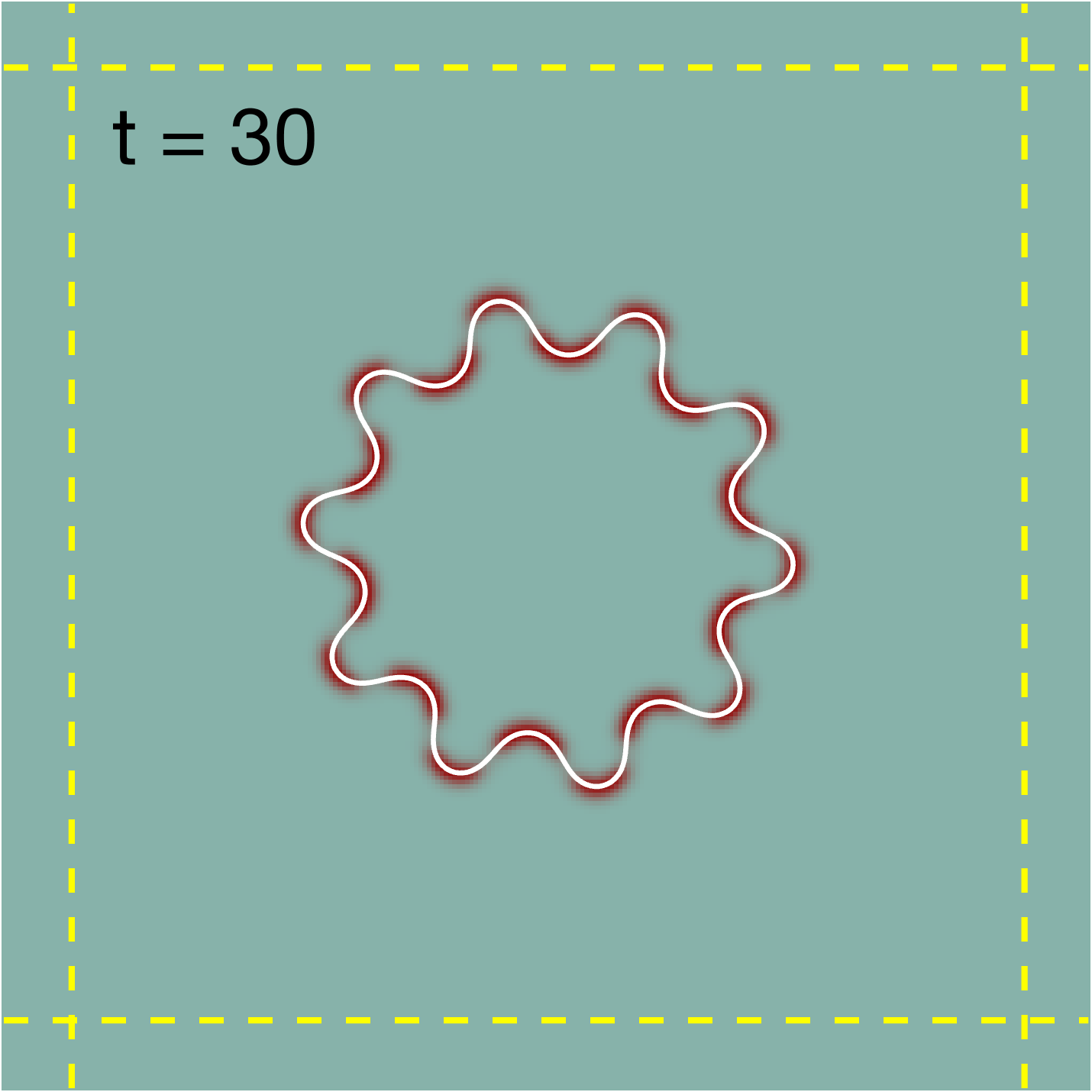}
    \includegraphics[width=0.18\textwidth]{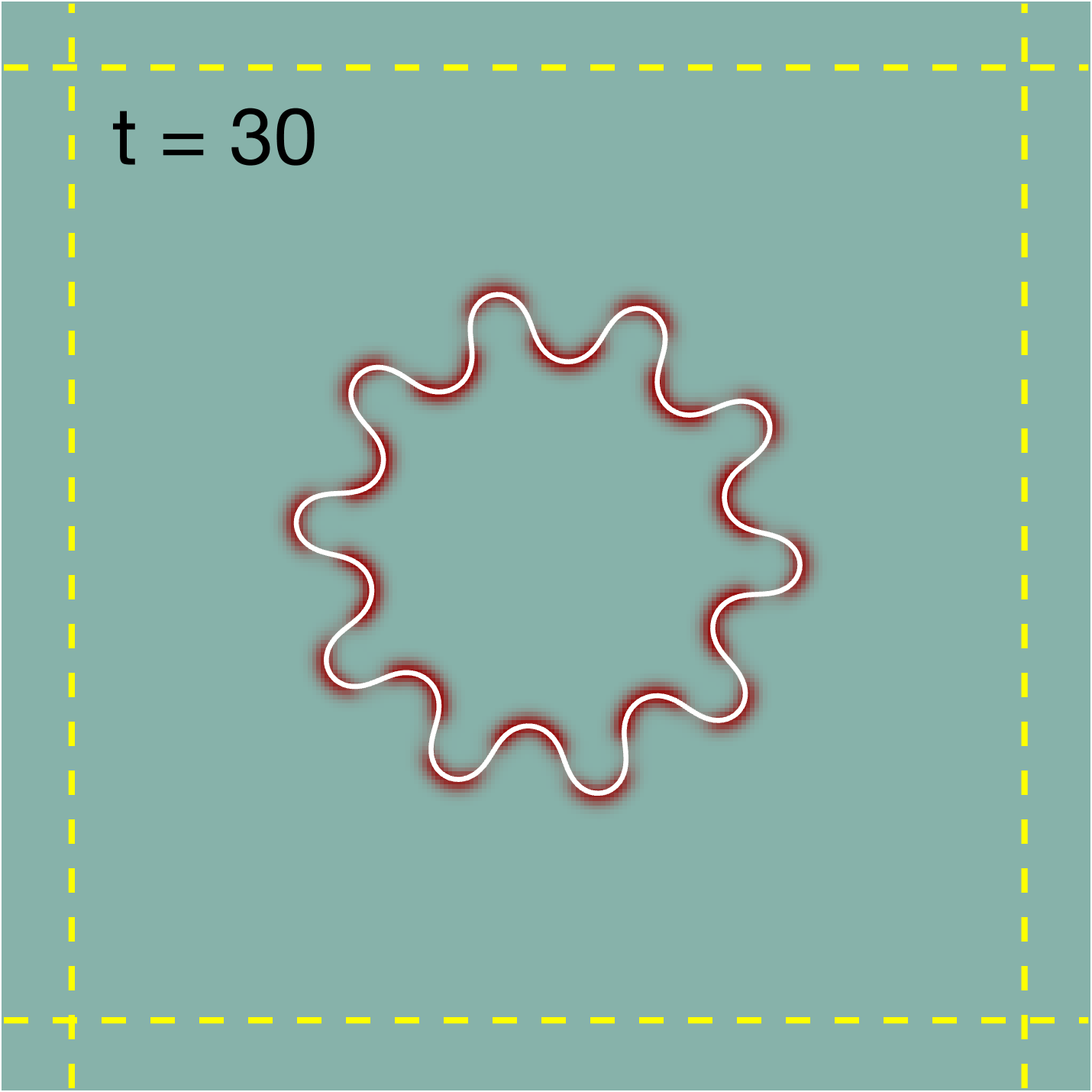}
    \includegraphics[width=0.18\textwidth]{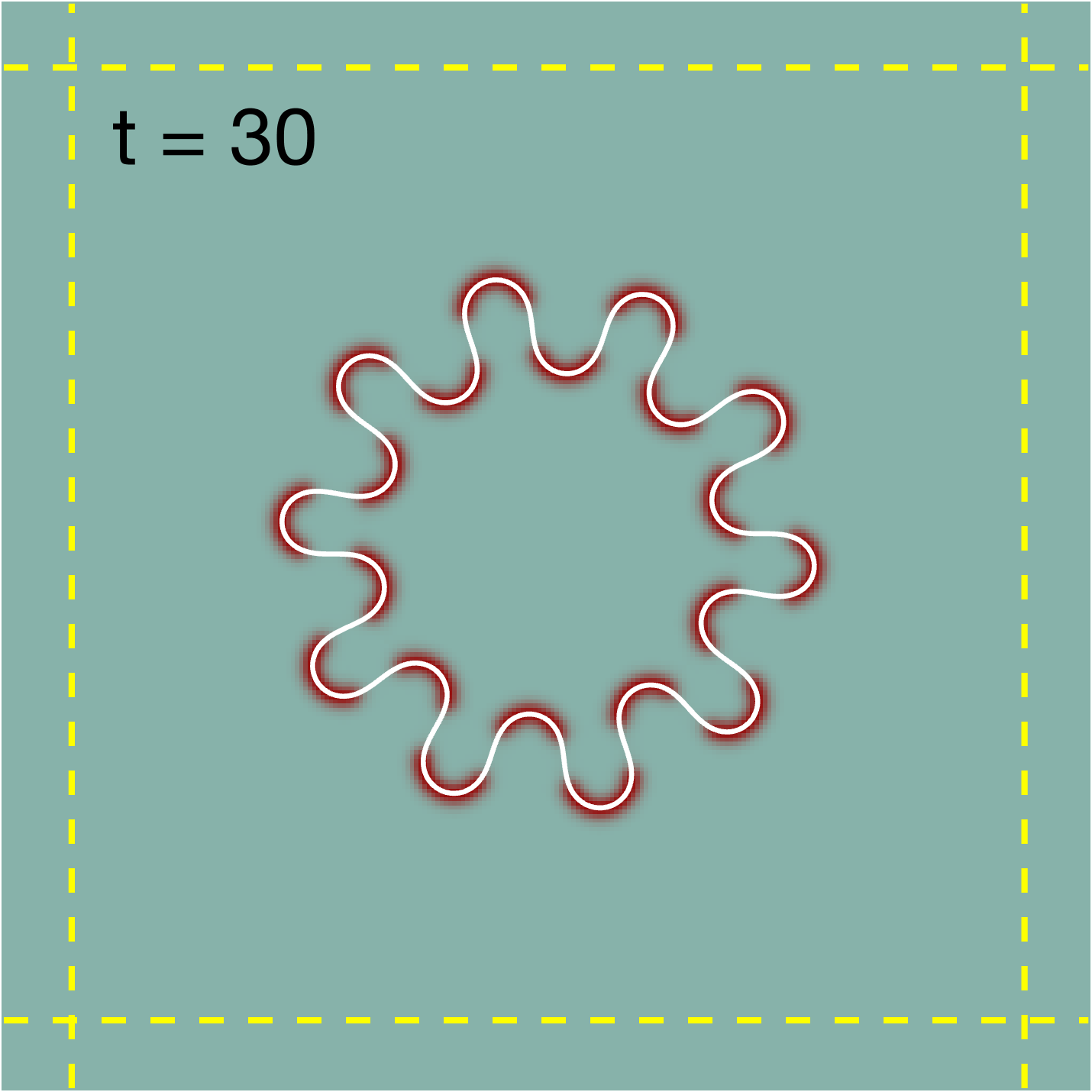}
    \end{center}
    \caption{Larger $\alpha$ induces more prominent fingering pattern. $\alpha = 6,7,8,9,10$ for the ten-finger membrane from left to right. Other parameters are the same as those in the bottom example of Figure \ref{fig:OK_multiplered}: $\lambda_{\mathrm{surf}} = 0.7$, $\lambda_{\mathrm{line}} = 3$, $\kappa=1$, $u_0 = 0$, $\bar{u} = 0.6$, $\gamma = 400$, and $\epsilon_u = 20h_x$. }
    \label{fig:OK_alphaeff}
\end{figure}

Finally, we examine how the parameter $\alpha$ affects the equilibrium morphology of cells with multiple protein-rich subdomains in Figure \ref{fig:OK_alphaeff}. Starting from the same initial as the bottom left one in Figure \ref{fig:OK_multiplered}, we take various values of $\alpha = 6, 7, 8, 9, 10$. Since increasing $\alpha$ amplifies both inward and outward biochemical forces, the cell exhibits greater deformation and more prominent fingering pattern.

\subsection{OK model on the deformable membrane in 3D case}

In this subsection, we extend the proposed model to 3D and compare the numerical results with the biological experiments reported in \cite{baumgart2003imaging}. Due to the high computational cost of 3D simulations, we carefully chose initial conditions for both the phase field cell $\phi$ and the protein density $u$.

\begin{figure}[b]
    \begin{center}
    \includegraphics[width=0.22\textwidth]{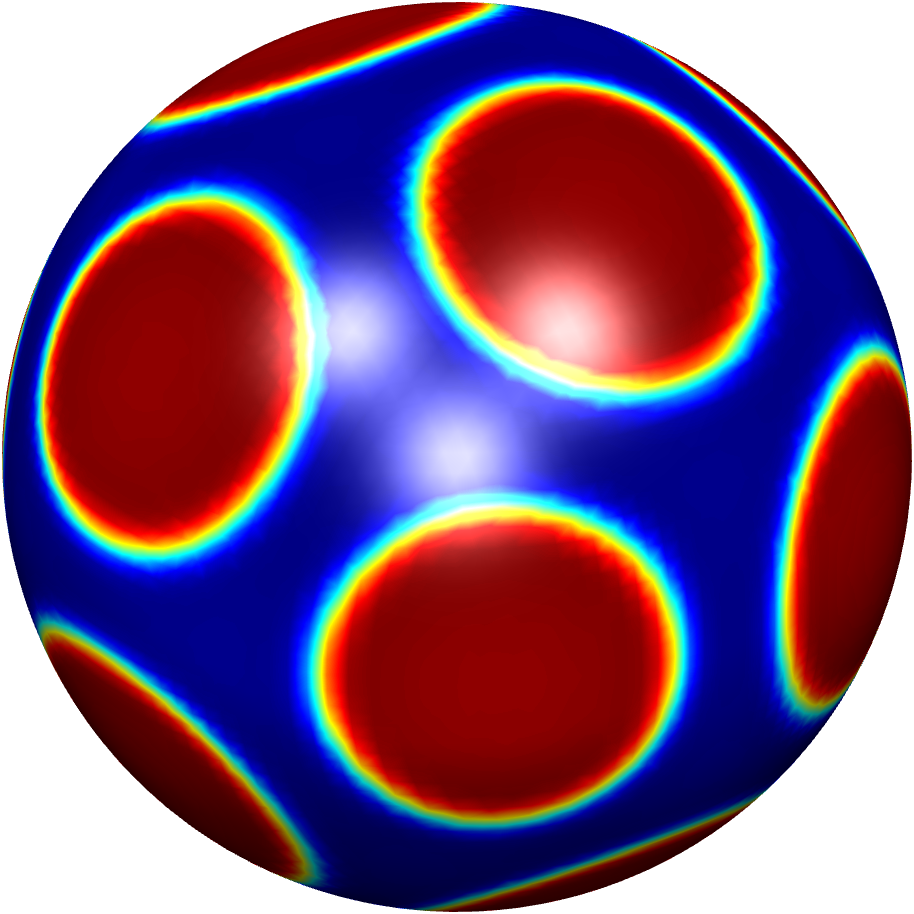}\quad\quad
    \includegraphics[width=0.22\textwidth]{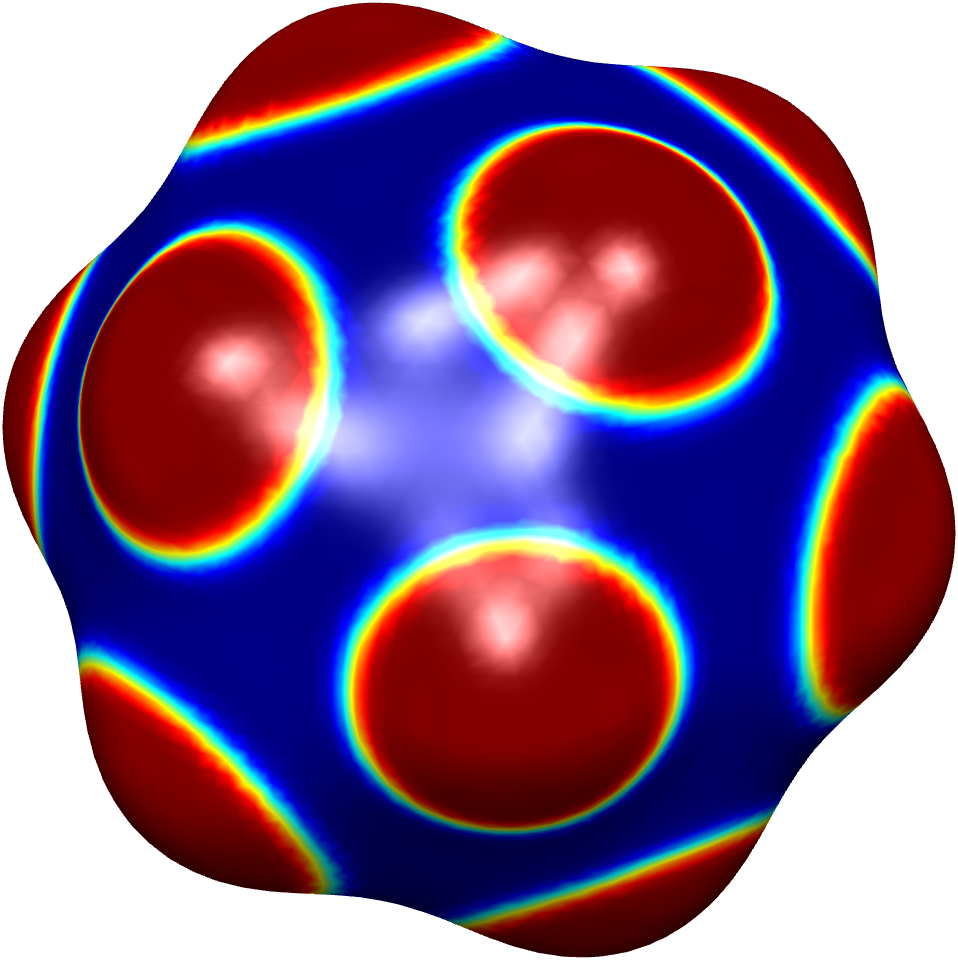}\quad\quad
    \includegraphics[width=0.22\textwidth]{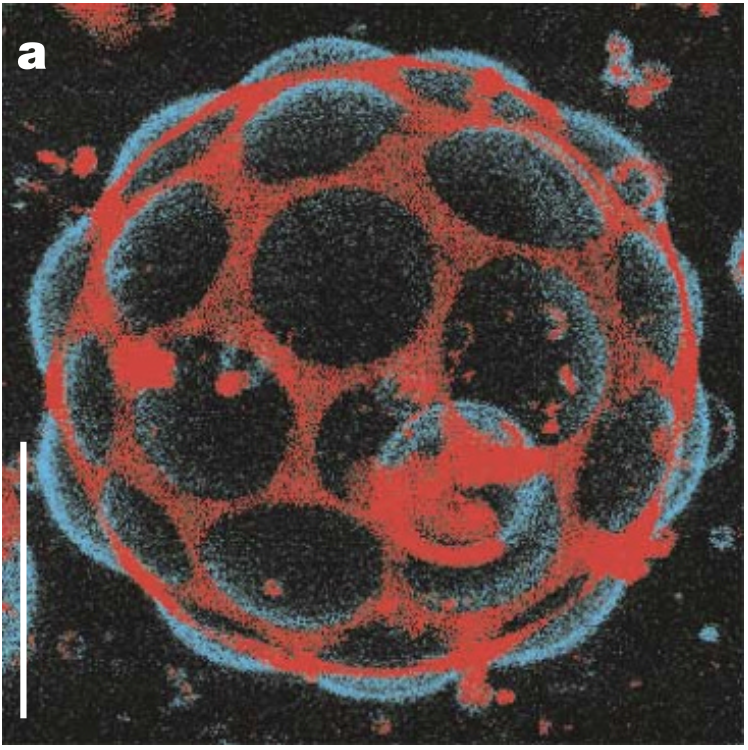}
    \end{center}
    \caption{3D vesicle membrane with multiple protein-rich subdomains (red patches). All patches are of equal size and equally distanced. The parameters are given as: $\lambda_{\mathrm{surf}} = 0.5$, $\lambda_{\mathrm{line}} = 3$, $\kappa=1$, $\alpha = 5$, $u_0 = 0$, $\bar{u} = 0.5$, $\gamma = 180$, and $\epsilon_u = 5h_x$.}
    \label{fig:OK_3dbub}
\end{figure}

For the numerical experiment shown in Figure \ref{fig:OK_3dbub}, the initial phase field variable $\phi^0$ is defined as a sphere centered at the origin with radius $r_0 = 4$:
\begin{align}
        \phi^0(x) = 0.5 + 0.5\tanh{\left(\frac{r_0 - \mathrm{dist}(x,0)}{\epsilon_{\phi}/3}\right)}. \label{eqn:phi_initial_3d}
\end{align}
The initial protein distribution $u^0$ consists of $12$ well-spaced spherical patches on the membrane, representing protein-rich domains. The OK model is first evolved on the fixed membrane \eqref{eqn:phi_initial_3d} up to $t = 1$ (left subfigure), after which the system is coupled with membrane deformation and evolved until $t = 5$ (middle subfigure). The resulting morphology closely resembles the experimental observation in Figure 2(a) of \cite{baumgart2003imaging}.

\begin{figure}[t!]
    \begin{center}
    \includegraphics[width=0.22\textwidth]{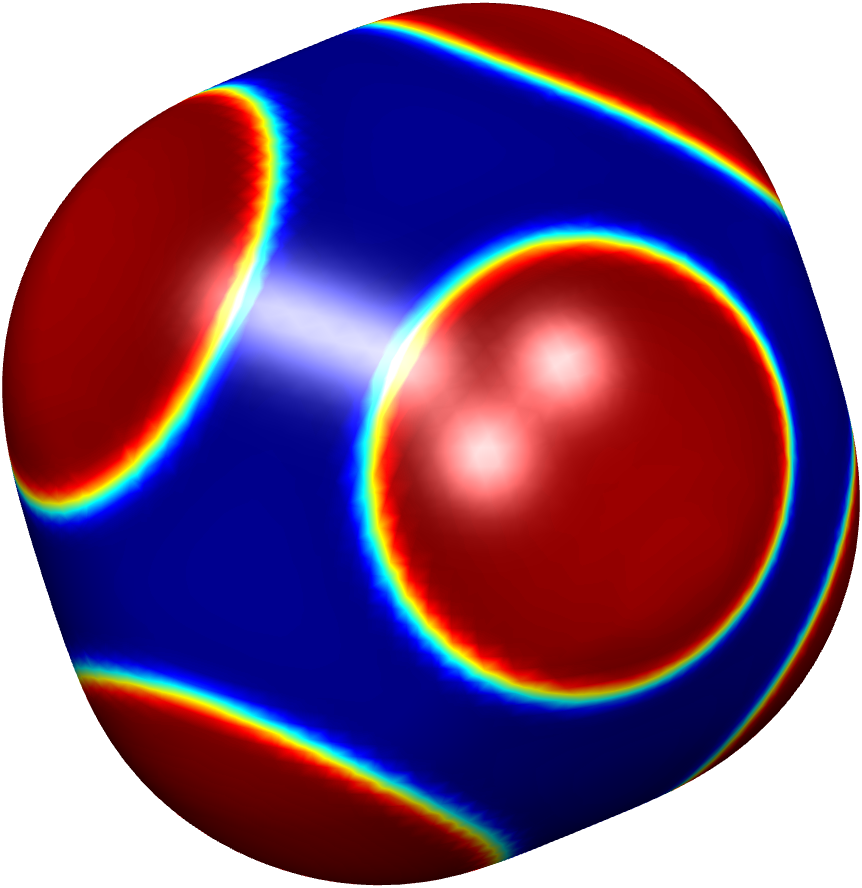}\quad\quad
    \includegraphics[width=0.22\textwidth]{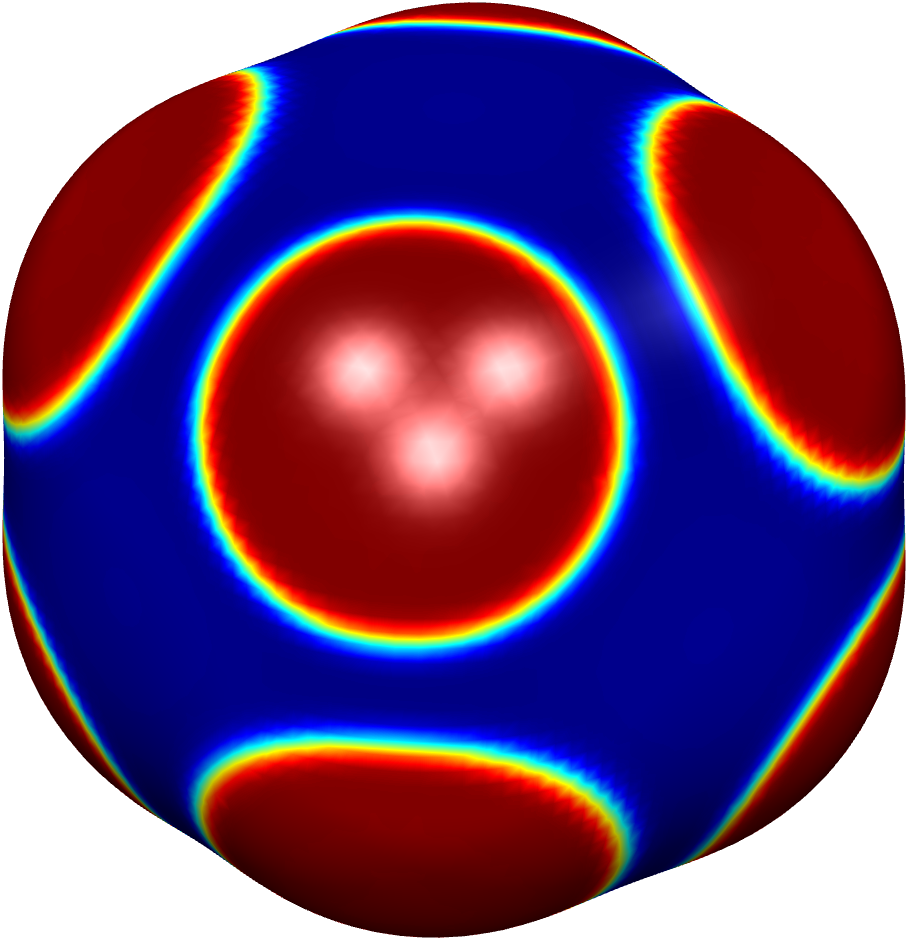}\quad\quad
    \includegraphics[width=0.22\textwidth]{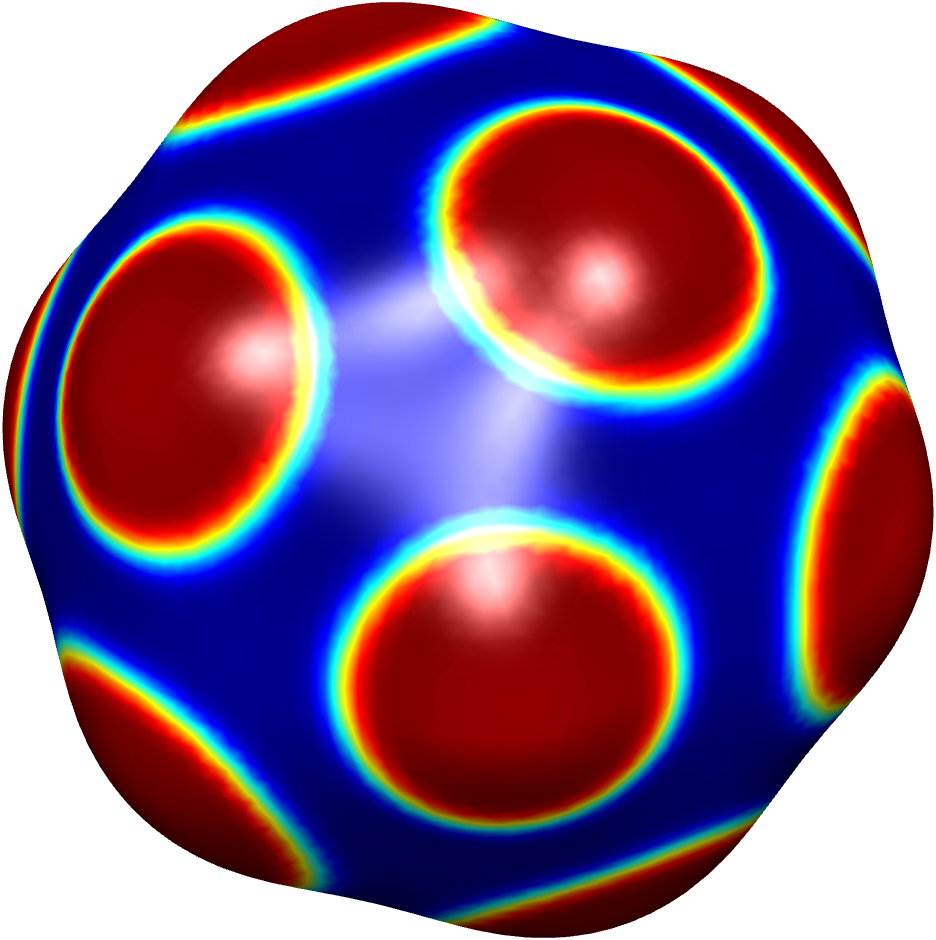}
    \end{center}
    \caption{Effect of $\gamma$ for 3D vesicle with various numbers of protein-rich subdomains (red patches) with $\gamma = 70, 100, 160$, respectively. The other parameters are given as: $\lambda_{\mathrm{surf}} = 1$, $\lambda_{\mathrm{line}} = 3$, $\kappa=1$, $\alpha = 3$, $u_0 = 0$, $\bar{u} = 0.5$, and $\epsilon_u = 5h_x$.}
    \label{fig:OK_gammaef_3d}
\end{figure}

Consistent with the two-dimensional results in Figure \ref{fig:OK_gammaeff}, Figure \ref{fig:OK_gammaef_3d} demonstrates how the repulsive strength $\gamma$ in the membrane-associated OK model influences protein segregation on the membrane surface in three dimensions. The equilibrium states from left to right correspond to $\gamma = 70, 100,$ and $160$, producing six, eight, and twelve protein-rich subdomains (red patches), respectively. Similar to the two-dimensional case, increasing the repulsive parameter $\gamma$ enhances protein segregation, resulting in a larger number of protein-rich domains distributed across the membrane. This behavior reflects the role of $\gamma$ as a measure of long-range repulsion between protein domains, where higher values inhibit coarsening and promote the formation of more microphase-separated patterns. 

\begin{figure}[b!]
    \begin{center}
    \includegraphics[width=0.16\textwidth]{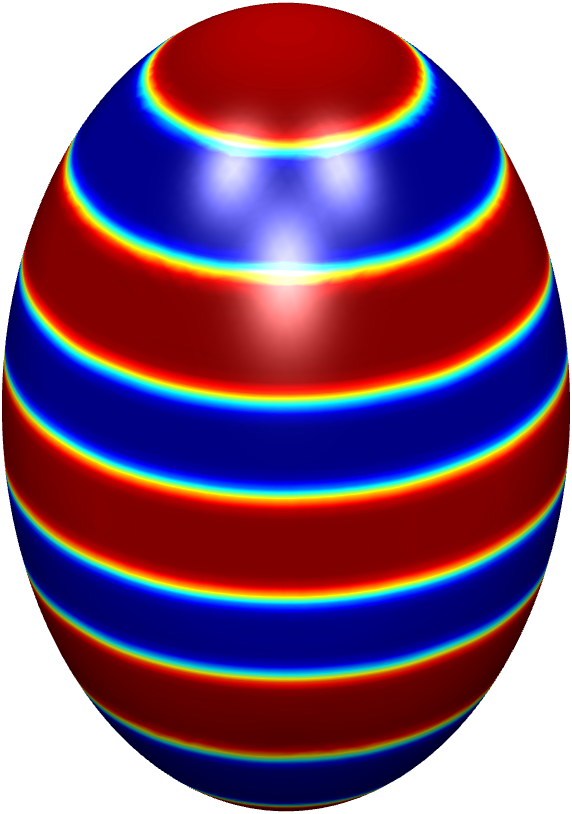}\quad\quad\quad
    \includegraphics[width=0.16\textwidth]{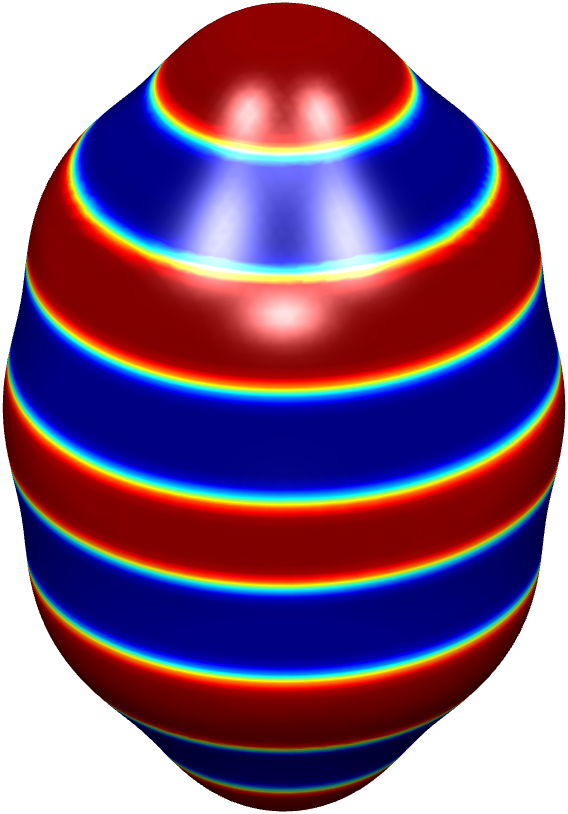}\quad\quad\quad
    \includegraphics[width=0.22\textwidth]{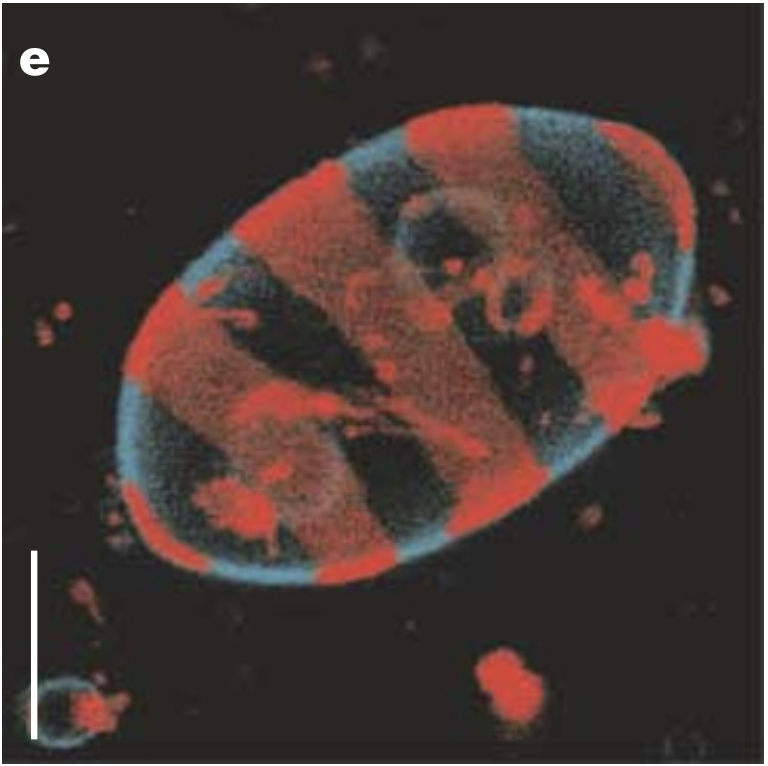}
    \end{center}
    \caption{A vesicle membrane with stripe-type protein-rich subdomains. The parameters are given as: $\lambda_{\mathrm{surf}} = 0.7$, $\lambda_{\mathrm{line}} = 3$, $\kappa=1$, $\alpha = 3$, $u_0 = 7$, $\bar{u} = 0.5$, $\gamma = 80$, and $\epsilon_u = 5h_x$. }
    \label{fig:OK_3dstripe}
\end{figure}

For the numerical experiment shown in Figure \ref{fig:OK_3dstripe}, the initial phase field variable $\phi^0$ is defined as an ellipsoid centered at the origin with radius $r_0 = 4$: 
\begin{align}
\phi^0(x) = 0.5 + 0.5\tanh{\left(\frac{r_0 - \mathrm{d}(x,0)}{\epsilon_{\phi}/3}\right)}, \label{eqn:phi_initial_3d_ellip}
\end{align}
where $\mathrm{d}(x,0)$ for $x = (x_1,x_2,x_3) \in \Omega$ is given by
\begin{align*}
\mathrm{d}(x,0) = \sqrt{x_1^2 + x_2^2 + (0.65x_3)^2}.
\end{align*}
The initial protein distribution $u$ consists of alternating stripe-like bands along the membrane surface, mimicking experimentally observed patterns on ellipsoidal vesicles. The OK model is first evolved on the fixed ellipsoidal membrane \eqref{eqn:phi_initial_3d_ellip} until $t = 1$ (left subfigure), after which the dynamics are coupled with membrane deformation and simulated until $t = 5$ (middle subfigure). The resulting shape agrees well with the experimental result in Figure 2(e) of \cite{baumgart2003imaging}.

\section{Concluding remarks and Outlooks}\label{sec:Remark}

In this paper, we have developed a new computational framework for simulating two-component vesicle membranes, motivated by a new mechanochemical perspective in which membrane-associated proteins laterally segregate on the surface and generate active biochemical forces that deform the membrane. In this framework, membrane evolution is governed by an equation \eqref{eqn:model_force} derived from the force balance between bending, surface tension, area, line tension, and protein-induced chemical forces, and the OK model on the membrane \eqref{eqn:model_OK} captures the microphase separation of membrane-associated proteins. For the numerical implementation, we employed a semi-implicit Fourier spectral method to solve the equation for the membrane deformation and a forward Euler scheme with central difference methods for the membrane-associated OK model. The numerical results are consistent with biological experiments in \cite{baumgart2003imaging} and with earlier diffuse-interface simulations \cite{du2005modeling}, successfully reproducing complex patterns in two-component vesicles. 

The proposed framework provides a powerful platform for studying multicomponent cell shape dynamics and suggests several promising directions for future research. First, we will extend the current formulation into a more comprehensive model by replacing the membrane equation by a transport equation $\partial_t\phi + \mathbf{v}\cdot \nabla \phi = 0$ coupled with a Stokes equation for $\mathbf{v}$. This extension will provide a more physically consistent description of multicomponent membranes interacting with their surrounding environments \cite{cherfils2019compressible}. Second, due to the high computational cost for numerical simulations, another important direction is to develop efficient numerical methods, such as exponential time-differencing schemes combined with adaptive time-stepping strategies, and boundary integral equation methods, together with rigorous stability and convergence analysis. Finally, theoretical studies of the sharp-interface limit of this diffuse-interface framework will be conducted in the future, potentially providing deeper analytical insights into the modeling of membrane morphology.

\section*{Acknowledgments} 

{W. Luo was partially supported by the CAS AMSS-PolyU Joint Laboratory of Applied Mathematics (No. JLFS/P-501/24).
Z. Qiao was partially supported by the Hong Kong Research Grants Council RFS grant RFS2021-5S03, GRF grant 15305624 and NSFC/RGC Joint Research Scheme (No. N\_PolyU5145/24).}

% \newpage

% %%%%%%%%%%%%%%%%%%%%%%%%%%%%%%%%%%%%%%%%%%%%%%%%%%%%%%%%%%%%%%%%%%%%%%%%%%%%%%%%%

%\section*{References}

% \bibliography{OhtaKawasaki}
\bibliography{citation}

\end{document}